\documentclass[a4paper,twocolumn,11pt,accepted=2024-02-19]{quantumarticle}
\pdfoutput=1
\usepackage{graphicx,amsmath,amsfonts,amssymb,amsthm,xr}
\usepackage{epsfig,amssymb,color,dsfont,upgreek,physics}
\usepackage{xkeyval,etoolbox,geometry,fancyhdr,ltxgrid,ltxcmds}
\usepackage{mathrsfs}
\usepackage{mathtools}
\usepackage{bbold}
\usepackage{comment}
\usepackage{float}
\usepackage{braket}
\usepackage[bookmarks=true,colorlinks,linkcolor=quantumviolet,urlcolor=quantumviolet,citecolor=quantumviolet]{hyperref}
\usepackage[dvipsnames]{xcolor}
\usepackage{tikz}
\usepackage[normalem]{ulem}
\usepackage[square,sort&compress,numbers]{natbib}
\usepackage[sectionbib]{bibunits}
\defaultbibliography{references}

\newcommand{\SP}{{\hat{\mathcal{L}}}}
\newcommand{\ZT}{{\hat{\mathcal{Z}}}}
\newcommand{\ST}{{\hat{\mathcal{T}}}}

\begin{document}
\title{Ergodicity Breaking Under Confinement in Cold-Atom Quantum Simulators}
\author{Jean-Yves Desaules}
\affiliation{School of Physics and Astronomy, University of Leeds, Leeds LS2 9JT, UK}
\orcid{0000-0002-3749-6375}

\author{Guo-Xian Su}
\affiliation{Hefei National Laboratory for Physical Sciences at Microscale and Department of Modern Physics, University of Science and Technology of China, Hefei, Anhui 230026, China}
\affiliation{Physikalisches Institut, Ruprecht-Karls-Universit\"at Heidelberg, Im Neuenheimer Feld 226, 69120 Heidelberg, Germany}
\affiliation{CAS Center for Excellence and Synergetic Innovation Center in Quantum Information and Quantum Physics, University of Science and Technology of China, Hefei, Anhui 230026, China}
\orcid{0000-0001-7936-762X}

\author{Ian P.~McCulloch}
\affiliation{School of Mathematics and Physics, The University of Queensland, St. Lucia, QLD 4072, Australia}
\orcid{0000-0002-8983-6327}

\author{Bing Yang}
\affiliation{Department of Physics, Southern University of Science and Technology, Shenzhen 518055, China}
\orcid{0000-0002-8379-9289}

\author{Zlatko Papi\'c}
\affiliation{School of Physics and Astronomy, University of Leeds, Leeds LS2 9JT, UK}
\orcid{0000-0002-8451-2235}

\author{Jad C.~Halimeh}
\email{jad.halimeh@physik.lmu.de}
\affiliation{Department of Physics and Arnold Sommerfeld Center for Theoretical Physics (ASC), Ludwig-Maximilians-Universit\"at M\"unchen, Theresienstra\ss e 37, D-80333 M\"unchen, Germany}
\affiliation{Munich Center for Quantum Science and Technology (MCQST), Schellingstra\ss e 4, D-80799 M\"unchen, Germany}
\orcid{0000-0002-0659-7990}

\begin{abstract}
The quantum simulation of gauge theories on synthetic quantum matter devices has gained a lot of traction in the last decade, making possible the observation of a range of exotic quantum many-body phenomena. In this work, we consider the spin-$1/2$ quantum link formulation of $1+1$D quantum electrodynamics with a topological $\theta$-angle, which can be used to tune a confinement-deconfinement transition. Exactly mapping this system onto a PXP model with mass and staggered magnetization terms, we show an intriguing interplay between confinement and the ergodicity-breaking paradigms of quantum many-body scarring and Hilbert-space fragmentation. We map out the rich dynamical phase diagram of this model, finding an ergodic phase at small values of the mass $\mu$ and confining potential $\chi$, an emergent integrable phase for large $\mu$, and a fragmented phase for large values of both parameters. We also show that the latter hosts resonances that lead to a vast array of effective models. We propose experimental probes of our findings, which can be directly accessed in current cold-atom setups.
\end{abstract}

\maketitle

\renewcommand{\baselinestretch}{0.7}\normalsize
\tableofcontents
\renewcommand{\baselinestretch}{1.0}\normalsize

\section{Introduction}
Gauge theories are quantum many-body models possessing gauge symmetries that dictate an intrinsic local relationship between matter and gauge fields \cite{Weinberg_book,Gattringer_book,Zee_book}. The quantum simulation of gauge theories has progressed tremendously in recent years across various platforms of synthetic quantum matter \cite{Martinez2016,Muschik2017,Bernien2017,Klco2018,Kokail2019,Schweizer2019,Goerg2019,Mil2020,Klco2020,Yang2020,Zhou2021,Nguyen2021,Wang2022,Mildenberger2022}. This offers the opportunity to probe high-energy physics on low-energy tabletop platforms \cite{Alexeev_review,klco2021standard,Dalmonte_review,Zohar_review,aidelsburger2021cold,Zohar_NewReview,Bauer_review,Catterall2022}, establishing a valuable complementary venue to dedicated high-energy experiments, such as the Large Hadron Collider, and to classical computational methods, such as Quantum Monte Carlo and tensor networks. As an example, a cold-atom quantum simulator has recently been successfully employed in probing gauge invariance and thermalization dynamics in a $\mathrm{U}(1)$ gauge theory \cite{Yang2020,Zhou2021}, with concrete proposals \cite{Halimeh2022tuning,Cheng2022tunable} of extending it to observe the confinement-deconfinement transition \cite{Buyens2016,Surace2020}. In particular, Ref.~\cite{Halimeh2022tuning} introduced a scheme that allows the realization of a tunable topological $\theta$-angle term, which emerges in quantum electrodynamics \cite{Byrnes2002,Buyens2014,Shimizu2014} on account of the topological structure of the vacuum, and has profound consequences on quantum phases, dynamical behavior, and inherent symmetries. Tuning this angle can lead to confinement, an extremely active topic of research in gauge theories \cite{Borla2020Confined,kebric2021confinement,Mildenberger2022} and other many-body spin systems \cite{Kormos2017Confinement,Liu2019Confined,Bastianello2022Fragmentation,Bastianello2022Prethermalization}, and thus a tunable topological $\theta$-angle term in a quantum simulator can allow the calculation of the time evolution of confined dynamics from first principles.

\begin{figure}[t!]
	\centering
	\includegraphics[width=\linewidth]{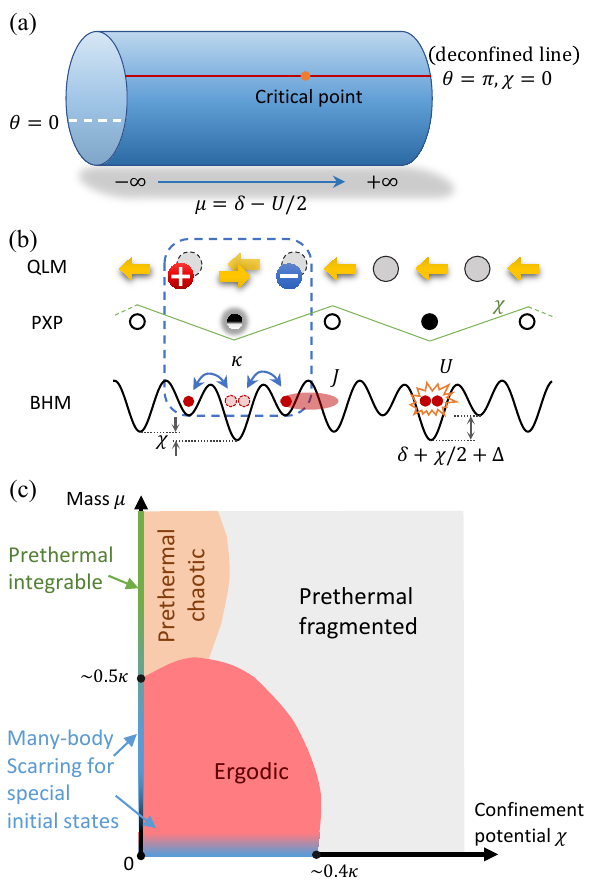}
	\caption{Tuning the topological $\theta$-angle in the Bose--Hubbard quantum simulator. (a) The $\mathrm{U}(1)$ quantum link model Hamiltonian~\eqref{eq:QLM} involves a topological $\theta$-angle, where charges are confined at $\theta{=}\pi$ in the limit of large negative mass, but otherwise deconfined. Also at $\theta=\pi$ there is Coleman's phase transition, which occurs at effective mass $\mu{=}0.3275\kappa$, corresponding to the spontaneous breaking of a global $\mathbb{Z}_2$ symmetry \cite{Coleman1976} that arises from the invariance of Hamiltonian~\eqref{eq:QLM} to transformations of parity and charge conjugation. (b) The staggered Bose--Hubbard model realizes the $\mathrm{U}(1)$ quantum link model with direct tunneling $J$, on-site interaction $U$, staggered potential $\delta$, and a small linear tilting potential $\Delta$. By further adding a period-$4$ potential $\chi$, the system is tuned away from $\theta{=}\pi$, leading to confinement. The $\mathrm{U}(1)$ quantum link model or equivalently the staggered Bose--Hubbard model can be further mapped onto the PXP model where spins reside on the gauge links, and doublons on the gauge links represent spin excitations (the mapping will be discussed in detail in Sec.~\ref{sec:model}). (c) Schematic representation of the rich phase diagram in terms of confinement potential $\chi$ and effective mass $\mu$ (see Appendix~\ref{app:rstat} for details). 
    }
	\label{fig:schematic}
\end{figure}

The understanding of non-equilibrium properties of gauge theories sheds light on a variety of dynamical phenomena across high-energy particle physics, condensed matter and even the evolution of the early universe. It is thus important to study the ergodicity-breaking mechanisms that can appear in these theories.
Gauge-theory quantum simulators offer the prospect of probing ergodicity-breaking phenomena that are relevant to condensed matter physics, such as disorder-free localization \cite{Smith2017,Brenes2018,smith2017absence,Metavitsiadis2017,Smith2018,Russomanno2020,Papaefstathiou2020,McClarty2020,hart2021logarithmic,Zhu2021,Sous2021,karpov2021disorder,Chakraborty2022,Halimeh2022TDFL}, quantum many-body scars (QMBS) \cite{Bernien2017,Surace2020,Moudgalya2018,Turner2018}, and Hilbert-space fragmentation (HSF) \cite{Sala2020,Khemani2020}, the understanding of which can provide a deeper generic picture of far-from-equilibrium dynamics in closed quantum many-body systems \cite{Rigol_review,Deutsch_review,Buca2023unified}. QMBS, a paradigm of weak ergodicity breaking, comprise states of anomalously low bipartite entanglement entropy that are usually equally spaced in energy across the entire spectrum of the Hamiltonian. Preparing a system in their subspace and quenching it will lead to persistent oscillations in local observables, an anomalously slow growth of bipartite entanglement entropy, and a corresponding significant delay in thermalization beyond any relevant timescales \cite{Bernien2017,Turner2018,Moudgalya2018,Surace2020}. Indeed, Bernien \textit{et al.}~\cite{Bernien2017} was the first experiment that observed QMBS in Rydberg arrays, and Surace \textit{et al.}~\cite{Surace2020} were able to show that the implemented quantum Ising-like model maps onto the spin-$1/2$ $\mathrm{U}(1)$ quantum link model, which is a regularization of the lattice Schwinger model where the infinite-dimensional gauge fields of the latter are represented by spin-$1/2$ operators \cite{Chandrasekharan1997,Wiese_review,Kasper2017}. More recently, it has been shown that QMBS are ubiquitous in gauge theories, occurring over an extensive set of initial states \cite{Su2022,Hudomal2022}, in $2{+}1$D \cite{Banerjee2020}, and for spin-$S$ representations of the gauge field when the system is prepared in highly excited vacua \cite{Desaules2022weak,Desaules2022prominent}. 
This raises the question about the gauge-theoretic origin of QMBS and motivates their further exploration in realistic gauge-theory quantum simulators.

On the other hand, HSF is a mechanism of ergodicity breaking emerging in models where the Hilbert space is fragmented into exponentially many invariant subspaces due to a exponentially large commutant algebra \cite{Sala2020,Khemani2020,Moudgalya2021}. In the case of dipole-conserving models \cite{Sala2020,Khemani2020}, HSF can be completely characterized by nonlocal integrals of motion \cite{Rakovszky2020}. More recently, HSF has also been shown to emerge in models with strict confinement \cite{DeTomasi2019,Yang2020fragmentation}, including in gauge theories \cite{Chen2021}. Experimental investigations of HSF have been carried out \cite{Scherg2021,Kohlert2023exploring}, but exploring its presence in experimentally relevant parameter regimes of gauge theories would be very appealing, given recent large-scale cold-atom realizations of spin-$1/2$ $\mathrm{U}(1)$ quantum link models \cite{Yang2020,Zhou2021}. In this vein, one wonders whether QMBS can occur in regimes of confinement in gauge theories where HSF also arises, or whether scars are a purely deconfined phenomenon. Indeed, while confinement has previously been linked to the appearance of non-thermal eigenstates in other theories with confinement \cite{James2019,Robinson2019} and to oscillations of local observables \cite{Mazza2019,Lerose2019}, a formal connection with QMBS has not been established. In the case of lattice gauge theories, Ref.~\cite{Surace2020} demonstrated a slowing down of the dynamics under confinement, while Ref.~\cite{Chen2021} showed the emergence of a symmetry in the limit of strong confinement at zero mass. However, a full analysis of the fate of scarring with both mass and confining potential at intermediate values of experimental interest is lacking.

In this work, we explore the regime of finite mass (denoted by $\mu$) and confining potential (denoted by $\chi$) using exact diagonalization and matrix product state (MPS) techniques~\cite{Uli_review,Paeckel_review}. 
We study the PXP model, which exactly corresponds to the spin-1/2 $\mathrm{U}(1)$ quantum link model (QLM)~\cite{Surace2020}. Both have been realized experimentally on a Bose--Hubbard quantum simulator \cite{Su2022,Zhou2021}. The mapping between the different models is sketched in Fig.~\ref{fig:schematic}(a)-(b).
We find prominent scarring regimes concomitantly with confinement, which we explain analytically and demonstrate numerically. We further map out the rich dynamical phase diagram where we find an ergodic phase for small values of the mass and confining potential. 
For larger values of these parameters, emergent conserved quantities at intermediate times appear, displaying a wide range of prethermal regimes with  an emergent integrable phase for large mass in the deconfined regime, in addition to a fragmented phase for large values of both the mass and the confining potential, as sketched in Fig.~\ref{fig:schematic}(c) (see Appendix~\ref{app:rstat} for more details). We note that the emergent conserved quantities are only truly conserved for infinite values of $\mu$ and $\chi$. However, for smaller values of these parameters, a long time can be needed to propagate between Hilbert space fragments, hence we used the prefix  ``prethermal'' to reflect this. We note that in smaller systems this timescale can already be too long to probe in experiment, leading to strong signs of fragmentation at all achievable times. 

We will focus on the experimentally relevant timescales, and in that context we will study two regions of Fig.~\ref{fig:schematic}(c). We will first assess the effect of confinement on quantum many-body scarring when the mass is set to zero, at the bottom left of Fig.~\ref{fig:schematic}(c). We will then turn to the top right region where the mass and confining potential are of similar order and Hilbert space fragmentation initially dominates the dynamics. We will show that this region is not homogeneous but hosts special dynamics at different resonances between $\mu$ and $\kappa$. We will also show that the integrable and chaotic prethermal regimes at the top left can be understood as such resonances.

The remainder of this paper is organized as follows: In Sec.~\ref{sec:model}, we introduce the $\mathrm{U}(1)$ quantum link model with a topological $\theta$-term and discuss its exact mapping onto the PXP model with a mass and staggered magnetization terms. In Secs.~\ref{sec:QMBS} and~\ref{sec:HSF}, we provide our analysis of scarring and fragmentation dynamics, respectively, in the presence of confinement. In Sec.~\ref{sec:experiment}, we discuss experimental probes in a state-of-the-art 
tilted Bose--Hubbard optical lattice that can facilitate the observation of our findings. We conclude and provide outlook in Sec.~\ref{sec:conc}. Several Appendices and the Supplemental Material (SM)~\cite{SM} contain supporting analytic and numerical results.

\section{Models and mapping}\label{sec:model}

For completeness, we start by introducing the models considered in this work and briefly outline the mappings between them. This section does not contain new results and the reader is referred to Refs.~\cite{Surace2020,Halimeh2022tuning} for derivations of the mappings.

\subsection{\texorpdfstring{$\mathrm{U}(1)$}{U(1)} quantum link model}

The Schwinger model on a lattice, which is $1+1$D quantum electrodynamics, can be conveniently represented through the quantum link formulation, where the gauge and electric fields are represented by spin-$S$ matrices \cite{Chandrasekharan1997,Wiese_review,Kasper2017}, thereby facilitating experimental feasibility. Upon restricting to $S=1/2$, and employing a Jordan-Wigner transformation to map the fermions onto Pauli operators and a particle-hole transformation on both the matter and gauge (electric) fields \cite{Yang2016,Halimeh2022tuning}, the resulting $\mathrm{U}(1)$ quantum link model (QLM) Hamiltonian takes the form
\begin{align}\nonumber
    \hat{H}_\text{QLM}=\sum_{\ell=1}^{L_m}\bigg[&{-}\frac{\kappa}{2}\big(\hat{\sigma}^-_\ell\hat{s}^+_{\ell,\ell+1}\hat{\sigma}^-_{\ell+1}+\text{H.c.}\big)\\\label{eq:QLM}
    &+\frac{\mu}{2}\hat{\sigma}^z_\ell-\chi(-1)^\ell\hat{s}^z_{\ell,\ell+1}\bigg],
\end{align}
where now the Pauli operator $\hat{\sigma}^z_\ell$ represents the matter field on site $\ell$, and the spin-$1/2$ operator $\hat{s}^{+(z)}_{\ell,\ell+1}$ represents the gauge (electric) field at the link between sites $\ell$ and $\ell+1$. We consider a lattice with $L_m$ sites and $L_m$ links under periodic boundary conditions (PBC). The deviation from the deconfined point $\theta{=}\pi$ is quantified by the \textit{confining potential} $\chi{=}g^2(\theta{-}\pi){/}(2\pi)$ \cite{Halimeh2022tuning}. 

At $\chi=0$, the $\mathrm{U}(1)$ QLM undergoes a second-order quantum (Coleman) phase transition at $\mu/\kappa=0.3275$ related to the spontaneous breaking of a global $\mathbb{Z}_2$ symmetry arising due to the invariance of Eq.~\eqref{eq:QLM} under a parity and charge (CP) transformation at $\chi=0$ \cite{Coleman1976,Rico2014,VanDamme2020}. When $\chi\neq0$, the global $\mathbb{Z}_2$ symmetry is explicitly broken by the last term in Eq.~\eqref{eq:QLM}, and the Coleman phase transition is no longer present. The generator of the $\mathrm{U}(1)$ gauge symmetry of Hamiltonian~\eqref{eq:QLM} is
\begin{align}\label{eq:G}
    \hat{G}_\ell=(-1)^\ell\bigg(\hat{s}^z_{\ell,\ell+1}+\hat{s}^z_{\ell-1,\ell}+\frac{\hat{\sigma}^z_\ell+\hat{\mathds{1}}}{2}\bigg),
\end{align}
which can be construed as a discretized version of Gauss's law. Throughout this work, we will work in the physical sector of Gauss's law, defined by the set of gauge-invariant states $\{\ket{\psi}\}$ satisfying $\hat{G}_\ell\ket{\psi}{=}0,\,\forall\ell$.

Despite being a regularization of the Schwinger model, the $\mathrm{U}(1)$ QLM~\eqref{eq:QLM} still captures a wealth of the physics of the Schwinger model, including Coleman's phase transitions at $\chi=0$ \cite{Coleman1976}, and the confinement-deconfinement transition \cite{Coleman1975}.

\subsection{PXP Hamiltonian}

The $\mathrm{U}(1)$ QLM with a topological $\theta$-term, described by Hamiltonian~\eqref{eq:QLM}, can be mapped onto the PXP model with a staggered-magnetization term \cite{Surace2020,Desaules2022weak,Desaules2022prominent}, described by the Hamiltonian
\begin{align}\nonumber
    \hat{H}_\mathrm{PXP}&{=}{-}\kappa\,\hat{\mathcal{P}}\bigg(\sum_{l=1}^{L_m} \hat{s}^x_{l}\bigg)\hat{\mathcal{P}}
    {-}\sum_{l=1}^{L_m} \left[2\mu+ (-1)^l \chi\right] \hat{s}^z_{l} \\\label{eq:H_spin}
    &{=}{-}\kappa\,\sum_{l=1}^{L_m} \hat{P}_{l-1}\hat{s}^x_{l}\hat{P}_{l+1}
    {-}\sum_{l=1}^{L_m} \left[2\mu+ (-1)^l \chi\right] \hat{s}^z_{l},
\end{align}
where we have assumed PBC, integrated out the degrees of freedom on the matter sites, and relabeled the gauge sites as $\hat{s}^\alpha_{l,l+1}\to \hat{s}^\alpha_{l}$ for notational brevity.\footnote{For open boundary conditions, Eqs.~\eqref{eq:QLM} and~\eqref{eq:H_spin} can also be mapped onto each other, where the results of the latter are similar to those obtained for the $\mathrm{U}(1)$ QLM, but with $\mu$ and $\chi$ halved on the two boundary spins.}
The projector $\hat{\mathcal{P}}$ annihilates any configuration with neighboring up-spins, in order to only allow configurations that respect Gauss's law. The Hamiltonian also admits a local formulation using the on-site projector $\hat{P}_l=\frac{1}{2}-\hat{s}^z_l$ that annihilates any component with $s^z_l=+1/2$. This Hamiltonian corresponds to the PXP model with detuning $2\mu$ and staggered detuning $\chi$. For $\chi=0$, the PXP model has been extensively studied in the context of QMBS both theoretically and experimentally \cite{Bernien2017,Turner2018,Surace2020,Su2022}. It has also been studied in the context of HSF in the limit of $\chi/\kappa\gg1$ \cite{Chen2021}.

In the following sections, we will explore both the ergodic and fragmented regimes of the PXP model with a staggered magnetization term. The approximate values of $\mu$ and $\chi$ to which these correspond for $L_m\approx 30$ are schematically illustrated in Fig.~\ref{fig:schematic} (see Appendix \ref{app:rstat} for actual data).

\section{Quantum many-body scarring in the presence of confinement}\label{sec:QMBS}

For simplicity, we shall set $\kappa=1$ as the overall energy scale and focus on the PXP model~\eqref{eq:H_spin} in this section. The PXP model is known to host QMBS associated with the N\'eel state $\ket{{\circ}{\bullet}\ldots {\circ}{\bullet}}$  and anti-N\'eel state $\ket{{\bullet}{\circ}\ldots {\bullet}{\circ}}$ at $\mu=\chi=0$~\cite{Bernien2017,Turner2018}. Scarring was also observed experimentally from the polarized state $\ket{{\circ}{\circ}\ldots {\circ}{\circ}}$ for $\mu\approx \pm 0.4$~\cite{Su2022}. Here, for historical purposes \cite{Bernien2017}, a filled (empty) circle denotes a Rydberg atom in the excited (ground) state, but one can equivalently think of it as a spin-$1/2$ particle in the up (down) eigenstate. 
The N\'eel and anti-N\'eel states correspond to the two vacua of the PXP model, i.e., its degenerate ground states at $\mu\to\infty$, while the polarized state corresponds to the charge-proliferated state, which is the non-degenerate ground state of the PXP model at $\mu\to-\infty$. These two realizations of QMBS occur in the ergodic regime at the bottom left of Fig.~\ref{fig:schematic}(c). This means that other initial states display the expected thermalizing behavior, while conventional probes of chaos, such as level spacing statistics, are consistent with those of a chaotic system. In this section, we focus on this region and discuss the impact of confinement on QMBS. One would generally expect confinement to strongly modify the dynamics and lead to the disappearance of revivals from scarred initial states. While this indeed happens for the polarized state at intermediate values of $\chi$,  we will show that this is not true for the N\'eel state, where the confining term has an interesting nonmonotonic effect and leads to a pronounced \emph{enhancement} of scarring at intermediate values. In this section, we first present the numerical results illustrating the effect of confinement on the QMBS dynamics and entanglement growth. We then provide analytical explanation of the observed enhancement of QMBS revivals. Finally, we discuss the effect of confinement on the dynamics of defects in QMBS initial states. 

\subsection{Effect of confinement on scarred revivals and entanglement growth}\label{sec:effect}

In order to assess the effect of the confining potential on QMBS dynamics, we compute the self-fidelity $\mathcal{F}(t){=}|\bra{\psi_0} e^{-i\hat{H}_\mathrm{PXP}t} \ket{\psi_0}|^2$, where $\ket{\psi_0}$ is the initial state. 
In a generic many-body system, the fidelity starts out at $\mathcal{F}(t=0)=1$ and then decays to a value exponentially small in system size, $\mathcal{F} \sim \exp(-L_m)$, after a time $\sim \hbar/\kappa$. In a QMBS system, following the initial decay, the fidelity periodically rises again to a value  $\sim O(1)$, corresponding to the revivals of the wave function. 
We will denote the first revival peak of the self-fidelity by $\mathcal{F}_1$, and the self-fidelity at half that time by $\mathcal{F}_{1/2}$. 
The signature of QMBS therefore is  $\mathcal{F}_1\approx 1$ and $\mathcal{F}_{1/2}\approx 0$. Hence, we will use the quantity $\mathcal{F}_1-\mathcal{F}_{1/2}$ as a probe of QMBS, where it expected to take values close to 1.
This quantity distinguishes QMBS from the trivial situation where the initial state is close to an eigenstate, in which case  $\mathcal{F}(t)\approx 1$ at all times and the fidelity difference is strongly suppressed, $\mathcal{F}_1-\mathcal{F}_{1/2} \ll 1$.  

\begin{figure}[tb]
	\centering
	\includegraphics[width=\linewidth]{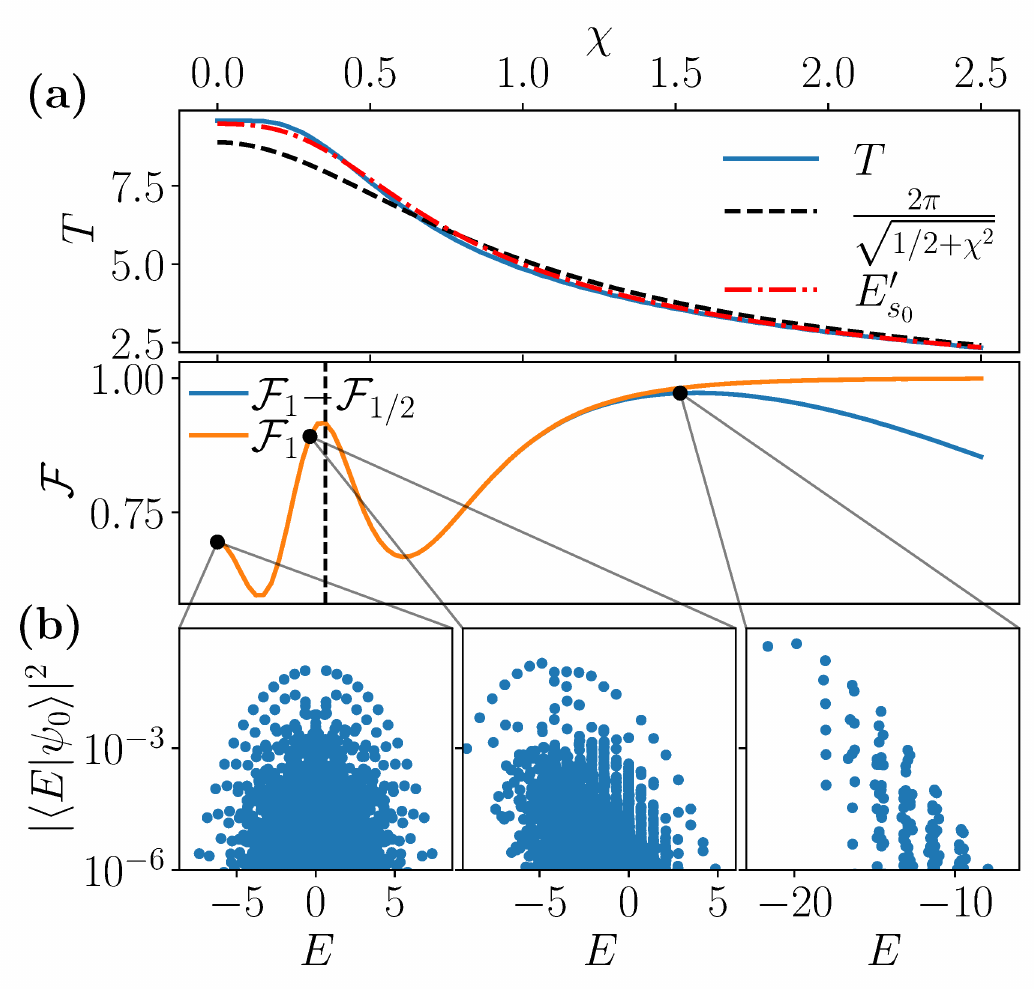}
	\caption{(a) Revival period and amplitude from the N\'eel initial state for the PXP model with staggered detuning and $L_m=26$. The first fidelity peak $\mathcal{F}_1$ is large for all values of $\chi$,  indicating the presence of revivals. However, as $\chi$ becomes large, the N\'eel state effectively becomes an eigenstate, leading to an increase of $\mathcal{F}_{1/2}$ and therefore a decrease of $\mathcal{F}_1-\mathcal{F}_{1/2}$. The revival period $T$ is in good agreement with the prediction based on spin precession, $2\pi/\sqrt{\chi^2+1/2}$, and first-order perturbation theory denoted by $E_{s_0}^\prime$, see  Sec.~\ref{sec:enh} for details. The latter predicts the optimal value $\chi=1/\sqrt{8}$, shown by the vertical black dashed line, which practically coincides with the optimal revival point from the numerics.
    (b) The nonmonotonic behavior of $\mathcal{F}$ as a function of  $\chi$ can be related to changes in the structure of the overlap between the Néel state and the energy eigenstates $\ket{E}$, here plotted for $\chi=0$, $\chi=0.3$ and $\chi= 1.52$.
    For the anti-N\'eel initial state or for $\chi<0$, the dynamics is the same but the spectrum is flipped, $E\leftrightarrow -E$.}
	\label{fig:FT_chi}
\end{figure}

In Fig.~\ref{fig:FT_chi}(a),  we plot both $\mathcal{F}_1$ and $\mathcal{F}_1-\mathcal{F}_{1/2}$ as a function of staggered magnetization for the PXP model with $\mu=0$ and the N\'eel initial state. We observe interesting nonmonotonic behavior as a function of $\chi$, whose mechanism will be explained in detail in the subsequent section. In Fig.~\ref{fig:FT_chi}(b) we see that this nonomonotic behavior of the fidelity is also reflected in the overlap between the Néel state and the energy eigenstates, plotted for several values of $\chi$. The enhancement of scarring at $\chi \approx 0.3$ can be related to a larger separation between the top band of eigenstates and the bulk of the spectrum, as well as due to the more even energy separation between eigenstates belonging to the top band, as shown in the middle panel of Fig.~\ref{fig:FT_chi}(b). At large values of $\chi$, fragmentation is clearly visible by eigenstates forming bands, separated in energy by multiples of $\chi$. This regime will be addressed in detail in Sec.~\ref{sec:HSF}.

\begin{figure}[t!]
	\centering
	\includegraphics[width=\linewidth]{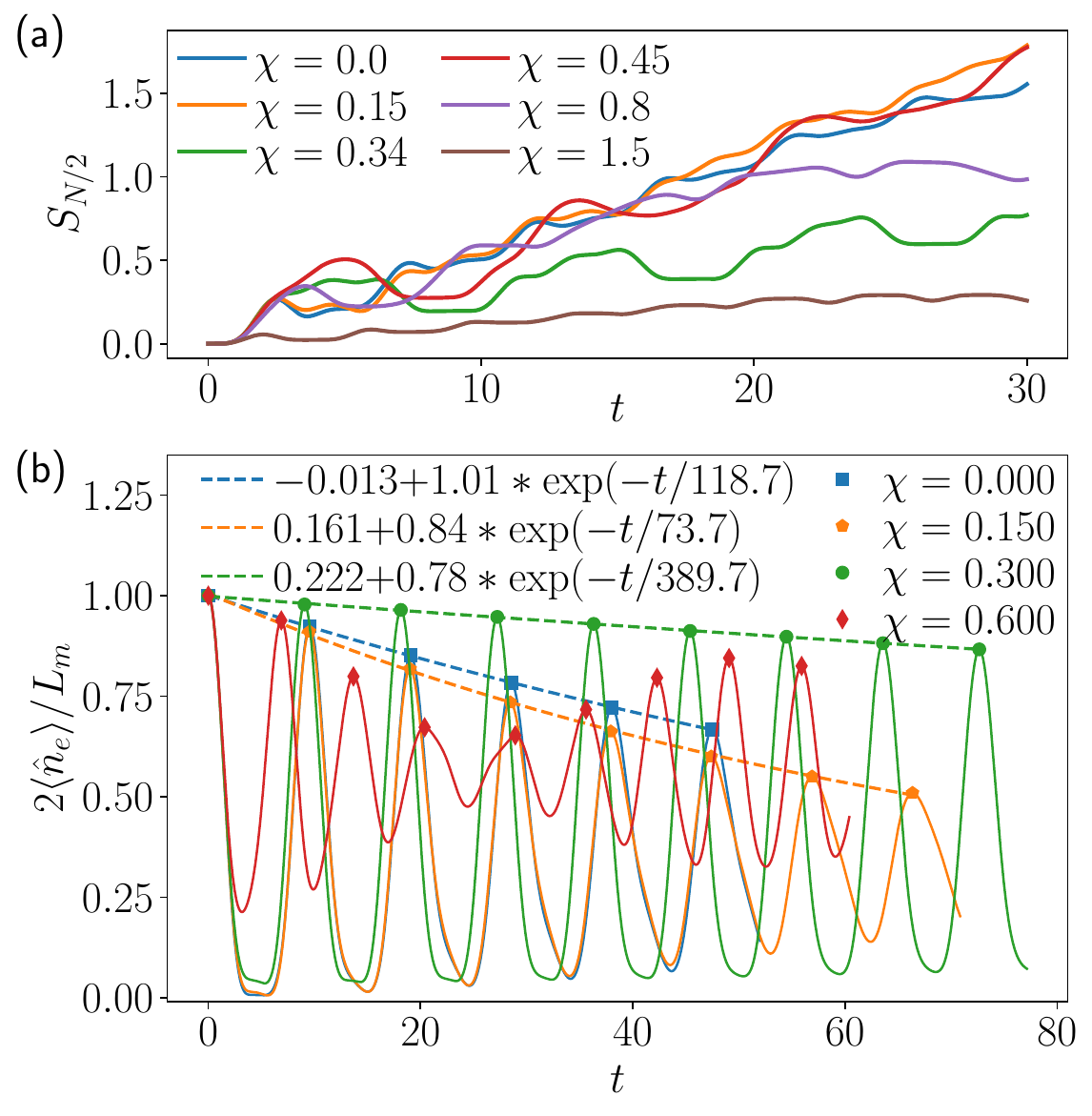}
	\caption{(a) Bipartite von Neumann entanglement entropy after a quench from the N\'eel state for $L_m=26$ and various values of $\chi$. The entropy growth is strongly suppressed around $\chi=0.35$. (b) Excitation density on even sites for $L_m=32$ and various values of $\chi$ and their exponential fit. Only times for which the expectation values are converged in system size are used. Once $\chi \geq 0.5$ oscillations are visible in the peaks and an exponential decay no longer describes their behavior with time. }
	\label{fig:chi_tau_multi}
\end{figure}

Focusing on the N\'eel state, we find that the growth of entanglement entropy is also highly suppressed at $\chi=0.35$, as shown in Fig.~\ref{fig:chi_tau_multi}(a). This nonmonotonicity in the post-quench behavior as a function of $\chi$ can be witnessed in local observables, measurable in experiment.
For a local observable, we choose the excitation density $\langle\hat{n}_e\rangle$ on the even sublattice. 
In order to avoid finite-size effects, we consider only the times where the dynamics is converged in system size $L_m=32$. As the sublattice density oscillates in time,
we first identify the local maxima and then fit its envelope. The details of this procedure are explained in the SM~\cite{SM}, and the results are shown in Fig.~\ref{fig:chi_tau_multi}(b) for several values of $\chi$. 
Overall, we find a clear enhancement of QMBS revivals around $\chi=0.3$ compared to other values.  
The decay time $\tau$ shows an approximately 3-fold increase
around $\chi=0.3$ compared to case without confining potential (see legend).
Finally, we see that for $\chi=0.6$ the peaks themselves show an oscillatory behavior and are no longer well approximated by a decaying exponential. We emphasize that this is not a finite-size effect but rather a sign that we enter the low-energy regime where only a handful of eigenstates participate in the dynamics.
The oscillation in the maxima is then a beating frequency linked to the mismatch in the energy spacing of these eigenstates.

\subsection{Mechanism of revival enhancement due to confinement}\label{sec:enh}

The effect of the staggered magnetization on its own in the PXP model has been studied in Ref.~\cite{Chen2021} in the limit of $\chi \gg \kappa$. In this regime, the Hilbert space fractures and an emergent symmetry appears due to an approximate su(2) algebra. The same spectrum-generating algebra has been linked to QMBS in the PXP model~\cite{Choi2018,Buca2019Non-stationary,IadecolaMagnons,Bull2020,Bhattacharjee2022Probing}.  In the algebraic picture, the PXP Hamiltonian can be thought of as the global $\hat{J}^x$ operator, while the N\'eel state is the highest-weight state of the corresponding global $\hat{J}^z$ operator. Thus, the scarred dynamics is interpreted as precession of a collective spin in a transverse magnetic field. If the algebra were exact, the N\'eel state would undergo perfect revivals.
The PXP model preserves this algebra but only approximately, hence the revivals are present but with a decaying fidelity amplitude. Here we will generalize this picture and identify the mechanism that explains the observed revival enhancement in Fig.~\ref{fig:FT_chi}(a).

At $\mu=0$, the operators defining the mentioned su(2) algebra can be explicitly constructed~\cite{Omiya22} by dimerizing the chain into 2-site unit cells and interpreting the states of each cell as basis states of a local spin-1, i.e., $\ket{\uparrow\downarrow} \equiv \ket{-}$, $\ket{\downarrow\downarrow} \equiv \ket{0}$, and $\ket{\downarrow\uparrow} \equiv \ket{+}$, where site $j$ of the spin-1 model maps to sites $2j-1$ and $2j$ of PXP\footnote{One can alternatively take a mapping where site $j$ of the spin-1 model maps to sites $2j$ and $2j+1$. In that case the results are the same except that the minus sign in front of the $\hat{\mathcal{P}}\hat{S}^z\hat{\mathcal{P}}$ term in Eq.~\ref{eq:PXP_S1} is replaced by a plus sign.}. The spin-1 representation automatically satisfies the constraint within each cell, as no $\ket{\uparrow\uparrow}$ state is admitted, however it can still violate the constraint \emph{between} the unit cells (e.g., $\ldots +- \ldots$). Hence, one needs to also add operators that annihilate the forbidden matrix elements. To obtain the algebra generators, one then simply maps the global spin-1 operators, $\hat{S}^\alpha$, into the spin-1/2 space.  The PXP Hamiltonian can then be written as (for $\mu=0$ and $\kappa=1$)
\begin{equation}\label{eq:PXP_S1}
    -\frac{1}{\sqrt{2}}\left(\hat{S}^x{+}\hat{H}_1{+}\hat{H}_2\right){-}\chi\hat{S}^z=\hat{H}_\mathrm{PXP}\oplus \hat{H}_{\perp},
\end{equation}
where $\hat{H}_{\perp}$ is the Hamiltonian acting on the part of the spin-$1$ Hilbert space that is annihilated by the PXP constraint~\cite{Omiya22}. The perturbations that take care of the constraint are
\begin{equation}\label{eq:H1H2}
   \hat{H}_1=\hat{H}_2^\dagger=-\frac{1}{2}\left(\ket{+,0}+\ket{0,-}\right)\bra{+,-}. 
\end{equation}
Without  $\hat{H}_1$ and $\hat{H}_2$, the Hamiltonian in Eq.~(\ref{eq:PXP_S1}) would be a free spin-1 model.  There would then be perfect revivals from the N\'eel state ($\ket{++\cdots ++}$ in the spin-1 language) with a period
\begin{align}
T=\frac{2\pi}{\sqrt{\chi^2+\frac{1}{2}}}.
\end{align}
This simple expression matches well the exact result for PXP in Fig. \ref{fig:FT_chi}(a).  The agreement improves with $\chi$, which is due to the fact that, as $\hat{S}^z$ becomes larger, the influence of $\hat{H}_1$ becomes weaker. 

\begin{figure}[t!]
	\centering
	\includegraphics[width=0.8\linewidth]{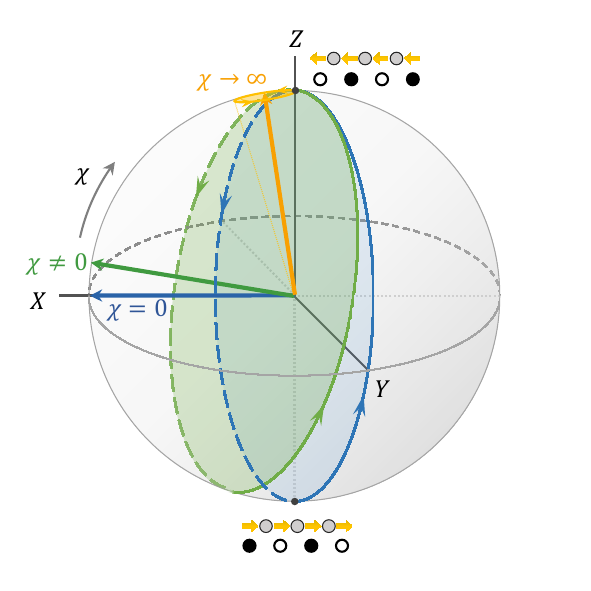}
	\caption{Schematic picture of the effect of the confining potential on the dynamics after a quench form the N\'eel (or vacuum) state. The $X$, $Y$ and $Z$ labels refer to the the approximate su(2) algebra, which matches with the spin-1 description.
    }
	\label{fig:Bloch_sphere}
\end{figure}

As $\chi$ becomes large, we saw in Fig.~\ref{fig:FT_chi}(a) that $\mathcal{F}_1-\mathcal{F}_{1/2}$ starts to decrease as the minimum fidelity between the revivals increases. This can be understood easily in the spin-precession picture with the help of the Bloch sphere, illustrated in Fig.~\ref{fig:Bloch_sphere}. The N\'eel state is at the North Pole, along the $Z$ axis. For $\chi=0$, the precession axis is the $X$ axis, lying in the equatorial plane. As $\chi$ is increased, the precession axis is tilted out of plane and moved closer to the $Z$ axis. Hence, the antipodal point
on the trajectory is no longer the anti-N\'eel state (at the South Pole), but another state placed on the opposite side of the precession axis. As $\chi$ is further increased, this opposite point gets closer and closer to the N\'eel state. In the limit $\chi\to\infty$, the precession axis becomes the $Z$ axis and the N\'eel state an exact eigenstate of the system.  The effect of this ``axis tilting'' can also be seen directly in the overlap between the N\'eel state and the corresponding PXP eigenstates at various values of $\chi$, as shown in Fig.~\ref{fig:FT_chi}(b). The top band of scarred states is always visible. However, as $\chi$ is increased, the eigenstates with the highest overlap shift from the middle of the spectrum towards its left edge, where the ground state resides. 

While the precession picture above gives a relatively good approximation of the revival period, it also predicts the revivals would increasingly get better as $\chi$ becomes larger. This is in stark contrast with the observed nonmonotonic behavior, in particular with the strong revival enhancement around $\chi\approx 0.35$. In order to understand the nonmonotonic behavior, we construct the states $\ket{S_n}$, which are the eigenstates of $-\hat{S}^x/\sqrt{2}{-}\chi\hat{S}^z$ projected into the constrained PXP space, with $n=-L_m/2, -L_m/2+1, \ldots, L_m/2$. This construction provides a good approximation of the scarred eigenstates in the case $\chi=0$ \cite{Omiya22}. By computing the energy variance of these approximate eigenstates, one observes nonmonotonic behavior, with local minima at $\chi=0$ and $\chi \to \pm \infty$ (see Appendix~\ref{app:spin1}). This suggest that the revivals will display better fidelity at these points and get worse as $\chi$ moves away from them. While this matches well with the numerical result in Fig.~\ref{fig:FT_chi}(a), it does not explain the behavior near $\chi=0.35$.

To understand why fidelity is maximal around that point, a finer analysis of the effects of the perturbations $\hat{H}_1$ and $\hat{H}_2$, Eq.~(\ref{eq:H1H2}), on the $\ket{S_n}$ states is needed. It is particularly important to understand how the energy spacing between eigenstates is modified by these terms, as a more regular spacing can lead to significantly better revivals.
We first note that as $\hat{\mathcal{P}}\hat{H}_2=0$ but $\hat{\mathcal{P}}\hat{H}_1=\hat{H}_1$, only $\hat{H}_1$ will contribute. 
Hence we can focus on computing $\Delta E_n$, the first order correction of the energy of the $\ket{S_n}$ states due to $H_1$. As the $\ket{S_n}$ are (projected) eigenstates of a non-interacting Hamiltonian, their structure can be written down and $\Delta E_n$ can be computed analytically for any $n$ and $\chi$. In particular, we find that the in the thermodynamic limit the energy spacing between two consecutive scarred eigenstates is given by (see Appendix \ref{app:spin1})
\begin{equation}\label{eq:Es_prime}
E^\prime_s{=}\frac{15+56\chi^2+64\chi^4-3s^2(1-8\chi^2)}{64(1/2 + \chi^2)^{3/2}},
\end{equation}
where $s \equiv (2k)/L_m \in [-1,1]$ can be thought of as energy density. From this, we can immediately see that for $\chi=1/\sqrt{8}\approx 0.354$ the energy spacing is \emph{independent} of $s$. At this special point, all eigenstates are exactly equally spaced at first order. This leads to largely improved revivals, accounting for the observed fidelity peak.

A few remarks are in order. The identification of the optimal $\chi$ from Eq.~(\ref{eq:Es_prime}) in the thermodynamic limit also applies equally well to finite systems. In finite systems, the N\'eel state is still far from the edges of the spectrum at $\chi=1/\sqrt{8}$ and the level statistics indicate a chaotic system, see Appendix \ref{app:rstat}. The revivals at this optimal $\chi$ converge very quickly in system size, and they are not visible for other initial states, which generically thermalize relatively fast~\cite{SM}. Moreover, the result in Eq.~\eqref{eq:Es_prime} also allows to get a much better estimate of the revival frequency. To achieve that, we only need to find the value of $s$ which maximizes the overlap with the N\'eel state. This can also be done analytically (see Appendix~\ref{app:spin1}) and yields 
\begin{equation}\label{eq:s0}
 s_0=-\frac{2\chi^2+\sqrt{2}\sqrt{1+2\chi^2}}{1+2\chi^2+\sqrt{2}\sqrt{1+2\chi^2}}   
\end{equation}
in the thermodynamic limit. The corresponding period $2\pi/E^\prime_{s_0}$ was shown in Fig.~\ref{fig:FT_chi}(a) and gives a very good approximation of the exact data for finite systems. Finally, in SM~\cite{SM}, we show that the optimal revival point $\chi\approx 0.35$ can also be witness within the semiclassical limit obtained by projecting the dynamics onto a variational manifold~\cite{Ho2019}. Interestingly, in that case the classical orbit linked to the scarred dynamics is not exactly periodic, in contrast with the vast majority of example of QMBS.

\subsection{Defects and confinement}
\begin{figure*}[htb!]
	\centering
	\includegraphics[width=\linewidth]{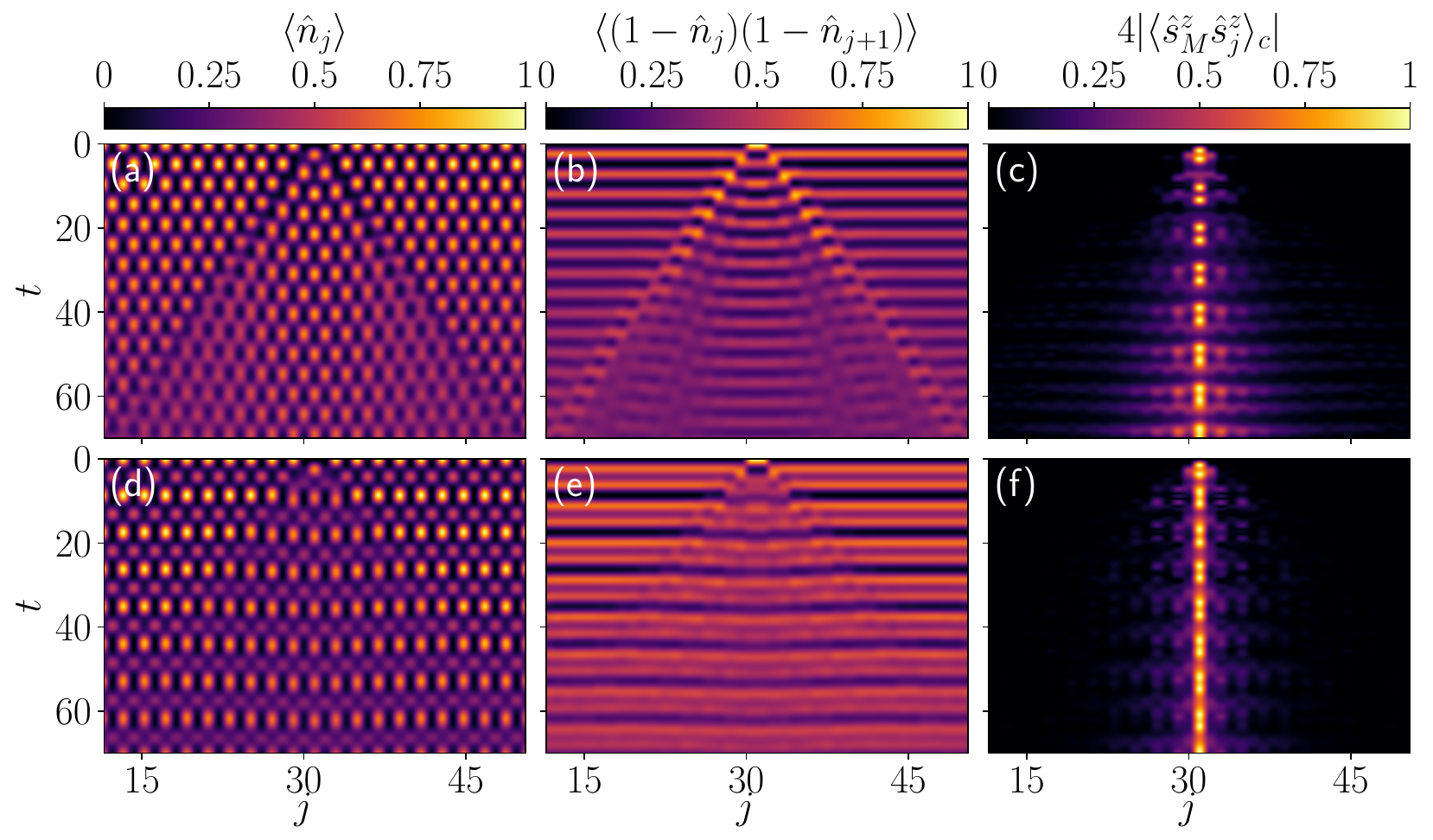}
	\caption{Observable dynamics after a quench from the N\'eel state with a defect for $L_m=61$ and OBC. The data is truncated to only focus on sites near the middle of the chain where the defect was initially localized. (a)-(c) $\chi=0$ (d)-(f) $\chi=0.35$. The effect of confinement is visible for the ZZ correlator in panels (c) and (f). At late times, the N\'eel and anti-N\'eel domains are still clearly distinguishable in panels (a) and (b) but not in panels (d) and (e).}
	\label{fig:defect_chi}
\end{figure*}

While adding a nonzero $\chi$ affects the algebra and the revivals from the N\'eel state, from a lattice gauge theory point of view it also causes confinement \cite{Buyens2016,Surace2020,Halimeh2022tuning,Cheng2022tunable}. Notably, each particle-antiparticle pair experiences an energy cost $\propto \chi d$, where $d$ is the distance between the particle and the antiparticle. For $\chi=0$, there is no cost and so the particles can spread ballistically in opposite directions, i.e., they are deconfined. As soon as $\chi>0$, any separation distance is penalized, leading to their confinement.

In the the PXP model, a configuration with neighboring unexcited sites $\ket{{\circ}{\circ}}$ implies the presence of a particle or antiparticle between them (depending on whether the sites are even-odd or odd-even). Similarly, an excited site next to an unexcited site ($\ket{{\circ}{\bullet}}$ or $\ket{{\bullet}{\circ}}$) means that there is no particle/antiparticle between them. Thus, a single particle-antiparticle pair on top of the vacuum takes the form of a single ``defect'' on top of the N\'eel (or anti-N\'eel) state. 
Specifically, we will use the state $\ket{{\circ}{\bullet}{\circ}{\bullet}\cdots {\circ} {\bullet}{\circ}{\circ}{\circ}{\bullet}\cdots {\circ}{\bullet}{\circ}{\bullet}}$, which corresponds to the N\'eel state with an excitation removed on site $M$ near the middle of the chain. This can also be understood as creating two domain-walls between N\'eel domains ${\circ}{\bullet}{\circ}{\bullet}$ and anti-N\'eel domains ${\bullet}{\circ}{\bullet}{\circ}$. We quench this state and monitor the excitation occupancy on each site $\langle \hat{n}_j \rangle$. We also monitor the presence of domain-walls using the quantity $\langle(1-\hat{n}_j)(1-\hat{n}_{j+1})\rangle=1-\langle\hat{n}_j\rangle-\langle\hat{n}_{j+1}\rangle$, where we have utilized that $\langle\hat{n}_j\hat{n}_{j+1}\rangle=0$. Finally, we track the ZZ connected correlator $\langle \hat{s}^z_M\hat{s}^z_j \rangle_c=\langle \hat{s}^z_M\hat{s}^z_j \rangle-\langle \hat{s}^z_M \rangle\langle \hat{s}^z_j \rangle$.
The corresponding results at $\chi=0$ and $\chi=0.35$ are shown in Fig.~\ref{fig:defect_chi}, which have been obtained by matrix product state (MPS) techniques with OBC. 
MPS allows us to probe large system sizes where the boundary is not reached even at late times (more details about this can be found in SM~\cite{SM}). 
The case $\chi=0$ was previously studied in Ref. [29] for a smaller system, and we find similar results here. There are oscillations both between the two domain walls (i.e., the anti-N\'eel domain in the middle of chain) and outside of the domain walls (the N\'eel domain). In the two regions, the dynamics returns to the N\'eel state with the same period but with a phase difference of $\pi/2$ ~\cite{Surace2020}. Meanwhile, for $\chi=0.35$ we find that the dynamics in the two domains synchronizes, with the phase difference quickly going close to 0. At later times it is difficult to see any clear trace of the original defects. While it is not certain if this is directly linked to the enhancement of revivals without any defect, this kind of ``self-correction'' of defects is also experimentally desirable.
Finally, the ZZ correlator shows a strong difference between the two values of $\chi$, with clear confinement at $\chi=0.35$.

\begin{figure}[hbt!]
	\centering
	\includegraphics[width=\linewidth]{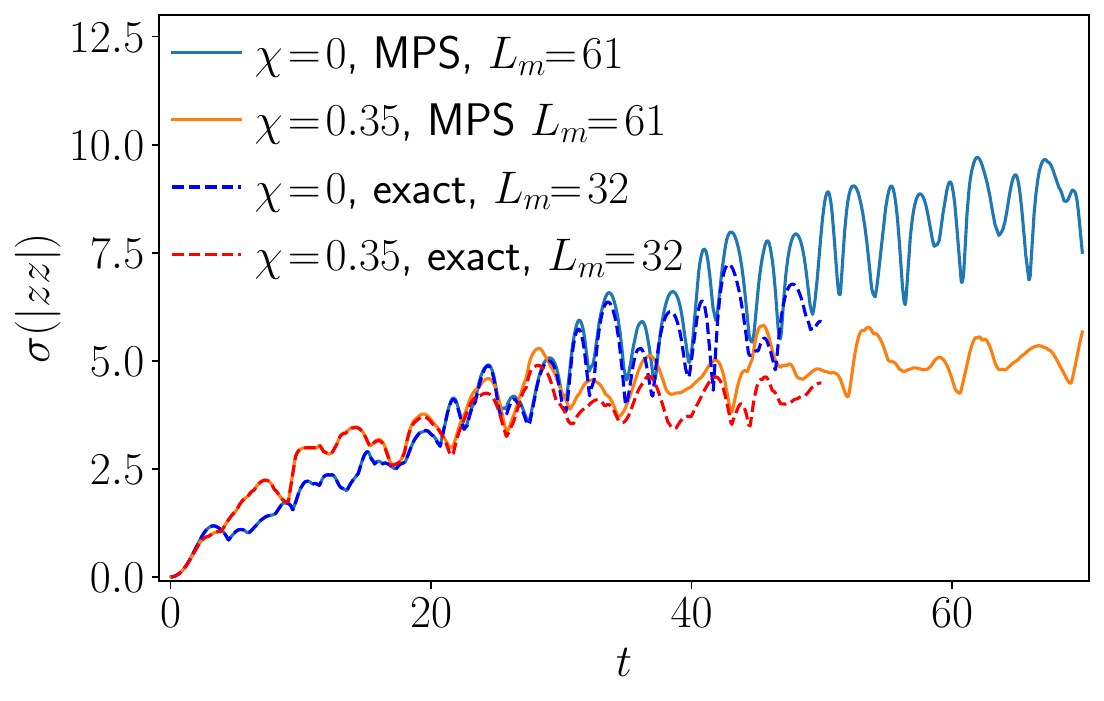}
	\caption{Root mean square of the spread of the connected correlator between Pauli $Z$ matrices at the middle site and other sites. The exact and MPS data coincide well at shorter times, when the influence of the chain boundaries are small. In both cases, the confinement induced by $\chi$ is clear.}
	\label{fig:defect_ZZ_RMS}
\end{figure}

In order to quantify confinement more precisely, we compute the ``root mean square spread'' of the ZZ correlator as
\begin{equation}
    \sigma(|zz|)=\sqrt{\frac{\sum_{j=1}^{L_m} |\langle \hat{s}^z_j\hat{s}^z_M\rangle_c| (j-M)^2}{\sum_{j=1}^{L_m} |\langle \hat{s}^z_j\hat{s}^z_M\rangle_c| }}.
\end{equation}
We plot this quantity at all times for $\chi=0$ and $\chi=0.35$ in Fig~\ref{fig:defect_ZZ_RMS}. There, we can see a clear difference between the two, as for $\chi=0.35$ the spreading reaches a plateau around $t\approx 30$, while for $\chi=0$ the correlations continue to spread until they reach the boundary of the system. We thus conclude that at the optimal point for QMBS $\chi=0.35$, there are clear signs of confinement even in finite systems.

\section{Hilbert space fragmentation induced by confinement}\label{sec:HSF}

The mass term in Eq.~\eqref{eq:H_spin} does not fit into the approximate algebra discussed above, and it generically destroys scarring and revivals from the N\'eel state.
However, the combination of nonzero mass and confining potential allows a more complex picture to emerge due to the interplay between them.

In the regime $\chi,\mu \gg \kappa$ at the top right corner of Fig.~\ref{fig:schematic}(c), different configurations can have large energy differences compared the dynamical term strength $\kappa$. Transitions between them is then heavily suppressed and resonant processes (between configurations with the same energy) dominate the dynamics. The effective Hamiltonian can be obtained by performing a Schrieffer-Wolff (SW) transformation. We will show that in many cases the leading order of that effective Hamiltonian shows exact Hilbert space fragmentation. While in general the higher-order terms can connect different sectors, they are suppressed by factors of approximately $\kappa/(\mu\pm\chi)$ and so will not matter until long-times. This means that at shorter times the wavefunction will be trapped in the sector in which it was initially, leading to a prethermal fragmentation regime. Our focus is on regimes that can be probed in quantum simulators. In these systems, only shorter times can be probed before decoherence starts to appear. So the impact of higher-order terms is minimal and we mostly focus on the leading order that will dominate the dynamics.

For general values of $\mu$ and $\chi$, we extend the approach of Ref.~\cite{Chen2021}, which studied the case $\mu=0$. We show the main results in this section, while all details are relegated to the SM~\cite{SM}. We find that the odd-order terms in the SW Hamiltonian are identically zero. The leading term is then the second-order one, which is diagonal and reads
\begin{align}\label{eq:H2_main}
    \hat{H}_\mathrm{eff}^{(2)}=\; - \frac{\kappa^2}{2}\sum_{j=1}^{L_m}\frac{\hat{P}_{j-1}\hat{s}^z_{j}\hat{P}_{j+1}}{2\mu+(-1)^j \chi}.
\end{align} 
As this term is diagonal, we need to go to the next nonzero order for dynamics to occur in the $Z$ basis. This happens at fourth order with an effective Hamiltonian given by
\begin{align}\nonumber
&\hat{H}_\mathrm{eff}^{(4)}=\frac{\kappa^4}{32} \sum_{j=1}^{L_m}4\frac{\hat{P}_{j-1}\hat{s}^z_j\hat{P}_{j+1}}{(2\mu+(-1)^j\chi)^3}\\\nonumber
&+2\frac{4\mu{-}(-1)^j\chi}{\left(4\mu^2{-}\chi^2\right)^2}\left(\hat{P}_{j-2}\hat{P}_{j-1}\hat{s}^z_{j}\hat{P}_{j+1}{+}\hat{P}_{j-1}\hat{s}^z_{j}\hat{P}_{j+1}\hat{P}_{j+2}\right) \\\label{eq:H4_main}
&+\frac{4\mu{+}(-1)^j\chi}{\left(4\mu^2{-}\chi^2\right)^2}\left(\hat{P}_{j-2}\hat{s}^+_{j-1}\hat{P}_{j}\hat{s}^-_{j+1}\hat{P}_{j+2}{+}{\rm H.c.}\right).
\end{align}
These results are similar to the $\mu=0$ case explored in Ref.~\cite{Chen2021}, in which the diagonal second-order term was understood as stemming from the approximate su(2) algebra. While the addition of the mass term ``breaks'' it, the additional terms due to a nonzero $\mu$ are non-resonant at low order except in the special case of $\chi=0$. So in general, the cases $\mu\neq 0$ and $\mu=0$ will only show different dynamical terms at higher orders.

\subsection{\texorpdfstring{$\chi=\pm 2\mu$}{chi=2 mu}  Resonance}

One clear feature of both Hamiltonians~\eqref{eq:H2_main} and~\eqref{eq:H4_main} is that some of their terms diverge if $\chi=\pm 2 \mu$ due to a resonance condition. Let us focus on the case $\chi=-2\mu$. Then we have that $2\mu+(-1)^j\chi=2\mu\left[1-(-1)^j\right]$ is equal to $4\mu$ for odd $j$ and to $0$ for even $j$. Thus, only spin-flips on odd sites lead to an energy change. In the limit $\chi,\mu \gg \kappa$, the odd sites are effectively frozen as the energy cost to change their state is too high. Hence, the Hilbert space fractures into $2^{L_m/2}$ sectors corresponding to all the possible combinations of the spins on odd sites. The sites that are frozen in the excited position also freeze their neighboring sites due to the PXP constraint. The remaining spins on even sites that have no up-spin neighbors can be excited freely and independently, and thus the effective model at leading order is that of an free spin-$1/2$ paramagnet.

We can extend our analysis to other values of $\mu$ and $\chi$ around the resonance such that $\chi=-2\mu+\gamma$ and $\mu,\chi \gg \kappa,\gamma$.
The leading Schrieffer-Wolff term is at first order and reads
\begin{align}\label{eq:res_n0_main}
\hat{H}_\mathrm{eff}^{(1)}{=}{-}\kappa \sum_{j\in \mathcal{K}} \hat{s}^x_{j}{-}\sum_{j=0}^{M}\gamma \hat{s}^z_{2j}=E_0{-}\sum_{j\in \mathcal{K}}\big(\kappa \hat{s}^x_{j}{+}\gamma \hat{s}^z_j\big),
\end{align}
where $\mathcal{K}$ is the set of all even sites with both neighbors frozen in the down position and $E_0$ is the (constant) contribution of all frozen even sites (so with at least one neighbor in the up position).
This Hamiltonian is then clearly integrable for all values of $\kappa$ and $\gamma$.

\begin{figure}[tb]
	\centering
	\includegraphics[width=\linewidth]{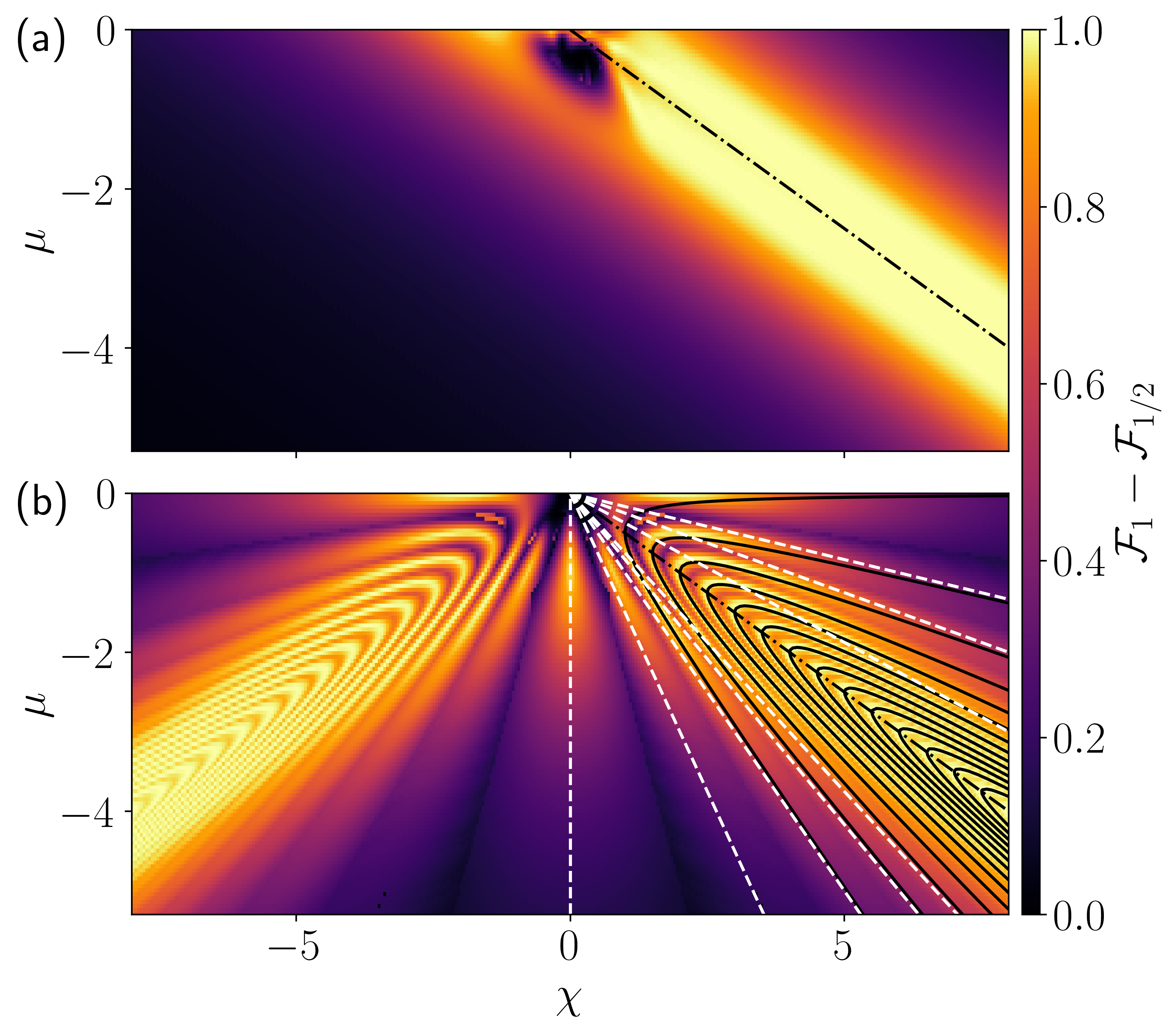}
	\caption{Self-fidelity from (a) the N\'eel state and (b) polarized state in the PXP model with $L_m=20$. The black dash-dotted line shows the resonance line $\chi=-2\mu$, while the black hyperbolas show the optimal revival around it. The white dashed lines show the other resonances $\mu=-\frac{n+1}{n-1}\chi$. }
	\label{fig:2D_frag}
\end{figure}

We show the effects of this resonance in quenching from the N\'eel and polarized states in Fig.~\ref{fig:2D_frag}. While this fragmentation affects all initial states, unlike QMBS discussed in Sec.~\ref{sec:QMBS}, we only show these two initial states for brevity as they are the most relevant one for experimental preparation. For both of them, strong revivals of the wavefunction can be seen around the resonance already at relatively small values of $\chi$ and $\mu$. For the N\'eel state, only $\chi= -2\mu$ leads to revival and not $\chi=2\mu$. Indeed, the N\'eel state only has excitations on even sites. So for $\chi= -2\mu$ these can be freely changed between the excited and unexcited states. Meanwhile, for $\chi=2\mu$ they cannot be deexcited, and thus also freeze the odd sites due to the constraints.

\begin{figure}[t!]
	\centering
	\includegraphics[width=.48\textwidth]{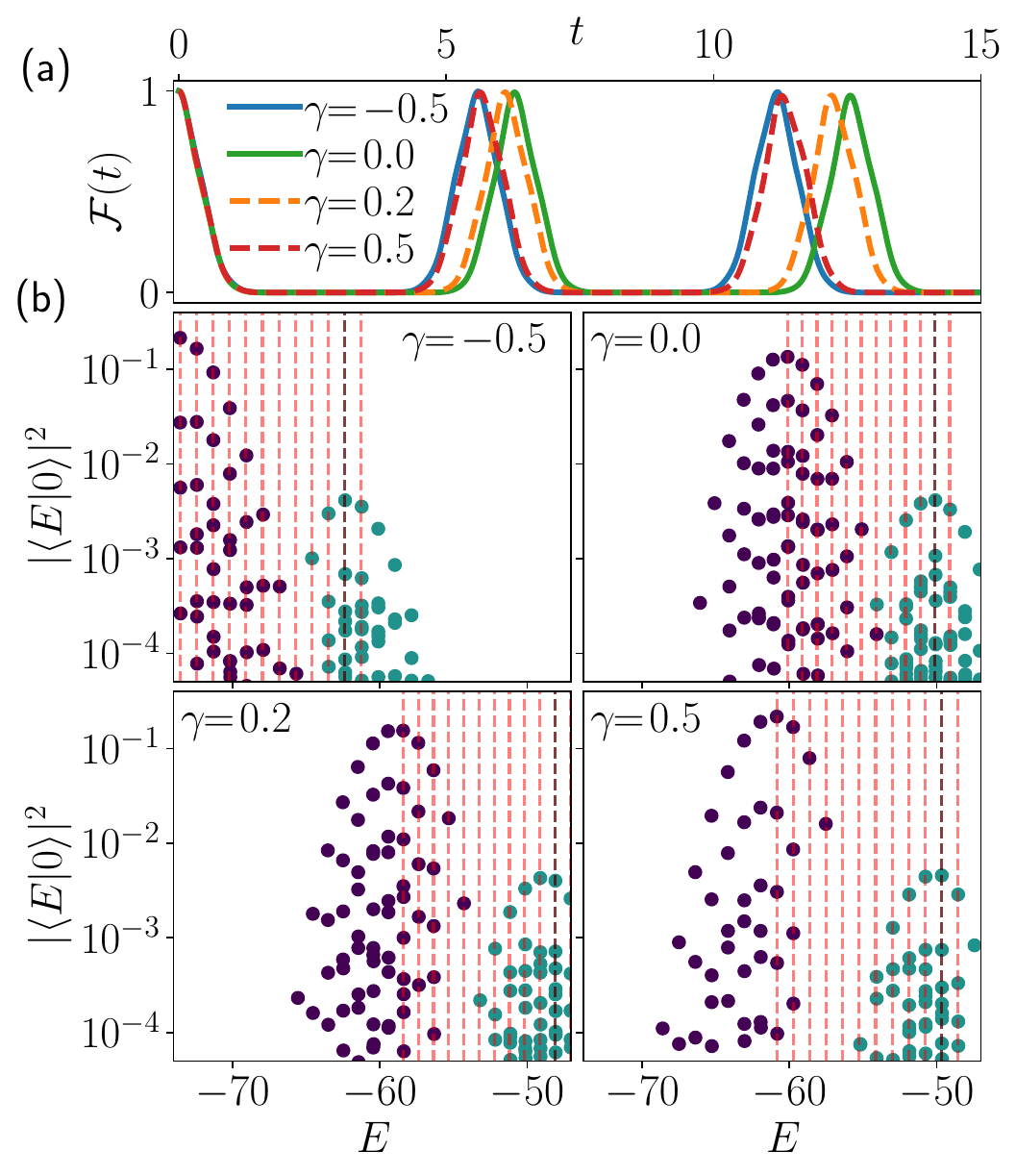}\\
	\caption{Quenches from the polarized state in the PXP model with $L_m=24$ around the resonance line as $\chi=5\sqrt{1+\gamma^2}$ and $\mu=\frac{\gamma-\chi}{2}$. (a) Self-fidelity after a quench. As $\gamma$ is varied the period changes but the revivals remain close to perfect (b) Overlap between polarized state and the eigenstates. The color indicates the occupation density of the odd sites, and and so the various sectors. The red dashed lines are all spaced in energy by $\sqrt{1+\gamma^2}$, while the gray dashed line is placed with energy difference 2$\chi$ from the highest overlap eigenstate. In all cases of $\gamma$ we see that all eigenstates show close to equal spacing.}
	\label{fig:OffResPolBands} 
\end{figure}

For the polarized state the pattern is more complicated.
This might seem surprising as this state is in the same Hilbert space sector as the N\'eel state. However, the terms in the Hamiltonian connecting the sectors (changing the state of odd sites) are much less suppressed for the polarized state than for the N\'eel state, as for the latter two even sites must first be deexcited before such a move can be done.  Hence, even for large values of $\mu$, the odd sites cannot be considered as totally frozen and we have some weak couplings between the Hilbert space sectors. This leads to further restriction for revivals in order to make all sectors aligned in energy, as shown in Fig.~\ref{fig:OffResPolBands}. The additional resonance condition can be cast as
\begin{equation}
\chi=\frac{n\sqrt{1+\gamma^2}}{2}, \ \mu=\frac{\gamma-\chi}{2}
\end{equation}
with $n$ an integer (see Appendix \ref{app:pol_res} for details). The case $\gamma=0$ yields the condition for $\chi$ to be half-integer or integer.  Alternatively, we can get rid of $\gamma$ and write the relation between $\mu$ and $\chi$ directly as
\begin{equation}
\frac{\left(2\chi\right)^2}{n^2}-\left(2\mu+\chi\right)^2=1.
\end{equation}
We recognize here the equation for a hyperbola, which was plotted in Fig.~\ref{fig:2D_frag}(b) for various values of $n$. 
Fig.~\ref{fig:OffResPolBands} illustrates how all eigenstates are approximately equally spaced as long as these equations are followed.

\subsection{Other resonances}

Beyond $\chi=\pm 2\mu$, there are other resonant ratios between $\mu$ and $\chi$. Exciting an even site leads to a change in energy of $\Delta E_1=-2\mu-\chi$ while on an odd site it costs $\Delta E_2=-2\mu+\chi$. In order to have resonances, we want $
\Delta E_1$ and $\Delta E_2$ to be commensurate. The two simplest cases are if $\Delta E_1$ is an integer multiple of $\Delta E_2$ or vice versa. This leads to the resonance condition 
\begin{equation}
\mu=\pm\frac{n+1}{2(n-1)}\chi,
\end{equation}
with $n$ an integer (which can also be negative or equal to 0). These resonances  are shown in white dashed lines in Fig.~\ref{fig:2D_frag}(b). The only notable ones that lead to changes in the effective Hamiltonian at order 4 or below, are $\chi=0$ ($n=1$) and $\mu=\frac{3}{2}\chi$ ($n=2$). Note that unlike the resonance at $\chi=\pm 2\mu$, in these two cases we do not expect to see revivals in Fig.~\ref{fig:2D_frag} at large values of $\chi$ and $\mu$, as the polarized and N\'eel states become eigenstates and the fidelity stays close to 1 at all times. Nonetheless, revivals can be seen at intermediate values along these resonances for the polarized state.

The effective Hamiltonian around $\chi=0$ also leads to interesting properties. By studying this special case we can understand the appearance of the effective chaotic and integrable regions at the top right of Fig.~\ref{fig:schematic}(c).
If $\mu \gg \chi$, the total number of excitations becomes conserved. The effective Hamiltonian at second order then gains a nearest-neighbor XY type term, and a small $\chi$ will add some additional diagonal term at first order. The effective Hamiltonian then becomes
\begin{align}\nonumber
\hat{H}_\mathrm{eff}^{(1,2)}&{=} {-
}\frac{\kappa^2}{8\mu}\sum_{j=1}^{L_m}\Big(2\hat{P}_{j-1}\hat{s}^z_{j}\hat{P}_{j+1}
+\hat{P}_{j-1}\hat{\sigma}^+_{j}\hat{\sigma}^-_{j+1}\hat{P}_{j+2}\\\label{eq:Heff12_res_mu}
&+{\rm H.c.}\Big)-\sum_{j=1}^{L_m}(-1)^l\chi \hat{s}^z_{l}.
\end{align}
There is no fragmentation for this Hamiltonian, as each $\mathrm{U}(1)$ sector is fully connected. The case  $\chi=0$ is known to be integrable~\cite{FSS}. It is interesting to note that the effective Hamiltonian resembles the lattice fermions $\mathcal{M}_1$ supersymmetric model introduced in Ref.~\cite{FendleySUSY}. The two terms that compose the Hamiltonian are the same, but their prefactors are different. Nonetheless, the same mapping to the an XXZ-type model can be performed, albeit with a different value of $\Delta$. Consequently, in each of the $\mathrm{U}(1)$ sectors the effective Hamiltonian is integrable~\cite{FSS}.
However, as soon as $\chi\neq 0$, this Hamiltonian no longer maps to an XXZ-type model, and when $\chi$ and $\frac{\kappa^2}{8\mu}$ are of comparable strength we find level spacing statistics close to Wigner-Dyson (see Appendix \ref{app:rstat}).
We emphasize that this does not mean that the full system is ergodic. It still effectively splits into various sectors, but in each of them the effective Hamiltonian is ergodic and there is no further fragmentation. As they are well separated in energy, the level spacing statistics is dominated by the spacings within each sector, which follow a Wigner-Dyson distribution.

Overall, the interplay of the mass and confining terms allows for a vast array of regimes. It should be noted that while our Schrieffer-Wolff transformations are performed assuming that $\mu$ or $\chi$ are much larger than $\kappa$, we observe the onset of fragmentation already for relatively small values of these parameters such as $\mu=0$, $\chi=\kappa/2$. This is likely a finite-size effect as larger values of $\chi$ and $\mu$ are needed to escape the ergodic region as $L_m$ is increased.  

\begin{figure*}[t!]
	\centering
	\includegraphics[width=\linewidth]{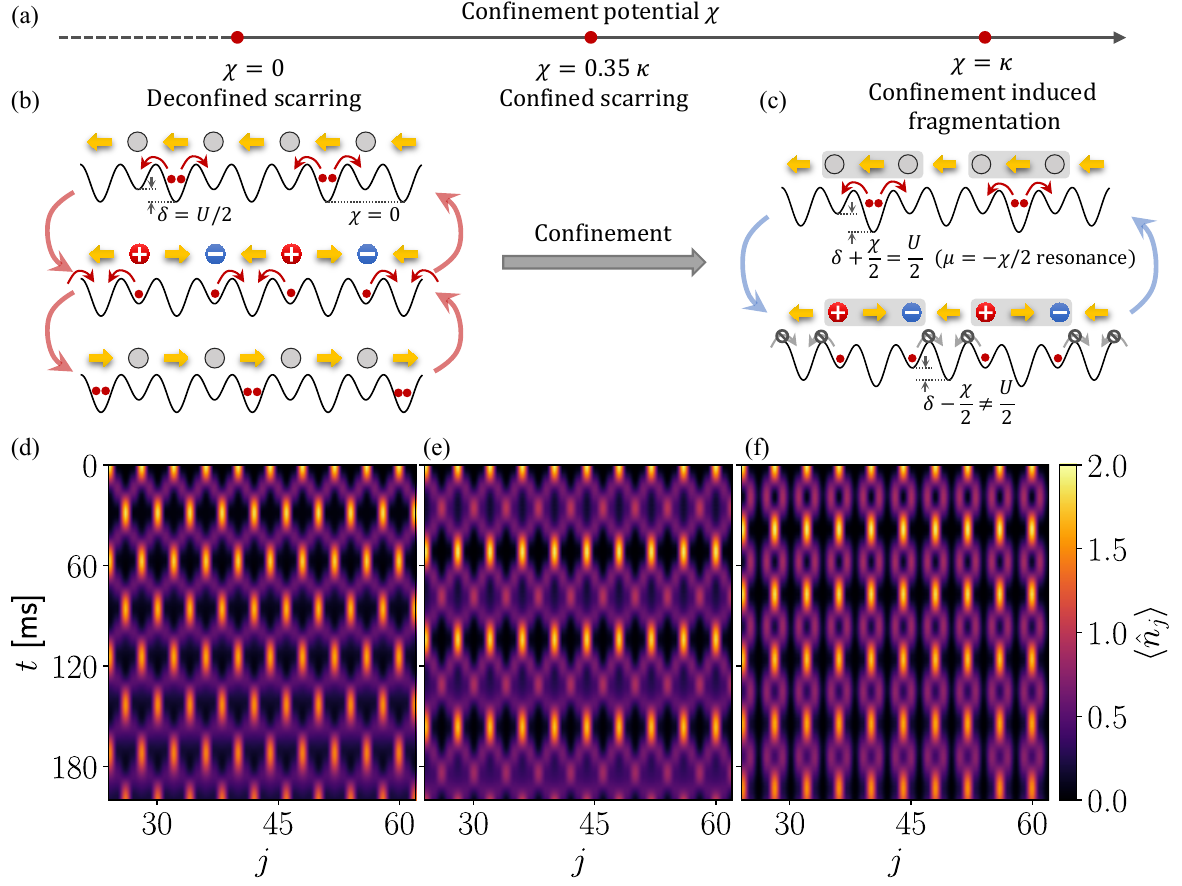}
	\caption{(Color online) Proposed experimental realization of the confinement-induced scarring and fragmentation in a Bose--Hubbard quantum simulator. (a) By tuning confinement potential $\chi$ from 0, regimes of the deconfined scarring ($\chi = 0$), confinement-enhanced scarring ($\chi = 0.35 \kappa$), and confinement-induced fragmentation (using the resonant case $\chi = \kappa$, $\mu = -0.5 \kappa$ as an example) can be realized. (b) and (d): Schematic and MPS simulation of the deconfined scarring, the ``state transfer'' between the two vacuums. (c) and (f): In the fragmented regime, at resonance, the Schwinger pair creation mechanism still happens, but dynamics are localized within each building block due to confinement. (e) MPS simulation of the confined scarring at $\chi = 0.35 \kappa$. Simulations are for $L=87$, with data near the edges truncated for clarity.}
	\label{fig:ConfinedScarring}
\end{figure*}

\begin{figure*}[t!]
	\centering
	\includegraphics[width=\linewidth]{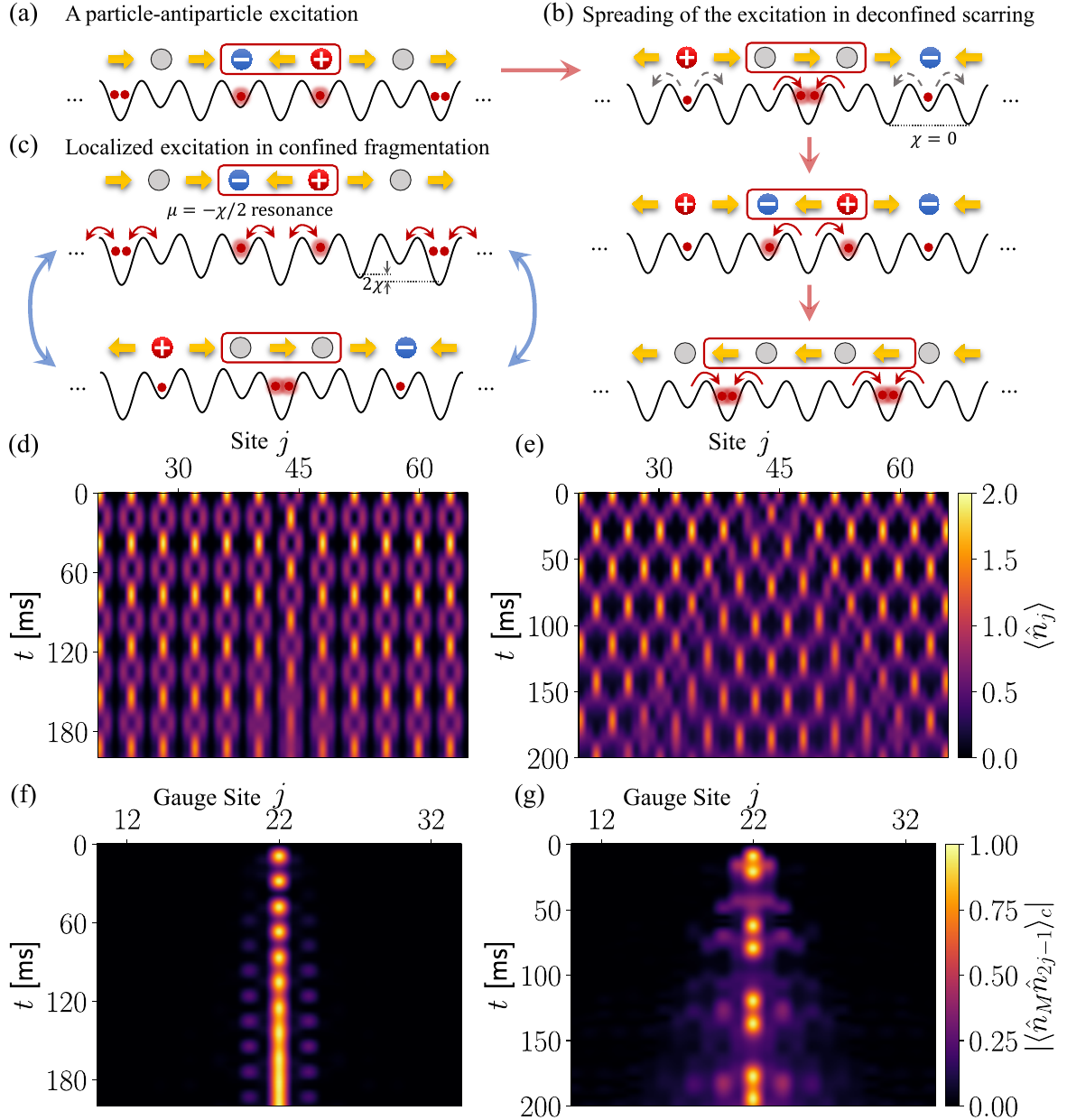}
	\caption{The experimental probe of a particle-antiparticle excitation in the vacuum background. (a) Preparation of the particle-antiparticle excitation in the Bose--Hubbard quantum simulator. (b) Schematic of the coupling between the excitation and the background in the deconfined case. Coupling between the unit cells leads to the spreading of the excitation. (e) and (g): MPS simulation of the light-cone spreading. (e) The particle density, (g) the density-density correlation between a center gauge site with the particle and the rest of the gauge sites. (c) In the fragmented case, at resonance $\chi=\kappa, \mu = -0.5 \kappa$, the defect is unable to couple to the neighboring unit cells due to the staggering $\chi$, and is hence unable to spread, resulting in decoupled dynamics between the excitation and the background. (d) and (f): Same simulation as for (e) and (g), for the fragmented case. The excitation is localized due to confinement. Simulations are for $L=87$, with data near the edges truncated for clarity.}
	\label{fig:defect_explanation}
\end{figure*}

\section{Experimental probes in a Bose--Hubbard simulator}\label{sec:experiment}

The $\mathrm{U}(1)$ QLM has been experimentally implemented on a Bose--Hubbard quantum simulator for the deconfined case of $\chi=0$, where gauge invariance was directly observed \cite{Yang2020}, and thermalization dynamics was probed \cite{Zhou2021}. More recently, an experimental proposal outlines how these experiments can be feasibly updated to study confinement in this model \cite{Halimeh2022tuning,Cheng2022tunable}. Using this proposal, we will show that our results in Sections~\ref{sec:QMBS} and \ref{sec:HSF} can be experimentally observed.

We begin by reviewing the Hamiltonian of the tilted Bose--Hubbard model employed in Refs.~\cite{Halimeh2022tuning,Cheng2022tunable}, which is
\begin{align}\nonumber
    \hat{H}_\text{BHM}=&-J\sum_{j=1}^{L-1}\big(\hat{b}_j^\dagger\hat{b}_{j+1}+\text{H.c.}\big)+\frac{U}{2}\sum_{j=1}^L\hat{n}_j\big(\hat{n}_j-1\big)\\\label{eq:BHM}
    &+\sum_{j=1}^L\bigg[(-1)^j\frac{\delta}{2}+j\Delta+\frac{\chi_j}{2}\bigg]\hat{n}_j,
\end{align}
where $J$ is the tunneling strength, $U$ is the on-site interaction strength, $\hat{b}_j$ and $\hat{b}_j^\dagger$ are bosonic ladder operators satisfying the canonical commutation relations $\big[\hat{b}_j,\hat{b}_r^\dagger\big]{=}\delta_{j,r}$, $\hat{n}_j{=}\hat{b}_j^\dagger\hat{b}_j$ is the bosonic number operator at site $j$, and $\Delta$ is an overall tilt. The staggering potential $\delta$ distinguishes between matter sites (odd $j$) and gauge links (even $j$). Connecting this to Eq.~\eqref{eq:QLM} means that $\ell$ corresponds to odd bosonic sites $j$, while the link between $\ell$ and $\ell{+}1$ corresponds to even bosonic sites $j$, where the bosonic model~\eqref{eq:BHM} hosts a total of $L{=}2 L_m-1$ sites, with $L_m$ the number of matter sites.  
The second staggering potential,
\begin{align}
    \chi_j=
    \begin{cases}
      0 & \text{if}\,\,\,j\,\mathrm{mod}\,2=1,\\
      \chi & \text{if}\,\,\,j\,\mathrm{mod}\,4=0,\\
      -\chi & \text{if}\,\,\,j\,\mathrm{mod}\,4=2,
    \end{cases}  
\end{align}
is related to the topological $\theta$-term, and in the bosonic lattice distinguishes between odd and even gauge links, but has no effect on the matter sites.

The mapping between the bosonic and QLM representations is such that on an odd bosonic site, which represents a site of the QLM, the presence of a single boson represents matter occupation, while no bosons means matter is absent. On an even bosonic site, which represents a link of the QLM, zero (two) bosons indicate that the local electric field points down (up). As such, we need to enforce these local occupations in an experiment. This is achieved in the regime of $U,\delta\gg J,\mu$, where Eq.~\eqref{eq:QLM} derives from Eq.~\eqref{eq:BHM} up to second order in perturbation theory \cite{Yang2020,Halimeh2022tuning}. 

The dominating terms of Eq.~\eqref{eq:BHM} are then diagonal in the on-site bosonic number operator, and can be collected as
\begin{align}\nonumber
    &\hat{H}_\mathrm{d}=\sum_\ell \bigg\{\frac{U}{2} \big[\hat{n}_\ell \big(\hat{n}_\ell-1\big) + \hat{n}_{\ell,\ell+1} \big(\hat{n}_{\ell,\ell+1}-1\big)\big]\\\label{eq:Hd}
    &+\Big[(-1)^\ell \frac{\chi}{2}-\delta\Big] \hat{n}_{\ell,\ell+1}+ \Delta \big[ 2\ell \hat{n}_\ell + (2\ell+1) \hat{n}_{\ell,\ell+1}\big]\bigg\},
\end{align}
where we have resorted to the QLM indexing, which relates to that of Eq.~\eqref{eq:BHM} as $\ell$ corresponding to an odd bosonic site $j$, while the link between sites $\ell$ and $\ell{+}1$ corresponds to $j{+}1$, the even (i.e., gauge) site between the odd (i.e., matter) sites $j$ and $j{+}1$. We can now define a ``proto-Gauss's law'' with the generator
\begin{align}\label{eq:proto}
\hat{\mathcal{G}}_\ell= (-1)^\ell\bigg[\frac{1}{2}\big(\hat{n}_{\ell-1,\ell}+\hat{n}_{\ell,\ell+1}\big) + \hat{n}_\ell -1\bigg].
\end{align}
Relating the configurations allowed by Gauss's law in both the bosonic and QLM representations, we find that $\mu=\delta-U/2$, which we can insert in Eq.~\eqref{eq:Hd} and utilize Eq.~\eqref{eq:proto} to get
\begin{align}\nonumber
    \hat{H}_\mathrm{d}=\sum_\ell \bigg\{&\frac{U}{2} \big[\hat{n}_\ell \big(\hat{n}_\ell-1\big) + \hat{n}_{\ell,\ell+1} \big(\hat{n}_{\ell,\ell+1}-2\big)\big]\\\label{eq:Hd_final}
    &+\Big[(-1)^\ell \frac{\chi}{2}-\mu\Big] \hat{n}_{\ell,\ell+1}+c_\ell\hat{\mathcal{G}}_\ell\bigg\},
\end{align}
up to an inconsequential energy constant, with $c_\ell{=}2\Delta(-1)^\ell\ell$. This formulation with the ``Stark'' coefficients $c_\ell$ has recently been shown to suppress couplings between different gauge sectors, and hence stabilize gauge invariance, up to all numerically accessible times \cite{Lang2022stark} based on the concept of \textit{linear gauge protection} \cite{Halimeh2020e}. Looking at Eq.~\eqref{eq:Hd_final}, we see that a large on-site potential constraints the local bosonic configurations on sites to $\{\ket{0}_\ell,\ket{1}_\ell\}$ and on links to $\{\ket{0}_{\ell,\ell+1},\ket{2}_{\ell,\ell+1}\}$, as desired. The latter correspond to the local eigenstates of the Pauli operator $\hat{\sigma}^z_\ell$ and $\hat{s}^z_{\ell,\ell+1}$ of Eq.~\eqref{eq:QLM}. In this regime, Eqs.~\eqref{eq:proto} and~\eqref{eq:G} become equivalent. Using degenerate perturbation theory, the parameters of Eqs.~\eqref{eq:BHM} and~\eqref{eq:QLM} can be related through
\begin{align}\label{eq:kappa}
    \kappa=2\sqrt{2}J^2\bigg[\frac{\delta}{\delta^2-\Delta^2}+\frac{U-\delta}{(U-\delta)^2-\Delta^2}\bigg],
\end{align}
where, as mentioned, $\mu=\delta-U/2$ is the fermionic mass.

The extended BHM Hamiltonian~\eqref{eq:BHM} can be realized by a three-period optical superlattice formed by standing waves of lasers. The main lattice laser with wavelength $\lambda$ forms a lattice with period $\lambda/2$, and two additional lattices with period $\lambda$ and $2\lambda$, respectively. These lasers can be phase stabilized with respect to each other to generate the desired superlattice potential as described in Ref.~\cite{Halimeh2022tuning}. The two vacuum states in the gauge theory correspond to states $\ket{000200020 \ldots}$ and $\ket{020002000 \ldots}$ in the BHM. They can be prepared with site-selective addressing techniques using the spin-dependent superlattice, also described in Ref.~\cite{Halimeh2022tuning}.

Now that we have reviewed the experimental implementation of the QLM with confinement in the tilted Bose-Hubbard model as proposed in Refs.~\cite{Halimeh2022tuning,Cheng2022tunable}, we can show how our findings have clear signatures in that model.
In keeping with experimental relevance, we set the microscopic parameters of Hamiltonian~\eqref{eq:BHM} to $U=1368$ Hz, $J=58$ Hz, and $\Delta=57$ Hz, which are close to the values of these parameters as employed in the experiment of Ref.~\cite{Zhou2021}.
We first investigate the crossover from deconfined scarring to fragmentation. In the deconfined case ($\chi=0$), when quenching the vacuum state to $\mu=0$, the system displays scarred dynamics in form of ``state transfer'' between the two vacuum states, see Fig.~\ref{fig:ConfinedScarring}(b). We obtain the numerical results for the BHM using the TenPy toolkit \cite{tenpy} (see \cite{SM} for details). The system shows persistent many-body oscillations between the two vacuum states, as can be seen in Fig.~\ref{fig:ConfinedScarring}(d). 

When the confinement potential is increased to $\chi=0.35\kappa$, the system remains ergodic, while scarring from the N\'eel state persists. This regime of confined scarring is discussed in Sec.~\ref{sec:QMBS}. The transfer to the opposite vacuum state is suppressed due to confinement, but dynamics between the unit cells are still present, as shown on Fig.~\ref{fig:ConfinedScarring}(e). At late times, we actually see an occupation of the gauge sites closer to 2 than in the $\chi=0$ case, indicating that the state returns nearer to the N\'eel state. 

The system becomes fragmented when confinement potential $\chi$ is above $0.5 \kappa$. Here, we use the resonant case with $\chi = \kappa$ and $\mu = -0.5 \kappa$ as an example. As shown in Fig.~\ref{fig:ConfinedScarring}(c), unit cells marked in gray shades are tuned on resonance with $\mu = - \chi/2$, while the adjacent unit cells are out-of-resonance. Therefore dynamics are restricted within the unit cells, see the numerical results in Fig.~\ref{fig:ConfinedScarring}(f).

We further probe confinement and fragmentation by investigating the spread of a particle-antiparticle excitation in the vacuum background~\cite{Surace2020,Halimeh2022tuning,zhang2023observation}. In the Bose--Hubbard quantum simulator, this state corresponds to a $\ket{\ldots 0101 \ldots}$ impurity at the center of the $\ket{\ldots 0020 \ldots}$ background, see Fig.~\ref{fig:defect_explanation}(a). This state can be prepared by a local addressing operation in the optical superlattice, as described in \cite{Halimeh2022tuning}. In the deconfined case, the excitation couples with neighboring unit cells, see Fig.~\ref{fig:defect_explanation}(b). The light-cone spreading can be seen in the numerical results in Fig.~\ref{fig:defect_explanation}(e) and (g), calculated numerically using MPS. Additional to the mean density $\braket{\hat{n}_j}$ in panel (e), in panel (g) we calculate the density-density correlation between site 43 (a gauge site between the two excitations) and the rest of the gauge sites, $\braket{\hat{n}_{\text{M}}\hat{n}_{j}}_c$. This quantity exactly maps to the ZZ correlator used in the PXP model.

In the fragmented case, however, the dynamics are restricted within each unit cell, as we demonstrate for the case $\chi = \kappa$ and $\mu = -0.5 \kappa$ in Fig.~\ref{fig:defect_explanation}(c). The excitation is disconnected from the rest of the system and goes on its own dynamics, see Fig.~\ref{fig:defect_explanation}(d), therefore, no spreading can be observed in the density-density correlation, see Fig.~\ref{fig:defect_explanation}(f).

We note that the PXP model with a confining potential can also be implemented in Rydberg atoms~\cite{Bernien2017,Surace2020} and in a tilted Bose-Hubbard simulators~\cite{Su2022, zhang2023observation}. While in this implementations only the physical sector with no background charges can be probed, it presents other advantages compared to the mapping discussed in this section. Notably, even with the confining potential only two lattices (with period $\lambda$ and $2\lambda$) are needed. As the matter degrees of freedom have been integrated out, with the same number of bosonic sites one can probe twice the number of gauge sites.

\section{Conclusions and outlook}\label{sec:conc}

Using a combination of analytic and numerical methods, we have calculated the dynamical phase diagram of the $\mathrm{U}(1)$ quantum link model with a topological $\theta$-term, or equivalently, the PXP model with mass and staggered magnetization terms~\cite{Surace2020}. By tuning the topological $\theta$-angle on a quantum simulator, as proposed in Refs.~\cite{Surace2020,Halimeh2022tuning} and realized in Ref.~\cite{zhang2023observation}, it is possible to controllably induce confinement in this model. We map out various ergodicity-breaking phases that can be observed experimentally by tuning the confining potential $\chi$ and the effective mass $\mu$ in a cold-atom setup. Starting from the ergodic phase at small values of the mass and confining potential, we study the crossover to the prethermal fragmented phase at large $\chi$. Additionally, we have identified various resonant processes in the $(\chi,\mu)$ dynamical phase diagram, which we have analyzed in detail. We further uncovered regimes of robust quantum many-body scarring in the presence of confinement.

Our results are readily accessible in modern cold-atom quantum simulators \cite{Yang2020,Zhou2021} by adding a staggering potential in the associated tilted Bose--Hubbard optical superlattice~\cite{Halimeh2022tuning,Cheng2022tunable}. Our findings highlight a very interesting phenomenological and technological aspect. They show that confinement and quantum many-body scarring, two paradigmatic phenomena of gauge theories, can arise concomitantly. Importantly, our results also show that this behavior can be captured on current quantum simulators, which is encouraging given the current drive of probing high-energy physics on table-top quantum devices. Our work also further sheds light on the question of the gauge-theoretic origin of quantum many-body scarring. Previously, scarring and confinement were thought to arise in different regimes, but here we show that they can coexist and give rise to interesting dynamics. Confinement is a fundamentally gauge-theoretic phenomenon, so it is interesting to further investigate the nature of confined scarring in more generic gauge theories such as $\mathbb{Z}_2$ lattice gauge theories \cite{Aramthottil2022} or quantum link models in higher spatial dimensions \cite{Banerjee2020}.

\begin{acknowledgments}
J.C.H.~acknowledges stimulating discussions with Pablo Sala. The authors are grateful to Aiden Daniel, Andrew Hallam, Philipp Hauke, Ana Hudomal, Jian-Wei Pan, and Zhen-Sheng Yuan for discussions and works on related topics. 
B.Y.~acknowledges support from National Key R$\&$D Program of China (grant 2022YFA1405800) and NNSFC (grant 12274199). I.P.M.~acknowledges support from the Australian Research Council (ARC) Discovery Project Grants No.~DP190101515 and DP200103760.
I.P.M.~and Z.P.~acknowledge support by the Erwin Schr\"odinger International Institute for Mathematics and Physics (ESI). J.-Y.D.~and Z.P.~acknowledge support by EPSRC grant EP/R513258/1 and by the Leverhulme Trust Research Leadership Award RL-2019-015. Statement of compliance with EPSRC policy framework on research data: This publication is theoretical work that does not require supporting research data. This research was supported in part by the National Science Foundation under Grant No. NSF PHY-1748958. J.C.H.~acknowledges funding within the QuantERA II Programme that has received funding from the European Union’s Horizon 2020 research and innovation programme under Grand Agreement No 101017733, support by the QuantERA grant DYNAMITE, by the Deutsche Forschungsgemeinschaft (DFG, German Research Foundation) under project number 499183856, funding by the Deutsche Forschungsgemeinschaft (DFG, German Research Foundation) under Germany's Excellence Strategy -- EXC-2111 -- 390814868, and funding from the European Research Council (ERC) under the European Union’s Horizon 2020 research and innovation programm (Grant Agreement no 948141) — ERC Starting Grant SimUcQuam.
\end{acknowledgments}

\newpage
\appendix

\section{Level statistics with \texorpdfstring{$\mu$ and $\chi$}{mu and chi}}\label{app:rstat}

In order to identify non-ergodic regimes, we compute the level-spacing statistics characterized by the $\langle r \rangle$ value~\cite{OganesyanHuse}. This is only meaningful once symmetries have been resolved, and we consider this quantity only in the most symmetric sector. The relevant symmetries and the sector will change depending whether $\mu$ and $\chi$ are zero or not. Let us denote the translation operator by $\ST$ and the spatial inversion operator by $\SP$. 
When $\chi=0$, the Hamiltonian commutes with both $\ST$ and $\SP$, and we consider the symmetry sector with momentum $k=0$ and spatial inversion eigenvalue $p=+1$. When $\chi\neq 0$, the Hamiltonian no longer commutes with either $\ST$ nor $\SP$ (as we consider even system sizes). However, it does commute with $\ST^2$ and $\ST \SP$. Just as $\ST$ and $\SP$ commute in the zero-momentum sector, $\ST^2$ and $\ST\SP$ commute when the momentum $ k_2$ (with respect to $\ST^2$) is 0. So in these cases we focus on the symmetry sector $k_2=0$ and $p_T=+1$. 
Finally, as shown in SM~\cite{SM}, when $\mu=0$ the Hamiltonian has a particle-hole symmetry and consequently a large number of zero modes. We always truncate the spectrum to remove these modes and we also exclude the edges of the spectrum with the low density of states.

\begin{figure}[tb]
	\centering
	\includegraphics[width=\linewidth]{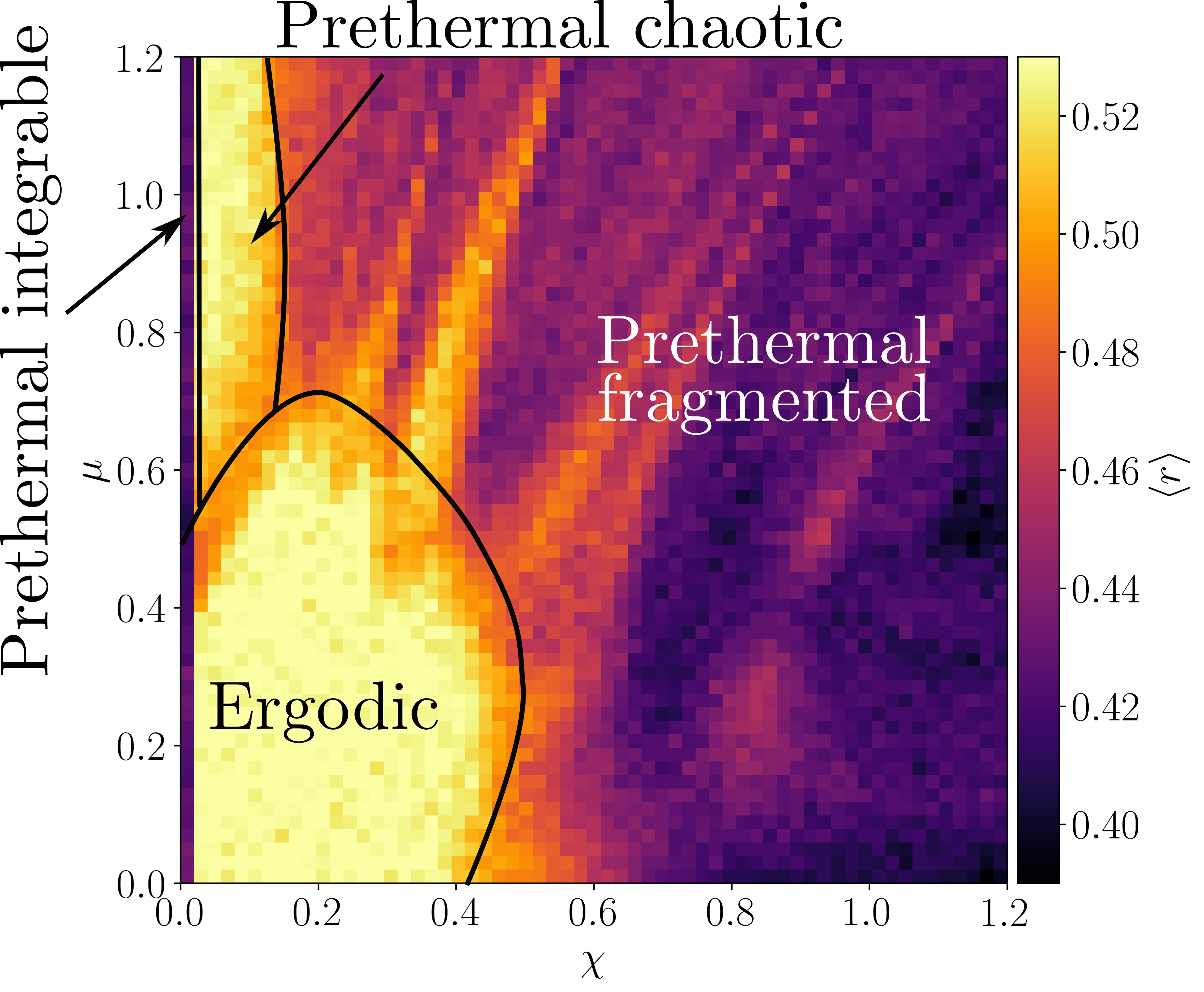}
	\caption{Level spacing statistics in the PXP model with PBC and $L_m=28$. All values  below $0.39$ are set to this value and all values above $0.53$ are set to $0.53$. The black lines represent the approximate boundaries between different regimes.
    }
	\label{fig:rstat_mu_chi}
\end{figure}

\begin{figure}[htb!]
	\centering
 	\includegraphics[width=\linewidth]{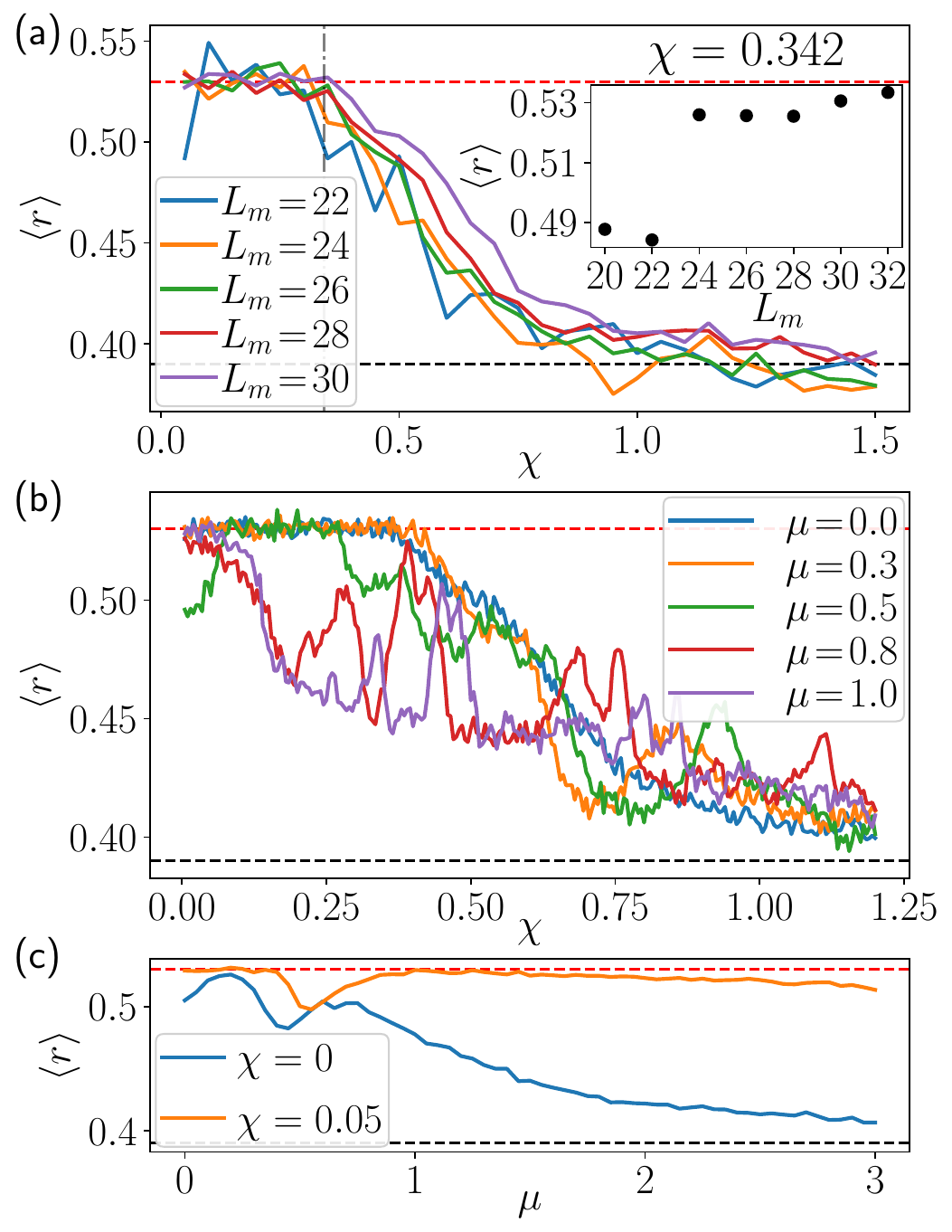}
	\caption{Level spacing statistics in the PXP model with PBC for various values of $\mu$, $\chi$ and $L_m$. The black and red dashed lines indicate the expected values for Poisson and Wigner-Dyson distributions, respectively. (a) Results for $\mu=0$, with each colored curve corresponding to a different system size. The inset shoes the system size scaling at the optimal value of $\chi$ that maximizes revivals from the N\'eel state. (b) Results for $L_m=30$, with each curve corresponding to a fixed value of $\mu$. The values for $\chi=0$ are not included as they have different symmetries. As $\mu$ is increased, a lower value of $\chi$ is needed to deviate from 0.53. For $\mu\geq 0.4$, one can see peaks in the $\langle r\rangle$ value. However, further inspection shows that these regimes are not truly ergodic (see the text). (c) Results for $L_m=30$ and two different values of $\chi$. A small $\chi$ is enough to break the emergent integrability that appears at large $\mu$.}
	\label{fig:rstat_fixed_mu}
\end{figure}

Once the symmetries have been resolved, for small values of $\mu$ and $\chi$ we expect the Hamiltonian to consist of a single connected component which is non-integrable. However, when fragmentation sets in, we have several disconnected sectors that do not have level repulsion between them. We have also seen that away from resonances the effective Hamiltonian at second order is non-interacting and so trivially integrable. As such we expect a quick departure from the Wigner-Dyson level statistics as soon as fragmentation starts to occur. Figure~\ref{fig:rstat_mu_chi} shows that a value of $\chi \sim0.5 \kappa$ is sufficient to break ergodicity for $L_m=28$ sites. The departure from ergodicity around $\chi/\kappa\approx 0.5$ is confirmed by Fig.~\ref{fig:rstat_fixed_mu} (a) that shows the scaling of the $r$ ratio for various values of $\chi$ and $L_m$. As expected, a larger perturbation is needed to induce signs of fragmentation in the level statistics as $L_m$ is increased. Nonetheless, spectrum statistics essentially probe infinite-time properties of the system. At the values where we see fragmentation in Figure~\ref{fig:rstat_mu_chi}, at finite times after a quench we still expect to see signatures of fragmentation in larger systems. Hence for all system sizes we can denote this regime as ``prethermal fragmented''. This allows us to use  Figure~\ref{fig:rstat_mu_chi} to sketch the phase diagram of the system in Figure~\ref{fig:schematic}(c).

\begin{figure}[tb!]
	\centering
	\includegraphics[width=\linewidth]{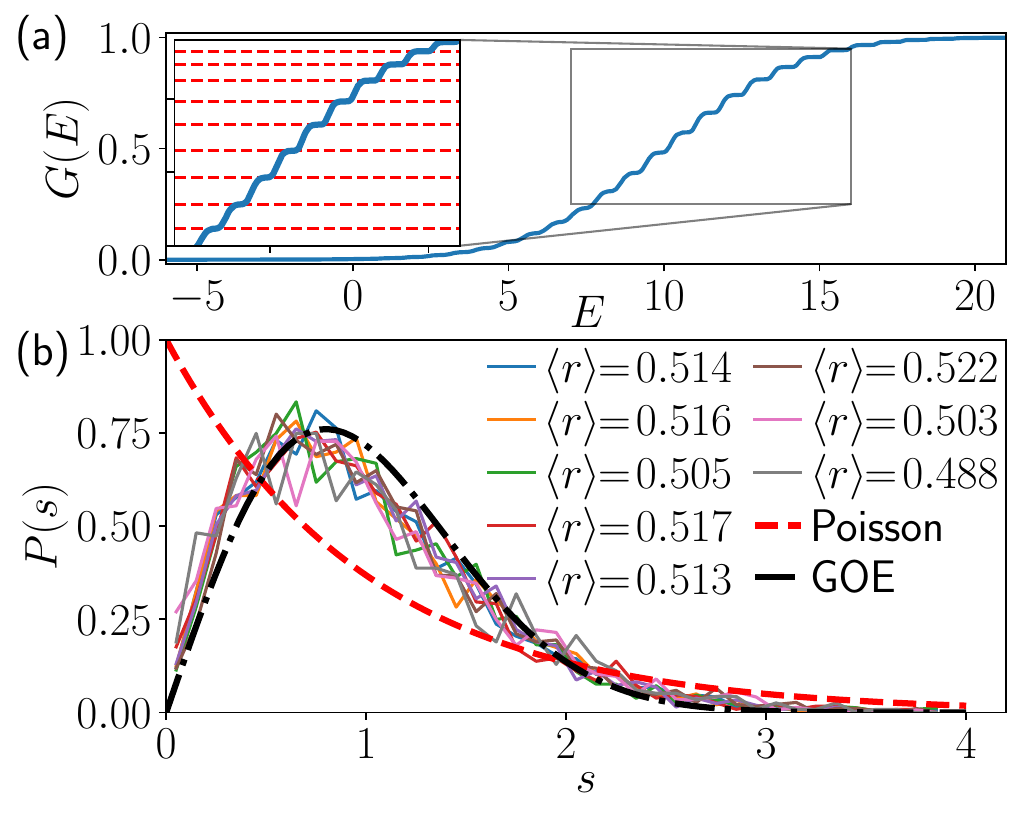}
	\caption{Level spacing statistics in the PXP model with $\mu=0.8$ and $\chi=0.39$ in a system size $L_m=28$ with PBC. The staircase-like behavior of $G(E)$ in (a) is emblematic of the presence of sectors with many energy levels separated by empty intervals where $G(E)$ does not grow. The red dashed lines in the inset indicate the fraction of eigenstates chosen as limits between sectors. In each sector the spectrum was then unfolded and the probability distribution of level spacings $s$ was plotted in (b).}
	\label{fig:rstat_split}
\end{figure}

Moving away from $\mu=0$, we see a nonmonotonic behavior of $\langle r \rangle$ as $\chi$ is increased. This is manifested by the lighter ``rays'' emanating from the ergodic region in Fig.~\ref{fig:rstat_mu_chi}. These seemingly special ergodic lines are also visible in Fig.~\ref{fig:rstat_fixed_mu}(b), which shows the influence of $\chi$ for various fixed $\mu$.  In order to show that these regions are not fully ergodic, we will concentrate on the case with $\mu=0.8$ and $\chi=0.39$, which corresponds to a local maximum of $\langle r \rangle$. We compute the integrated density of states 
\begin{equation}
    G(\epsilon)=\frac{1}{\mathcal{D}}\sum_E \Theta (\epsilon-E),
\end{equation}
which simply counts the fraction of energy levels below energy $\epsilon$.
For a fully chaotic system in which symmetries have been resolved, we expect $G(\epsilon)$ to be a smooth function. However, plotting $G$ for $\mu=0.8$ and $\chi=0.39$ in Fig.~\ref{fig:rstat_split}(a) shows visible plateaus, indicating gaps in the spectrum where no levels are found. If we now consider each set of eigenstates between the plateaus as independent spectra, we can unfold them and plot the distribution of level spacings as in Fig.~\ref{fig:rstat_split}(b). Individually, they show good agreement with the Wigner surmise, and consequently the $\langle r \rangle$ value is close to 0.53. However, due to this splitting into bands, the full system is clearly not ergodic. 
Thus, we understand the non-uniformity of the ``prethermal fragmented" region in Fig.~\ref{fig:rstat_mu_chi} as the result of interplay between disconnected components. Indeed, the first few orders of the effective Hamiltonian for generic $\mu$ and $\chi$ do not connect the entire Hilbert space. Thus, for large enough values of these parameters, we end up with an extensive number of disconnected components that can be diagonalized independently. Each of these will have energies centered around some energy $\overline{E}$, which depends on $\mu$ and $\chi$. Around resonances, several disconnected components can end up close in energy. This can happen without them becoming connected by the low-order terms of the effective Hamiltonian. In that case, their energy levels will overlap, thus impacting the energy spacing and the average $\langle r \rangle$ value. This will lead to lower $\langle r \rangle$ value, as in the case of unresolved symmetries.

For $\mu$ the picture is different, as even for $\mu=1.2$ and $\chi=0.1$ the $\langle r \rangle$ value is close to $0.53$. However, further investigation shows that this does not indicate the Hamiltonian is fully ergodic. Indeed, for a small $\chi$ and a large $\mu$ the dominant Hamiltonian is in the one in Eq.~(\ref{eq:Heff12_res_mu}), which is non-integrable for any nonzero $\chi$. We still have $L_m/2+1$ $\mathrm{U}(1)$ sectors (corresponding to the total excitation number), but they are very far apart energy due to $\mu$ being large. So their energies do not overlap and as such do not create any accidental degeneracies (or near-degeneracies). The level spacing statistics is dominated by the spacings \emph{within} each sector, which all obey Wigner-Dyson distribution. Hence starting in a sector will lead to prethermalization within that sector for an exponential long time depending on $\mu/\kappa$, before then starting to explore the other sectors. 
This means that, at short and intermediate times, the dynamics will only explore a part of the system and so it is non-ergodic. If instead $\chi=0$, then the Hamiltonian in each $\mathrm{U}(1)$ sector is integrable and we see much faster convergence towards Poisson statistics with $\mu$, as shown in Fig.~\ref{fig:rstat_fixed_mu}(c).
Nonetheless, as we have noted above, for $\mu\gg \chi, \kappa$ there is an emergent $\mathrm{U}(1)$ conservation law at short and intermediate time but no fragmentation within each $\mathrm{U}(1)$ sector.

\section{Spin-1 parent Hamiltonian}\label{app:spin1}

\begin{figure}[t!]
	\centering
	\includegraphics[width=\linewidth]{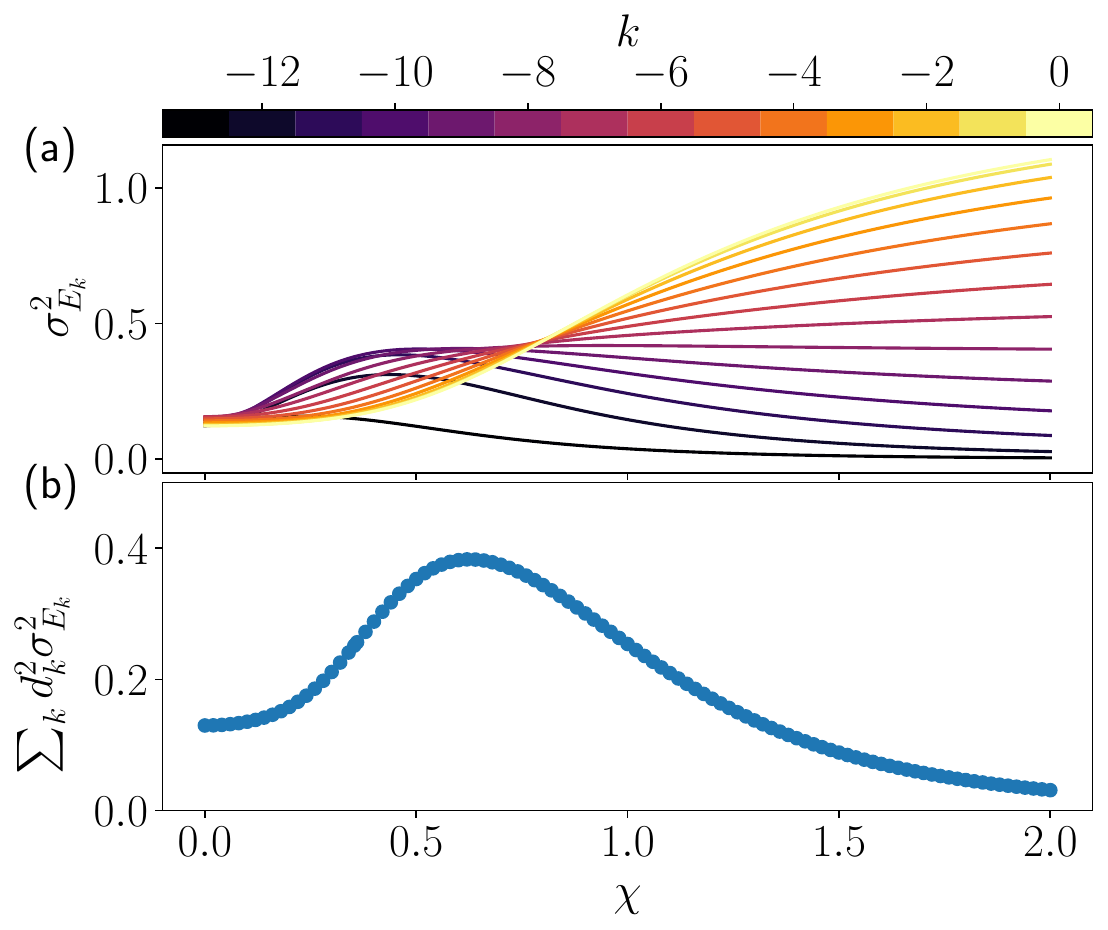}
	\caption{Energy variance of the $\ket{S_k}$ states in the PXP model with $L_m=26$. (a) Data for individual states. (b) Data weighted by the overlap with the N\'eel state, which accounts for their relevance in the dynamics. The nonmonotonic behavior in $\chi$ with local minima at $\chi=0$ and $\chi \to \infty$ is clearly visible. This explains why these points are local maxima of revival fidelity.
     }
	\label{fig:Sk_var}
\end{figure}

In this section we use the spin-1 picture of the PXP model~\cite{Omiya22} to derive the optimal $\chi$ that enhances the QMBS revivals from the N\'eel initial state.  We consider the eigenstates $\ket{\tilde{S}_k}$ of the integrable part of the spin-1 parent model of PXP, $-\frac{1}{\sqrt{2}}\hat{S}^x-\chi\hat{S}^z$. If we denote by $N_b=L_m/2$  the number of sites in the spin-1 model, the index $k$ runs between $-N_b$ and $N_b$. The states $\ket{\tilde{S}_k}$ are all eigenvalue of angular momentum with total spin and energy $E_k=\sqrt{1/2+\chi^2}$.
We can define their projected counterparts $\ket{S_k}$, and compute their energy variance in the PXP model, 
$\sigma_{E_k}^2 \equiv \langle S_k | \hat H_\mathrm{PXP}^2 | S_k\rangle - \langle S_k | \hat H_\mathrm{PXP} | S_k\rangle^2$. 
The variance is shown in Fig.~\ref{fig:Sk_var}(a), where we observe very different behavior depending on the value of $k$.
In order to pick out the relevant states, we need to understand their overlap with the N\'eel state. At $\chi=0$, the N\'eel state can be written as a superposition of the $\ket{\tilde{S}_k}$ states as~\cite{Omiya22}
\begin{equation}\label{eq:olap_Neel0}
\ket{\mathbb{Z}_2}=\sum_{k=-N_b}^{N_b} d_k \ket{\tilde{S}_k}, \quad d^2_k=\binom{2N_b}{k+N_b}.
\end{equation}
For arbitrary $\chi$, a similar expansion holds as 
\begin{equation}\label{eq:olap_Neel}
    d_k^2=\frac{\binom{2N_b}{k+N_b}}{4^{N_b}}\frac{(\sqrt{2}\chi-\mathrm{sgn}(k)\sqrt{1+2\chi^2})^{2|k|}}{(1+2\chi^2)^{N_b}}.
\end{equation}
A simple argument as to why this is possible is that the N\'eel state is the highest-weight eigenstate of $S^z$ in the spin-1 picture. This implies that it also has maximal total spin, and the $\ket{\tilde{S}_k}$ always form a basis of maximal total spin states. We can then use this to compute the average of the $\sigma^2_{E_k}$ weighted by it. This is shown in Fig.~\ref{fig:Sk_var}(b), where we recover the anticipated nonmonotonic behavior in $\chi$. The local minima at $\chi=0$ and $\chi \to \infty$ are visible, and this explain why as we go away from these points the fidelity of the revivals starts to decay. However, the peak at $\chi=1/\sqrt{8}$ is not visible here. To understand its appearance we need to take into account the spacing of the eigenvalues.

For that we need to compute the perturbation of the energies of the $\ket{\tilde{S}_k}$ states due to the presence of $\hat{H}_1$ in the full Hamiltonian in Eq.~(\ref{eq:PXP_S1}). Thanks to the simple structure of the $\ket{\tilde{S}_k}$, it is possible to compute analytically the first-order correction as $\Delta E_k=\langle \tilde{S}_k |\hat{H}_1 | \tilde{S}_k \rangle $. The derivation is similar to Appendix~D of \cite{Omiya22} and can be found in SM~\cite{SM}.
The final result is
\begin{equation}
\Delta E_k=-k\frac{(1-3N_b)+k^2(1-8\chi^2) +N_b^2(1+8\chi^2)}{8\sqrt{2} (1 + 2 \chi^2)^{3/2}(N_b-1)(2N_b-1)}.
\end{equation}
In general, this cannot be factorized further. However,  in the special case $\chi=1/\sqrt{8}$, we find it reduces to $-k/(5\sqrt{10})$.
This is a remarkable result: for any state $\ket{\tilde S_k}$, at first order its energy will be given by 
\begin{equation}
E_k=k\left[\sqrt{\frac{5}{8}}-\frac{1}{5\sqrt{10}}\right]=\frac{23k}{10\sqrt{10}}.
\end{equation}
As a consequence, the energy spacing between two consecutive eigenstates will simply be $\frac{23}{10\sqrt{10}}$, which is independent of both $k$ and $N_b$. Thus, at first order, all scarred states are exactly equidistant for any system size. This means that any superposition of these states will have good revivals, which explains the peak in fidelity around this value of $\chi$.

This special property of the spacing at $\chi=1/\sqrt{8}$ is linked to a change in the behavior of $\Delta E_k$. Namely, for $|\chi |<1/\sqrt{8}$, $\Delta E_k$ is larger near the edges of the spectrum, while for $|\chi |>1/\sqrt{8}$,  $\Delta E_k$ is at its largest in the middle of the spectrum. This becomes much clearer after taking the thermodynamic limit, $N_b \to \infty$. We can the define $s=k/N_b$, where $s\in [-1,1]$, to get
\begin{equation}
\begin{aligned}
\Delta E_s=\;- N_bs \frac{(1+8\chi^2)+s^2(1-8\chi^2)}{16\sqrt{2} (1 + 2 \chi^2)^{3/2}}
    \end{aligned}
\end{equation}
Adding the nominal reference energy of $sN_b\sqrt{1/2+\chi^2}$, we get
\begin{equation}
    E_s=sN_b\frac{15+56\chi^2+64\chi^4-s^2(1-8\chi^2)}{64(1/2 + \chi^2)^{3/2}}
\end{equation}
As the period of revivals is determined by the energy difference between the eigenstates, we need to compute $E_{k+1}- E_k=\frac{1}{N_b}\frac{\dd E_s}{\dd s}$, which gives
\begin{equation}\label{eq:Es_prime_app}
E^\prime_s=\frac{1}{N_b}\frac{\dd E_s}{\dd s}{=}\frac{15+56\chi^2+64\chi^4-3s^2(1-8\chi^2)}{64(1/2 + \chi^2)^{3/2}}.
\end{equation}
The change of sign of the term linked to $s^2$ at $\chi=1/\sqrt{8}$ is now fully apparent. 

In order to predict the revival frequency, we need to target the value of $s$ which maximizes the overlap with the N\'eel state. Eq.~(\ref{eq:olap_Neel}) gives us the overlap with $\ket{\tilde{S}_k}$; this is not the same as  $\ket{S_k}$, however the difference between the two is polynomial in $N_b$. As we will show later, this means we can neglect this as a subleading correction in the limit $N_b \to \infty$. In the same limit, we can use the Stirling approximation
\begin{equation}
\binom{2N_b}{k+N_b}\approx\frac{e^{-2N_b H(\frac{k+N_n}{2N_b})}}{\sqrt{\pi (N_b+k)(1-k/N_b)}},
\end{equation}
where $H(x)=-x\log(x)-(1-x)\log(1-x)$ is the Shannon entropy function. We can then rewrite the $d^2_k$ using $s=k/N_b$ to get
\begin{equation}
    d_s^2=\frac{e^{-2N_b H(\frac{1+s}{2})}}{\sqrt{\pi N_b(1-s^2)}}\frac{(\sqrt{2}\chi-\mathrm{sgn}(s)\sqrt{1+2\chi^2})^{2N_b|s|}}{(1+2\chi^2)^{N_b}}
\end{equation}
Let us first focus on the case $\chi>0$. In that regime, $\sqrt{2}\chi -\sqrt{1+2\chi^2}$ goes to 0 as $\chi$ becomes large while $\sqrt{2}\chi +\sqrt{1+2\chi^2}$ increases with $\chi$. This is expected as the N\'eel state gets closer to the ground state for larger positive $\chi$. A consequence is that for positive $\chi$ the maximum value will always be in the range $[-1,0]$. We can then restrict to this range and drop the signum function and absolute values.
In the end, this allows us to rewrite
\begin{equation}
    d_s^2=\frac{e^{-N_b f(s)}}{4^{N_b}(1+2\chi^2)^{N_b}\sqrt{\pi N_b(1-s^2)}}
\end{equation}
with 
\begin{equation}
\begin{aligned}
    f(s)=&(1+s)\log(1+s)+(1-s)\log(1-s)\\
    &+2s\log\left(\sqrt{2}\chi+\sqrt{1+2\chi^2}\right).
\end{aligned}
\end{equation}
We recognize here that in the thermodynamic limit the most important term is the exponential $e^{-N_b f(s)}$ and we use a saddle point approximation. This means that the maximum of $d_s^2$ will simply be the minimum of $f(s)$, which yields $s_0$ that was given in Eq.~(\ref{eq:s0}) in the main text.
This justifies why we can neglect the difference in norm between the $\ket{\tilde{S}_k}$ and $\ket{S_k}$: as it is polynomial in $N_b$ and $s$, it will irrelevant in the thermodynamic limit and lead to the same $s_0$.

\begin{figure}[t!]
	\centering
	\includegraphics[width=.48\textwidth]{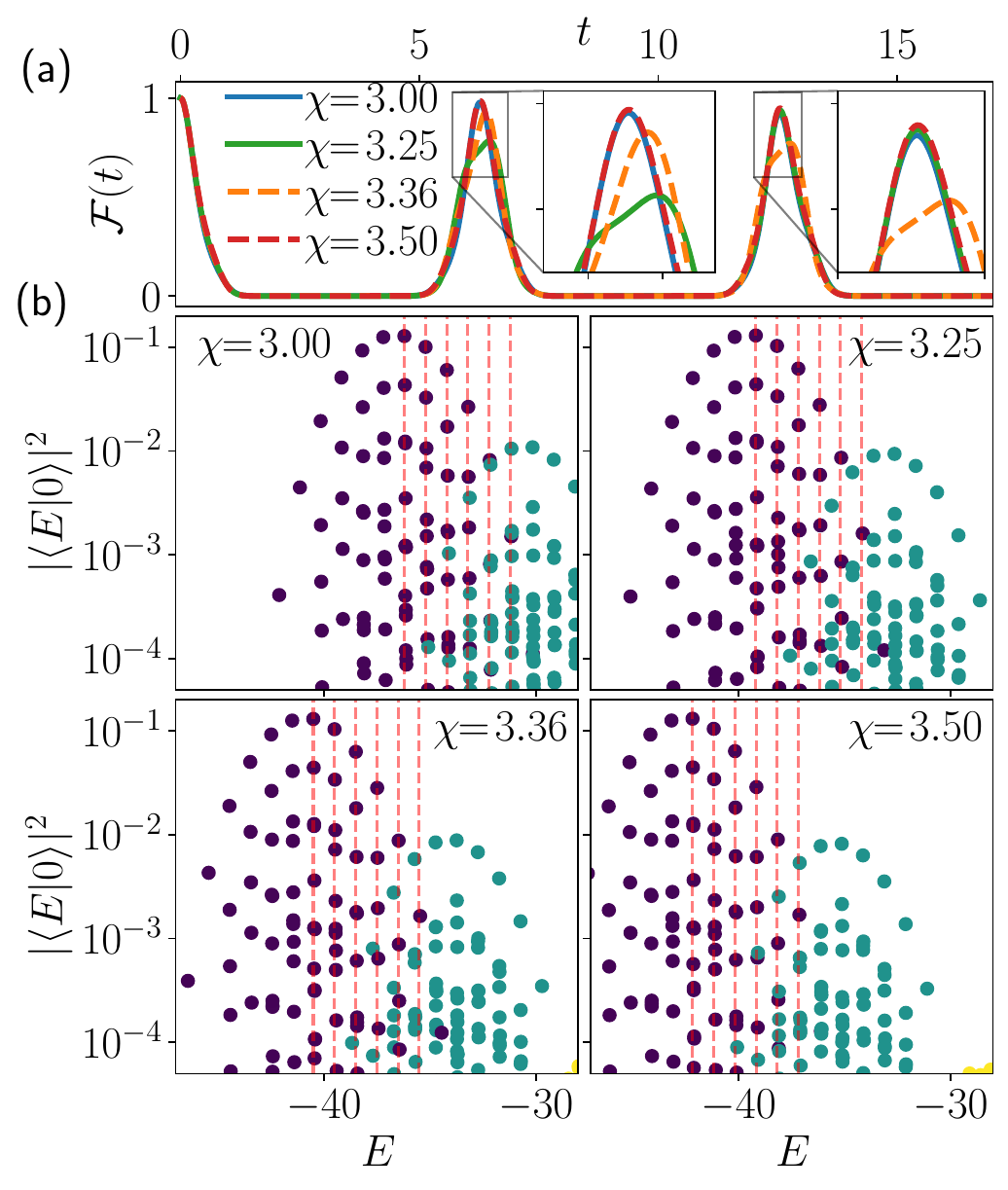}\\
	\caption{Quenches from the polarized state in the PXP model with $L_m=24$ along the resonance line $\chi=-2\mu$. (a) Self-fidelity after a quench. When $\chi$ is integer or half-integer, the first peak is significantly higher. (b) Overlap between polarized state and the eigenstates. The color indicates the number of excitations on odd sites and so differentiates states in different sectors. The red dashed-lines are all equally spaced by one unit of energy. When $\chi$ is integer or half-integer, the spacing between sectors is a multiple of the energy spacing within each of them and all eigenstates are are approximately equally spaced.}
	\label{fig:ResPolBands} 
\end{figure}

A quick sanity check shows that Eq.~(\ref{eq:s0}) correctly yields $s_0=0$ at $\chi=0$ and $s_0 \to -1$ as $\chi\to \infty$. A similar derivation can be done for $\chi<0$, yielding the same formula but without the minus sign in front.
This allows us to estimate the frequency of oscillations for a given $\chi$ after a quench from the N\'eel state by plugging $s_0$ into Eq.~\eqref{eq:Es_prime}. While this derivation is done in the thermodynamic limit, Fig.~\ref{fig:FT_chi} in the main text shows a good agreement in a finite system with $N_b=13$.

\section{Polarized state at \texorpdfstring{$\chi\approx \pm 2\mu$}{chi close to 2 mu}}\label{app:pol_res}
At the resonance point $\chi\approx -2\mu$, the Hilbert space fractures when $\chi, \ \mu \gg \kappa$ as creating an excitation on an odd site is extremely costly in energy. However, when $1<\chi/\kappa<10$ such moves are still possible from the polarized state. As a consequence, the dynamics is not restricted to the Hilbert space sector with no excitations on the odd sites, but the sector with one excitation of these sites also has some nonzero contribution. For a detuning $\gamma=2\mu+\chi$, the Hamiltonian in each fragment is given by Eq.~\eqref{eq:res_n0_main} in the main text. It is straightforward to see that the energy spacing between eigenstates in that system is $\sqrt{1+\gamma^2}$ for $\kappa=1$. Meanwhile, the energy spacing between the ``center'' of fragments with respectively 0 and 1 excitations on odd sites is largely unaffected by $\gamma$ and is approximately $2\chi$. As a consequence, in order for all energies to be spaced by a regular amount we require one of these quantities to be a multiple of the other. As $\chi \gg \gamma$, it must be that $2\chi > \sqrt{1+\gamma^2}$ and so the condition can be expressed as
\begin{equation}
\chi=\frac{n\sqrt{1+\gamma^2}}{2}, \ \mu=\frac{\gamma-\chi}{2}
\end{equation}
with $n$ an integer. Alternatively, one can combine both equations to remove  $\gamma$ and get
\begin{equation}
\frac{\left(2\chi\right)^2}{n^2}-\left(2\mu+\chi\right)^2=1,
\end{equation}
which parameterizes a hyperbola. There is such a hyperbola for any $n$, although the agreement gets better as $n$ increases as the condition $\mu,\chi \gg \gamma,\kappa$ is better satisfied.
At the same time, points between hyperbolas also get better revivals between as $\mu$ and $\chi$ increase. This is best seen on the resonance line $\gamma=0$. In that case, the additional condition for energies to align is simply to require $\chi$ to be half-integer or integer multiples of $\kappa$, as the spacing within each sector is simply $\kappa$. The effect of this can be seen on Fig.~\ref{fig:ResPolBands} for $\kappa=1$. When the condition is met, the polarized state revives with period $2\pi$.

\begin{figure}[t!]
	\centering
	\includegraphics[width=\linewidth]{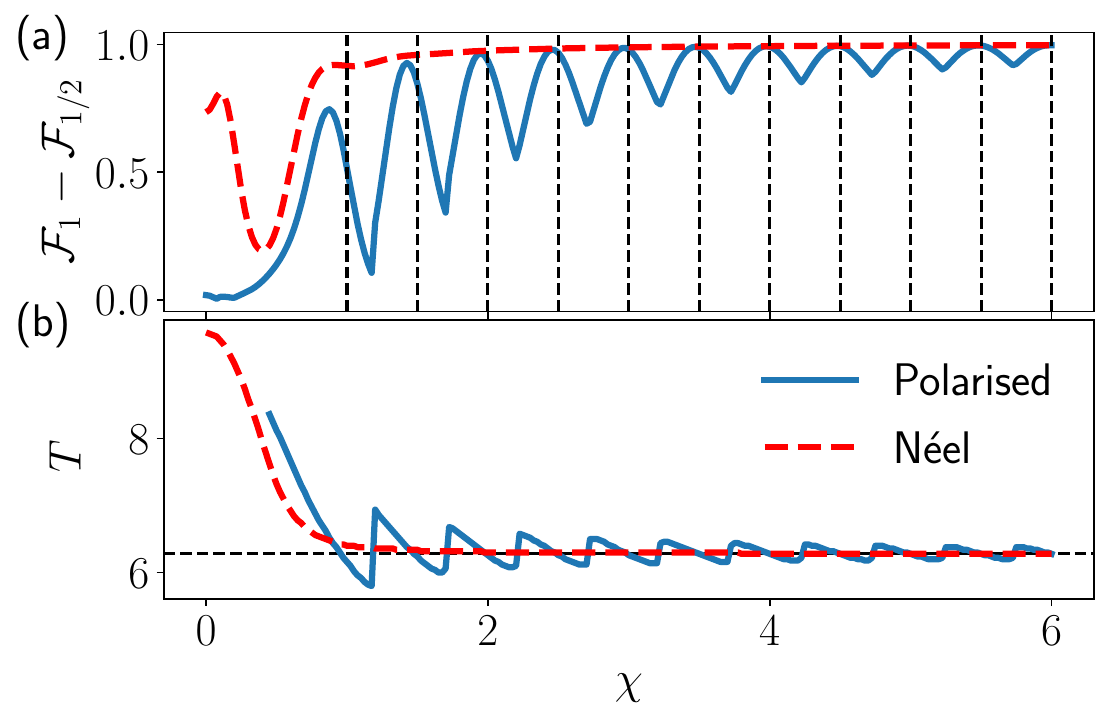}
	\caption{Self-fidelity and period of revival in the PXP model with $L_m=22$ along the resonance line $\chi=-2\mu$ for the polarized and N\'eel states. As $\chi$ gets large, contributions from sectors with excitations on odd sites become smaller and the further resonance condition for $\chi$ to be integer or half-integer matters less.}
	\label{fig:res_range}
\end{figure}

One can monitor that period as well as the fidelity at the first revival as $\chi=2\mu$ is increased. Figure~\ref{fig:res_range} shows this for the N\'eel and polarized states. As expected, the fidelity gets better as $\chi$ is increased because other Hilbert space sectors are more suppressed.  

\bibliographystyle{quantum}
\bibliography{biblio}

\clearpage

\onecolumngrid 
\newpage

\begin{center}
	{\bf \large Supplementary Online Material for  
		``Ergodicity Breaking Under Confinement in Cold-Atom Quantum Simulators''}
\end{center}
\begin{center}
	
	Jean-Yves Desaules$^1$, Guo-Xian Su$^{2,3,4}$, Ian P.~McCulloch$^5$, Bing Yang$^6$, Zlatko Papi\'c$^1$, and Jad C.~Halimeh$^{7,8}$\\
	\vspace*{0.1cm}
	{\footnotesize
		$^1$School of Physics and Astronomy, University of Leeds, Leeds LS2 9JT, UK\\
		$^2$Hefei National Laboratory for Physical Sciences at Microscale and Department of Modern Physics, University of Science and Technology of China, Hefei, Anhui 230026, China
		$^3$Physikalisches Institut, Ruprecht-Karls-Universit\"at Heidelberg, Im Neuenheimer Feld 226, 69120 Heidelberg, Germany \\
		$^4$CAS Center for Excellence and Synergetic Innovation Center in Quantum Information and Quantum Physics, University of Science and Technology of China, Hefei, Anhui 230026, China\\
		$^5$School of Mathematics and Physics, The University of Queensland, St. Lucia, QLD 4072, Australia\\
		$^6$Department of Physics, Southern University of Science and Technology, Shenzhen 518055, China\\
		$^7$Department of Physics and Arnold Sommerfeld Center for Theoretical Physics (ASC), Ludwig-Maximilians-Universit\"at M\"unchen, Theresienstra\ss e 37, D-80333 M\"unchen, Germany\\
		$ ^8$Munich Center for Quantum Science and Technology (MCQST), Schellingstra\ss e 4, D-80799 M\"unchen, Germany
	}
\end{center}

\setcounter{section}{0}
\setcounter{subsection}{0}
\setcounter{equation}{0}
\setcounter{figure}{0}
\renewcommand{\theequation}{S\arabic{equation}}
\renewcommand{\thefigure}{S\arabic{figure}}
\renewcommand{\thesubsection}{S\arabic{subsection}}
\renewcommand{\thesection}{S\arabic{section}}

{\footnotesize
	In this Supplementary Material, we provide further analysis and background calculations to support the results in the main text. 
	In Sec.~S1, we provide additional details on scarring from the N\'eel state at low values of $\chi$.
	In Sec.~S2, we derive the first-order energy shift from the free spin-1 parent model.
	In Sec.~S3, we discuss the particle-hole symmetry at $\mu=0$ and its consequences.
	In Sec.~S4, we derive the effective Hamiltonians in the non-ergodic regimes using a Schrieffer-Wolf transformation.
	In Sec.~S5, we provide additional details of the MPS simulation of the spreading of defects in the PXP and Bose-Hubbard models. 
	In Sec.~S6, we show that the nonmonotonicity of the revivals from the N\'eel state as $\chi$ is increased can also be seen in the semi-classical limit using the time-dependent variational principle.
}

\vspace*{1cm}

\twocolumngrid

\addtocontents{toc}{\protect\setcounter{tocdepth}{0}}
\section*{S1 \quad Additional results on scarring from the N\'eel state with \texorpdfstring{$\mu=0$}{mu=0}}
\begin{figure}[htb!]
	\centering
	\includegraphics[width=\linewidth]{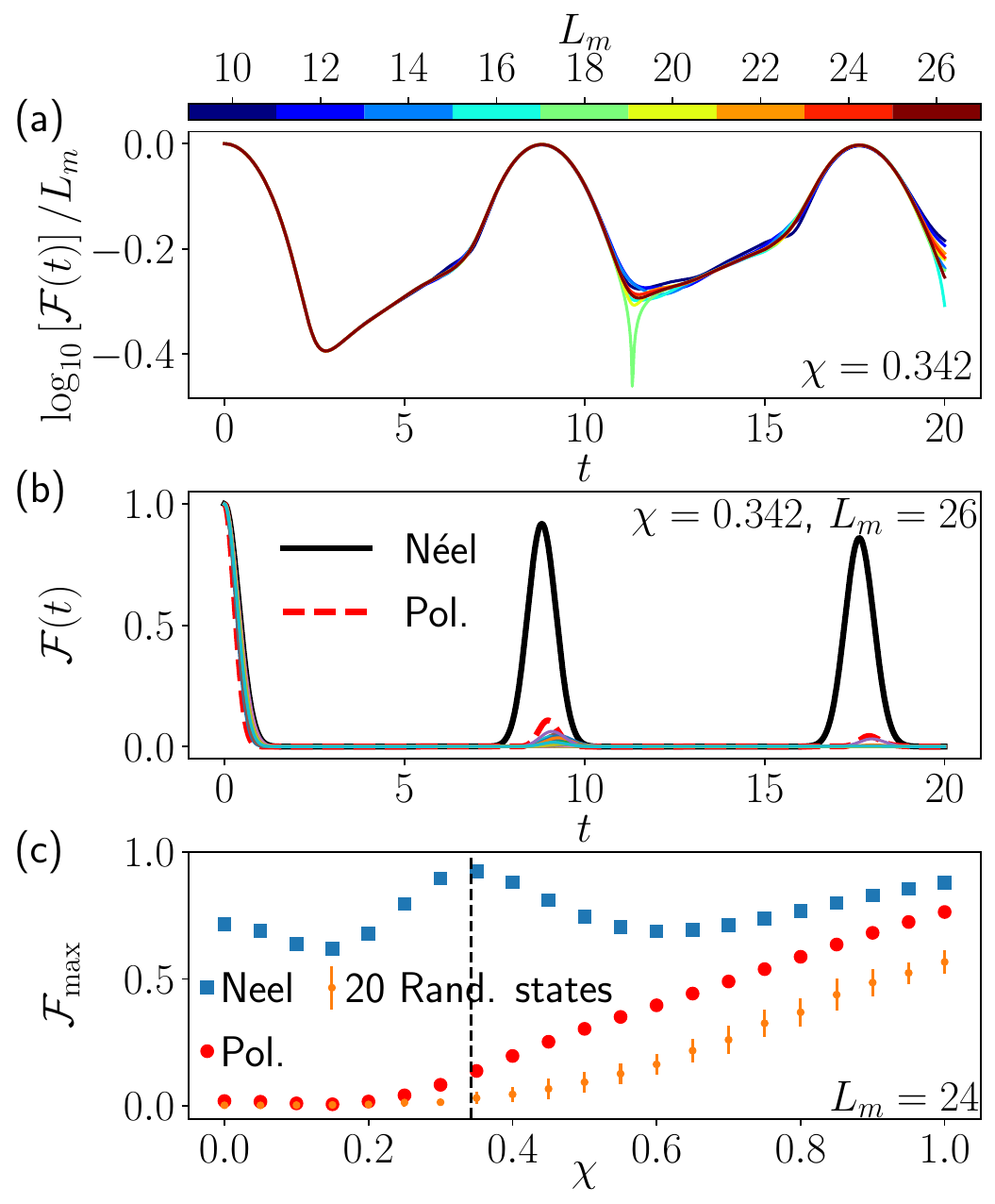}
	\caption{ (a) Fidelity after quenches from the N\'eel state $\chi=0.342$ and various system sizes.  (b) fidelity dynamics after quenches from various product states with $\chi=0.342$ and $L_m=26$. (c) Fidelity at the first revival from various product states with $\chi=0.342$ and $L_m=24$.}
	\label{fig:chi_revs_Neel_N}
\end{figure}

\addtocontents{toc}{\protect\setcounter{tocdepth}{0}}
In this section we provide additional results on quenches from the N\'eel state with $\mu=0$. 
Fig.~\ref{fig:chi_revs_Neel_N} (a) shows that after quenches from the N\'eel state at the optimal point $\chi=0.342$, the fidelity density $\log_{10}\left[\mathcal{F}(t)/\right]/L_m$ is well converged already at small system sizes $L_m\approx 14$.
Fig.~\ref{fig:chi_revs_Neel_N} (b) and (c) show that the nonmonotonic fidelity behavior witnessed for this state is not visible when quenching from other initial product states, and that at the optimal point for the N\'eel state (around $\chi=0.342$) other states display relatively quick thermalization in large systems.
\begin{figure}[htb!]
	\centering
	\includegraphics[width=\linewidth]{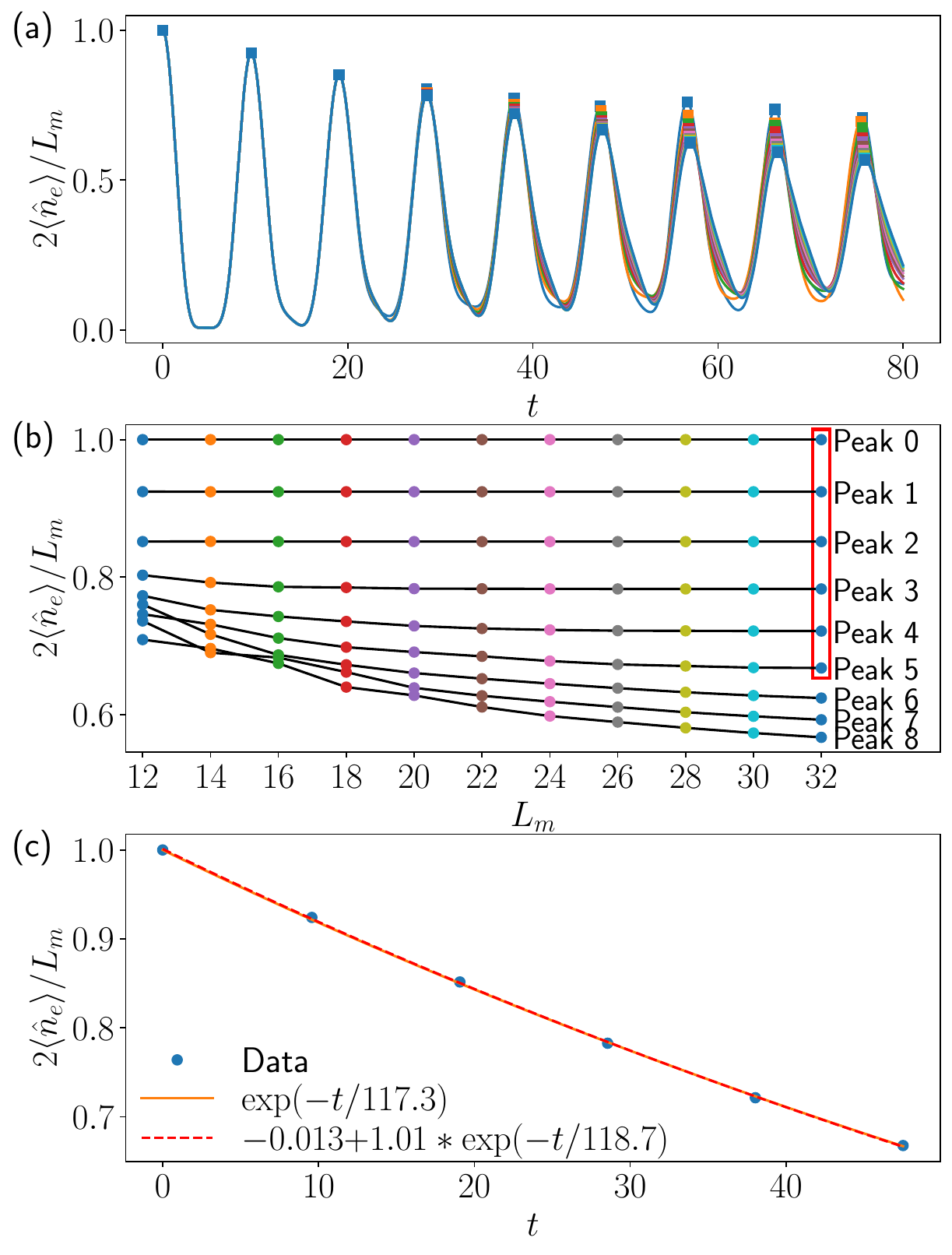}
	\caption{Occupation density on the even sublattice after quenches from the N\'eel state for $\chi=0$ and various system sizes. (a) data for all system sizes. The squares indicate the local maxima. (b) Local maxima as a function of system size. Each line correspond to a different peak and each color to a different system size. The red rectangle highlights the peaks that are converged in system size at $L_m=32$ and that will be used to extract a decay timescale. (c) Exponential fit of the converged peaks for $L_m=32$ to get the decay exponent of the envelope. }
	\label{fig:chi_tau_0}
\end{figure}
These results confirm that the revivals observed around $\chi=0.342$ for the N\'eel state are genuine scarring and are not due to fragmentation (which would affect most initial states) or to finite- size effects.

We now show more details on how Fig.~3(b) in the main text is obtained. We perform quenches from the N\'eel state and monitor the excitation density on the even sublattice. We do this for several system sizes and identify the local maxima for each of them. We can then see which of these are converged in system size and discard the other ones. We finally fit these maxima with a decaying exponential of the form $f(t)=a+be^{-t/\tau}$. Fig.~\ref{fig:chi_tau_0} shows the raw data for various system sizes, the scaling of the height of each peak with system sizes, and finally the plot of the converged peaks for $\chi=0$. 
We reproduce the same procedure for $\chi=0.15$, $\chi=0.3$ and $\chi=0.6$. For brevity we do not show the full data here but only the converged peaks and their exponential fit (see Fig.~3(b)in the main text). We note that as $\chi$ gets larger the finite-size effects occur at increasingly longer times. Overall we find a clear difference between $\chi=0.3$ and the other values already after 3-4 revivals. The decay time $\tau$ also shows a much larger value around $\chi=0.3$. 

\section*{S2 \quad Correction to the energy spacing in the spin-1 parent Hamiltonian}
In this section we compute the first-order perturbation to the energy eigenvalue of the $\ket{\tilde{S}_n}$ states due to $\hat{H}_1$.
Our derivation follows closely the one in appendix D of Ref.~\cite{Omiya22}.
We start from the spin-1 Hamiltonian 
\begin{align}\label{eq:PXP_S1sm}
	-\frac{1}{\sqrt{2}}\hat{S}^x-\chi\hat{S}^z
\end{align}
with spin 1 and $N_b=L_m/2$ sites and consider its eigenstates with maximal total spin. We denote these states by $\ket{\tilde{S}_n}$, with $n=0$ to $2N_b=N$ (where we use $N=2N_b=L_m$ to keep the notation of Ref.~\cite{Omiya22}). As this Hamiltonian is the same as for $\chi=0$ up to a rescaling and a rotation around $\hat{S}^y$, the $\ket{\tilde{S}_n}$ will also be identical up to a rotation. The eigenstate $n$ will have energy 
\begin{equation}\label{eq:En}
	E_n=(N_b-n)\sqrt{1/2+\chi^2}
\end{equation}
Due to the rescaling, $\hat{H}_1$ is also modified and becomes $-\frac{1}{2}\left(\ket{+,0}+\ket{0,-}\right)\bra{+,-}$. Apart from that, the construction from Ref.~\cite{Omiya22} is not fundamentally modified, meaning that the state $\ket{\tilde{S}_{N-n}}$ can still be computed by applying a raising operator on $\ket{\tilde{S}_N}$. We can then also target a pair of spin-1 sites and decompose it in the basis of the local $-\frac{1}{\sqrt{2}}\hat{S}^x-\chi\hat{S}^z$ operator with total spin-2. Let us denote these local eigenstates as $\ket{\hat{T}_2,m}$, where $m$ is the magnetization quantum number. It is straightforward to compute
\begin{equation}
	\delta e^m=\frac{1}{2}\langle \hat{T}_2,m |(\ket{+,0}+\ket{0,-})\bra{+,-} |\hat{T}_2,m\rangle,
\end{equation}
and we get
\begin{align}
	\delta E_2&=\frac{\sqrt{2}}{16}\frac{1}{(1+2\chi^2)^{3/2})},\\
	\delta E_1&=\frac{\sqrt{2}}{4}\frac{\chi^2}{(1+2\chi^2)^{3/2})},\\
	\delta E_0&=0,\\
	\delta E_{-1}&=-\frac{\sqrt{2}}{4}\frac{\chi^2}{(1+2\chi^2)^{3/2})},\\
	\delta E_{-2}&=-\frac{\sqrt{2}}{16}\frac{1}{(1+2\chi^2)^{3/2})}.
\end{align}
The next step is to compute the coefficients $c_m$ that describe the weight of the state $\ket{\tilde{S}_{N-n}}$ on $\ket{\hat{T}_2,m}$. Since this computation only depends on the number of times the raising operator has been applied, it is completely unaffected by our rescaling and rotation. Hence we get the same values as in Ref.~\cite{Omiya22}:
\begin{align}
	c_2&=\prod_{M=N_b{-}n{-}1}^{N_b{-}2}\alpha (N_b,M),\\
	c_1&=2n\prod_{M=N_b{-}n}^{N_b{-}2}\alpha (N_b,M),\\
	c_0&=\sqrt{6}n(n{-}1)\hspace{-0.6cm}\prod_{M=N_b{-}n+1}^{N_b{-}2}\hspace{-0.6cm}\alpha (N_b,M),\\
	c_{{-}1}&=2n(n{-}1)(n{-}2)\hspace{-0.6cm}\prod_{M=N_b{-}n{+}2}^{N_b{-}2}\hspace{-0.6cm}\alpha (N_b,M),\\
	c_{{-}2}&=n(n{-}1)(n{-}2)(n{-}3)\hspace{-0.6cm}\prod_{M=N_b{-}n{+}3}^{N_b{-}2}\hspace{-0.6cm}\alpha (N_b,M),
\end{align}
with $\alpha (N_b,M)=\sqrt{(N_b{-}1)(N_b{-}2){-}M(M{-}1)}$.
The overall normalization of the state is also unaffected and is given by
\begin{equation}
	\mathcal{N}_{N-n}=\prod_{M=N_b-n+1}^{N_b} \left[N_b(N_b+1)-M(M-1)\right]
\end{equation}
To simplify computations, we can perform an additional step that is not used in Ref.~\cite{Omiya22}. These products can be recast is a more similar way, that allows to get cancellations between their terms. Indeed, by performing the change of variable $M=N_b-K$ we get  
\begin{equation}
	\mathcal{N}_{N-n}=\prod_{K=0}^{n-1} (K+1)(2N-K).
\end{equation}
Similarly, using $M=N_b-2-K$ we get
\begin{equation}
	\begin{aligned}
		|c_2|^2&=\prod_{K=0}^{n-1}(K+1)\\
		&=\prod_{K=0}^{n-1}(K+1) \prod_{K=0}^{n-1}(2N-K-4) \\
		&=\prod_{K=0}^{n-1}(K+1) \prod_{K=4}^{n+3}(2N-K), \\
	\end{aligned}
\end{equation}
which leads to the simple expression
\begin{equation}
	\frac{|c_2|^2}{\mathcal{N}_{N-n}}=\prod_{K=0}^{3}\frac{2N-K-n}{2N-K}.
\end{equation}
The same kind of transformation can be done for the other $c_m$ to get
\begin{align}
	\frac{|c_2|^2}{\mathcal{N}_{N-n}}&=\prod_{K=0}^{3}\frac{2N-K-n}{2N-K}\\
	\frac{|c_1|^2}{\mathcal{N}_{N-n}}&=\frac{4n}{2N-3}\prod_{K=0}^{2}\frac{2N-K-n}{2N-K}\\
	\frac{|c_0|^2}{\mathcal{N}_{N-n}}&=\frac{6n(n-1)(2N-n)(2N-n-1)}{\prod_{K=0}^{3} (2N-K)}\\
	\frac{|c_{-1}|^2}{\mathcal{N}_{N-n}}&=\frac{4n(n-1)(n-2)(2N-n)}{\prod_{K=0}^{3} (2N-K)} \\
	\frac{|c_{-2}|^2}{\mathcal{N}_{N-n}}&=\prod_{K=0}^{3}\frac{n-K}{2N-K}.
\end{align}
This simplification allows to directly compute the overall energy shift of $\ket{\tilde{S}_{N-n}}$ for any $N_b$, $n$ and $\chi$ as 
\begin{equation}
	\begin{aligned}
		&\Delta E_{N{-}n}\\
		&{=} N_b\sum_{m={-}2}^2 \frac{|c_{m}|^2}{\mathcal{N}_{N_b-n}}\delta E_m \\
		&{=}\frac{(N_b{-}n) [1{+}n^2{-}3N_b {-}2nN_b {+}2N_b^2 {-}8n(n {-} 2 N_b)\chi^2]}{8\sqrt{2} (1 + 2 \chi^2)^{3/2}(N_b-1)(2N_b-1)} \\
		&{=}(N_b{-}n)\frac{(1{-}3N_b){+}(N_b{-}n)^2(1{-}8\chi^2){+}N_b^2(1+8\chi^2)}{8\sqrt{2} (1 + 2 \chi^2)^{3/2}(N_b-1)(2N_b-1)}.
	\end{aligned}
\end{equation}
This finally allows to get 
\begin{equation}
	\frac{\Delta E_{n}}{n{-}N_b}{=}\frac{(1{-}3N_b){+}(n{-}N_b)^2(1{-}8\chi^2){+}N_b^2(1+8\chi^2)}{8\sqrt{2} (1 + 2 \chi^2)^{3/2}(N_b-1)(2N_b-1)}.
\end{equation}

We can perform a change of variable $k=N_b-n$, where $k$ runs between $-N_b$ and $+N_b$.
The state $k$ then has energy 
\begin{equation}\label{eq:Ek}
	E_k=k\sqrt{1/2+\chi^2},
\end{equation}
with energy deviation 
\begin{equation}
	\Delta E_{k}{=}-k\frac{(1{-}3N_b){+}k^2(1{-}8\chi^2){+}N_b^2(1+8\chi^2)}{8\sqrt{2} (1 + 2 \chi^2)^{3/2}(N_b-1)(2N_b-1)}.
\end{equation}

\section*{S3 \quad Particle-hole symmetry with \texorpdfstring{$\chi{\neq} 0$}{chi not equal to 0} }

In the PXP model with an even system size, the N\'eel state $\ket{\mathbb{Z}_2}$ and anti-N\'eel state $\ket{\overline{\mathbb{Z}_2}}$ are degenerate for any value of $\mu$ as long as $\chi=0$. 
A non-zero value of $\chi$ will break that as one of the state goes towards the ground state and the other one towards the ceiling state. Nonetheless, when $\mu=0$ they still show identical dynamics for any $\chi$ due to a symmetry of the spectrum. 
Let us start with the PXP model with $\mu=0$ and an even number of sites $L_m=2M$ and rewrite it as
\begin{equation}
	\hat{H}_{PXP}=\hat{H}_0+\hat{H}_{\chi}
\end{equation}

For $\mu=\chi=0$, it is well known that the spectrum has a particle-hole symmetry with respect to the operator
\begin{equation}
	\ZT=\prod_{j=1}^{L_m}\hat{s}^z_j
\end{equation}
as $\left\{\hat{H}_0 ,\ZT \right\}=0$. At the same time, the Hamiltonian is also symmetric under the spatial inversion operator $\SP$ that takes site $j$ to $L_m+1-j$ and so $\left[\hat{H}_0,\SP\right]=0$.
Interestingly, the staggered potential term $\hat{H}_{\chi}$ has the exact inverse relations with these two operators and $\left[\hat{H}_{\chi},\ZT \right]$=0. while $\left\{\hat{H}_{\chi},\SP\right\}=0$. 
Both these operators are hermitian and square to the identity: ${\SP}^2={\ZT}^2=\mathbb{1}$. They also commute with each other.
It is then straightforward to prove that $\hat{H}_0+\hat{H}_{\chi}$ anti-commutes with $\SP \ZT$. Indeed, we have that
\begin{equation}
	\begin{aligned}
		\left(\hat{H}_0+\hat{H}_{\chi}\right)\SP \ZT&=\SP \left(\hat{H}_0-\hat{H}_{\chi}\right) \ZT \\
		&=\SP \ZT\left(-\hat{H}_0-\hat{H}_{\chi}\right) \\
		&=\; -\SP \ZT\left(\hat{H}_0+\hat{H}_{\chi}\right).
	\end{aligned}
\end{equation}
This has a few direct consequences, such as the presence of the same number of zero-modes as for $\chi=0$ and the fact that the spectrum is symmetric around $E=0$.

It also has less obvious consequences for the N\'eel and anti-N\'eel states, as both $\SP$ and $\ZT$ have special relations with these two states. They are eigenstates of $\ZT$ with eigenvalue $(-1)^M$, while $\SP$ takes one to the other as $\ket{\overline{\mathbb{Z}_2}}=\SP \ket{\mathbb{Z}_2}$.
These properties allow us to directly show that for any value of $\chi$, the return fidelity of both states will be exactly identical as long as $\mu$ is 0.
Let us first consider the energy expectation value of both states:
\begin{equation}
	\begin{aligned}
		\braket{\overline{\mathbb{Z}_2}|\hat{H}_{PXP}|\overline{\mathbb{Z}_2}}&=\braket{\mathbb{Z}_2|\SP \left(\hat{H}_0+\hat{H}_{\chi}\right)\SP|\mathbb{Z}_2} \\
		&=\braket{\mathbb{Z}_2| \left(\hat{H}_0-\hat{H}_{\chi}\right)\SP\SP|\mathbb{Z}_2} \\
		&=\braket{\mathbb{Z}_2|\ZT \ZT \left(\hat{H}_0-\hat{H}_{\chi}\right)|\mathbb{Z}_2} \\
		&=\braket{\mathbb{Z}_2|\ZT \left(-\hat{H}_0-\hat{H}_{\chi}\right)\ZT|\mathbb{Z}_2}  \\
		&=(-1)^{2M+1}\braket{\mathbb{Z}_2|\left(\hat{H}_0+\hat{H}_{\chi}\right)|\mathbb{Z}_2} \\
		&=\; -\braket{\mathbb{Z}_2|\hat{H}_{PXP}|\mathbb{Z}_2}.
	\end{aligned}
\end{equation}
We indeed find that the degeneracy between them is broken. The exact same procedure can be done for any arbitrary power of the Hamiltonian to get
\begin{equation}
	\braket{\overline{\mathbb{Z}_2}|\hat{H}_{PXP}^k|\overline{\mathbb{Z}_2}}= (-1)^k\braket{\mathbb{Z}_2|\hat{H}_{PXP}^k|\mathbb{Z}_2}.
\end{equation}
This allows us to easily compute the return fidelity by doing the power expansion of the exponential of the Hamiltonian as
\begin{equation}
	\begin{aligned}
		\braket{\overline{\mathbb{Z}_2}|e^{-it\hat{H}_{PXP}}|\overline{\mathbb{Z}_2}}&=\sum_{k=0}^\infty (-it)^k \braket{\overline{\mathbb{Z}_2}|\hat{H}_{PXP}^k|\overline{\mathbb{Z}_2}} \\
		&=\sum_{k=0}^\infty (it)^k \braket{\mathbb{Z}_2|\hat{H}_{PXP}^k|\mathbb{Z}_2} \\
		&=\braket{\mathbb{Z}_2|e^{it\hat{H}_{PXP}}|\mathbb{Z}_2}\\
		&=\left(\braket{\mathbb{Z}_2|e^{-it\hat{H}_{PXP}}|\mathbb{Z}_2}\right)^\star.
	\end{aligned}
\end{equation}
From there it is trivial to see that
\begin{equation}
	|\braket{\overline{\mathbb{Z}_2}|e^{-it\hat{H}_{PXP}}|\overline{\mathbb{Z}_2}}|^2=|\braket{\mathbb{Z}_2|e^{-it\hat{H}_{PXP}}|\mathbb{Z}_2}|^2.
\end{equation}
Similar equivalences can be proven for the expectation of an operator $\hat{O}$ after quenches from the N\'eel and anti-N\'eel states as long as $\left[\hat{O},\SP \ZT \right]=0$. So while the spatial symmetries of the Hamiltonian do not guarantee any equivalence between the N\'eel and anti-N\'eel states after a quench, the particle-hole symmetry does. 

This immediately breaks down when a non-zero $\mu$ is added. The simplest way to see this is by checking the relation between the chemical potential term $\hat{H}_{\mu}$ and $\SP$ and $\ZT$. A quick calculation yields $\left[\hat{H}_\mu,\SP \right]=\left[\hat{H}_\mu,\ZT \right]=0$. So in the procedure used in this section neither $\SP$ nor $\ZT$ can be used to change the sign of $\hat{H}_\mu$ and it will not be possible to get a term equal to $-\hat{H}_{PXP}$.

\section*{S4 \quad Large \texorpdfstring{$\mu$ and $\chi$}{mu and chi} limit in the PXP model}

In the limit where the mass and confining potential are much larger than the hopping term $\kappa$, the Hilbert space fractures and we can perform a Schrieffer-Wolf (SW) transformation \cite{Bravyi2011} by writing the PXP Hamiltonian as $\hat{H}_0+\kappa \hat{V}$ with
\begin{align}
	\hat{H}_0 &=\;{-}\sum_{l=1}^{L_m} \left[2\mu+ (-1)^l \chi\right] \hat{s}^z_{l}, \\ 
	\hat{V}&= {-}\,\sum_{l=1}^{L_m} \hat{P}_{l-1}\hat{s}^x_{l}\hat{P}_{l+1}.
\end{align}
Each sector is characterized by a different expectation value of $H_0$, which we will denote by $h_0$. As this operator is diagonal in the Fock basis, it is convenient to express each subspace as the span of all Fock states $\ket{\psi}$ with $\braket{\psi|H_0|\psi}=h_0$. Let us denote the number of excitations on the odd (even) sublattice by $n_o$($n_e$). Let us also assume an even number of sites $L_m=2M$, so we have $M$ sites on each sublattice.
Adding an excitation on an odd site lead to an energy change of $-2\mu+\chi$, and on an even site to an energy change of $-2\mu-\chi$.
We can finally write  
\begin{equation}
	h_0=(n_o-\frac{M}{2})(-2\mu+\chi)+(n_e-\frac{M}{2})(-2\mu-\chi).
\end{equation}
In order to perform the SW transformation, we can follow the same procedure as in Ref.~\cite{Chen2021} in which the case $\mu=0$ was investigated. We first need to find the Schrieffer-Wolf generator $\hat{S}^{(1)}$ that obeys
\begin{equation}
	[\hat{S}^{(1)},\hat{H}_0]+\hat{V}=0.
\end{equation}
This is straightforward to do and leads to 
\begin{equation}\label{eq:S1}
	\hat{S}^{(1)}=\;  i\sum_{j=1}^{L_m}\frac{\hat{P}_{j-1}\hat{s}^y_j\hat{P}_{j+1}}{2\mu+(-1)^j\chi},
\end{equation}
which is well defined as long as $\chi\neq \pm 2\mu$. The case $\chi= \pm 2\mu$ will be treated separately in the next section, and for now we assume $\chi\neq \pm 2\mu$.
The effective Hamiltonian at order $n$ can then be computed as \begin{equation}\label{eq:Heff}
	\hat{H}_\mathrm{eff}^{(n)}=\kappa^n\frac{n-1}{n!}\hat{\mathcal{P}}_0\underbrace{[\hat{S}^{(1)},[\hat{S}^{(1)},\ldots,[\hat{S}^{(1)},\hat{V}]\ldots ]}_{n-1}\hat{\mathcal{P}}_0
\end{equation}
for $n>1$, where $\hat{\mathcal{P}}_0$ is the projector that projects each term into sectors with fixed $h_0$.
The first order Hamiltonian can also be directly computed as
\begin{equation}\label{eq:Heff1}
	\hat{H}_\mathrm{eff}^{(1)}=\kappa \hat{\mathcal{P}}_0\hat{V}\hat{\mathcal{P}}_0.
\end{equation}
However, as $\hat{V}$ has no term that conserves $h_0$, $\hat{H}_\mathrm{eff}^{(1)}$ is identically zero. This is also true for all odd order terms.

To compute the effective Hamiltonian up to 4th order we need to compute 
\begin{equation}\label{eq:S1c1}
	\begin{aligned}
		&\left[\hat{S}^{(1)},\hat{V}\right]{=} {-
		}\sum_{j=1}^{L_m}\frac{\hat{P}_{j-1}\hat{s}^z_{j}\hat{P}_{j+1}}{2\mu+(-1)^j \chi}\\
		&{-
		}\mu\sum_{j=1}^{L_m}\frac{\hat{P}_{j-1}\hat{\sigma}^+_{j}\hat{\sigma}^-_{j+1}\hat{P}_{j+2}+\hat{P}_{j-1}\hat{\sigma}^-_{j}\hat{\sigma}^+_{j+1}\hat{P}_{j+2}}{4\mu^2-\chi^2},
	\end{aligned}
\end{equation}
\begin{equation}\label{eq:S1c2}
	\begin{aligned}
		&\left[\hat{S}^{(1)},\left[\hat{S}^{(1)},\hat{V}\right]\right]{=} {-
		}\sum_{j=1}^{L_m}\frac{\hat{P}_{j-1}\hat{s}^x_{j}\hat{P}_{j+1}}{(2\mu+(-1)^j \chi)^2}\\
		&{+
		}\sum_{j=1}^{L_m}(4\mu{-}({-}1)^j\chi)\frac{\hat{P}_{j-2}\hat{P}_{j-1}\hat{s}^x_{j}\hat{P}_{j+1}{+}\hat{P}_{j-1}\hat{s}^x_{j}\hat{P}_{j+1}\hat{P}_{j+2}}{2(4\mu^2-\chi^2)(2\mu{-}({-}1)^j\chi)} \\
		&{-
		}\mu\sum_{j=1}^{L_m}\frac{\hat{P}_{j{-}2}\hat{\sigma}^+_{j{-}1}\hat{\sigma}^-_{j}\hat{\sigma}^+_{j{+}1}\hat{P}_{j{+}2}{+}{\rm h.c}}{(4\mu^2-\chi^2)(2\mu-(-1)^j\chi)},
	\end{aligned}
\end{equation}
and finally
\begin{equation}\label{eq:S1c3}
	\begin{aligned}
		&\left[\hat{S}^{(1)},\left[\hat{S}^{(1)},\left[\hat{S}^{(1)},\hat{V}\right]\right]\right]{=} \sum_{j=1}^{L_m}\frac{\hat{P}_{j-1}\hat{s}^z_j\hat{P}_{j+1}}{(2\mu+(-1)^j\chi)^3}\\ 
		&{+}\frac{4\mu{-}({-}1)^j\chi}{2\left(4\mu^2{-}\chi^2\right)^2}\left(\hat{P}_{j-2}\hat{P}_{j-1}\hat{s}^z_{j}\hat{P}_{j+1}{+}\hat{P}_{j-1}\hat{s}^z_{j}\hat{P}_{j+1}\hat{P}_{j+2}\right) \\
		&{+}\frac{4\mu+(-1)^j\chi}{4\left(4\mu^2-\chi^2\right)^2}\left(\hat{P}_{j-2}\hat{\sigma}^+_{j-1}\hat{P}_{j}\hat{\sigma}^-_{j+1}\hat{P}_{j+2}+{\rm h.c}\right)\\
		&{+}\frac{6\mu+(-1)^j\chi}{4\left(4\mu^2-\chi^2\right)^2}\left(\hat{P}_{j-2}\hat{\sigma}^+_{j-1}\hat{P}_{j}\hat{\sigma}^+_{j+1}\hat{P}_{j+2}+{\rm h.c}\right)\\
		&-\frac{\mu}{\left(4\mu^2-\chi^2\right)^2}\left(\hat{P}_{j-1}\hat{\sigma}^+_{j}\hat{\sigma}^-_{j+1}\hat{\sigma}^+_{j+2}\hat{\sigma}^-_{j+3}\hat{P}_{j+4}+{\rm h.c}\right) \\
		&{+}\mu\frac{8\mu^2-\chi^2}{\left(4\mu^2-\chi^2\right)^3}\left(\hat{P}_{j-1}\hat{\sigma}^+_{j}\hat{\sigma}^-_{j+1}\hat{P}_{j+2}+{\rm h.c}\right) \\
		&{+}\frac{\left(8\mu+(-1)^j\chi\right)\left(\hat{P}_{j-1}\hat{\sigma}^+_{j}\hat{\sigma}^-_{j+1}\hat{P}_{j+2}\hat{P}_{j+3}+{\rm h.c}\right)}{8\left(2\mu+(-1)^j\chi\right)^2\left(4\mu^2-\chi^2\right)} \\
		&{+}\frac{\left(8\mu-(-1)^j\chi\right)\left(\hat{P}_{j-2}\hat{P}_{j-1}\hat{\sigma}^+_{j}\hat{\sigma}^-_{j+1}\hat{P}_{j+2}+{\rm h.c}\right)}{8\left(2\mu-(-1)^j\chi\right)^2\left(4\mu^2-\chi^2\right)}.
	\end{aligned}
\end{equation}

\subsection*{General case, \texorpdfstring{$\mu/\chi$}{mu/chi} irrational}

If $\chi/\mu$ is irrational, each sector is characterized by the number of excitations on each sublattice, $n_o$ and $n_e$. 
Using Eq.~(\ref{eq:S1}), it is straightforward to derive the next order terms in the SW transformation. The leading order is then the second order one which is diagonal, giving
\begin{equation}\label{eq:H2}
	\hat{H}_\mathrm{eff}^{(2)}=\; - \frac{\kappa^2}{2}\sum_{j=1}^{L_m}\frac{\hat{P}_{j-1}\hat{s}^z_{j}\hat{P}_{j+1}}{2\mu+(-1)^j \chi}.
\end{equation}
The next non-zero term is at 4th order and reads
\begin{equation}\label{eq:H4}
	\begin{aligned}
		&\hat{H}^{(4)}=\frac{\kappa^4}{32} \sum_{j=1}^{L_m}4\frac{\hat{P}_{j-1}\hat{s}^z_j\hat{P}_{j+1}}{(2\mu+(-1)^j\chi)^3}\\ 
		&{+}2\frac{4\mu{-}({-}1)^j\chi}{\left(4\mu^2{-}\chi^2\right)^2}\left(\hat{P}_{j-2}\hat{P}_{j-1}\hat{s}^z_{j}\hat{P}_{j+1}{+}\hat{P}_{j-1}\hat{s}^z_{j}\hat{P}_{j+1}\hat{P}_{j+2}\right) \\
		&{+}\frac{4\mu+(-1)^j\chi}{\left(4\mu^2-\chi^2\right)^2}\left(\hat{P}_{j-2}\hat{s}^+_{j-1}\hat{P}_{j}\hat{s}^-_{j+1}\hat{P}_{j+2}+{\rm h.c}\right).
	\end{aligned}
\end{equation}
These terms are all similar to the ones studied in Ref.~\cite{Chen2021} for $\mu=0$, and match them exactly in that limit. As the off-diagonal terms are the same (up to an overall factor), the connectivity of the Hilbert space is unaltered. Thus, we will not focus more on this case as it has already been studied extensively, but instead, we will show that adding the mass term $\mu$ allows to engineer different types of fragmentation due to resonances between the $\chi$ and $\mu$ terms.

\subsection*{Resonances}
There is technically an infinite number of resonances between $\mu$ and $\chi$. 
The simplest case in which they happen is when the energy difference on one sublattice is equal to $n$ times ($n$ being integer) the energy change on the other sublattice. Hence two configurations can have a different number of excitations on their sublattices but still the same $h_0$. 
The resonance condition can be expressed as 
\begin{equation}\label{eq:res_n}
	2\mu\pm\chi{=}n\left(2\mu \mp \chi\right) \rightarrow \mu{=}\pm\frac{n+1}{2(n-1)}\chi,
\end{equation}
and setting these values can lead to a resonance at order $|n|+1$ at the minimum. Indeed, the allowed processes that match the resonance require adding $n$ excitations on one sublattice and removing one on the other, so they require $n+1$ moves. 
We can also chose $n$ to be a negative integer. In that case we need to do the same operation (add or remove) on both sublattices.
In general, as we have a dynamical term at 4th order for any value of $\mu$ and $\chi$, anything beyond $|n|=3$ will never be the dominant off-diagonal term. 

More complex resonances can also occur, in which the ratio between the energy cost of an excitation on each sublattice is not an integer but a rational number. In this case we have the same condition as in Eq.~\ref{eq:res_n} but with $n$ a rational number. Let us denote by $a$ and $b$ the numerator and denominator of the irreducible form of $n$. In that case the resonance can only lead to new terms in the effective Hamiltonian at order $a+b$.  As integers can be represented with $b=1$, this includes the integer-$n$ case. Let us also note that taking a fraction with $a=1$ will lead to the same result as the integer case with $n=-b$. With that in mind, the simplest case that does not fit in the integer limit is $a=2$ and $b=3$ or vice-versa. As this can only lead to new effective terms at order 5 or more, we will not consider these and thus we focus on integer $n$.

\subsubsection*{\texorpdfstring{$n=0$: $\chi=\pm 2\mu $}{n=0 chi=2 mu}}

For this particular relation between the two parameters, we have that $\mu$ and $\chi$ add up on one sublattice and cancel on the other one. For brevity, we will discuss the case $\chi=-2\mu$, for which $h_0=2\chi(n_o-\frac{M}{2})$. Here we see that the sectors are much larger than in the trivial case.

The approach used to compute the SW transformation terms in Eq.~\ref{eq:Heff} fails as $\hat{V}$ now has terms within individual sectors. This results in an operator $\hat{S}^{(1)}$ with some infinite matrix elements (see Eq.~\ref{eq:S1}). However, it also means that the first order effective Hamiltonian, Eq.~(\ref{eq:Heff1}), is non-zero. It is the leading order Hamiltonian and is given by
\begin{equation}
	\hat{H}_\mathrm{eff}^{(1)}=-\kappa \sum_{j=0}^{M}\hat{P}_{2j-1} \hat{s}^x_{2j}\hat{P}_{2j+1} 
\end{equation}
Thus, all the odd sites are frozen while even sites can flip freely if their neighbors are frozen in the unexcited position.

As the odd sites are frozen, two  configurations with the same $n_o$ but different arrangements of them will be disconnected. Therefore, in each sector with a fixed $h_0$, the Hilbert space shows further fragmentation. The number of subsectors is given by the number of unique configurations on the odd sites: 
\begin{equation}
	\mathcal{N}_{n_0}=\binom{L_m/2}{n_0}.
\end{equation}
The size of each sector is then $2^k$, where $k$ denotes the number of even sites with unexcited neighbors. This is not uniquely determined by $L_m$ and $n_0$, as even with these two parameters fixed different configurations on the odd sites can lead to different values $k$.   

Note that all sites are either completely frozen (odd sites and the neighbors of excited odd sites) or allowed to flip freely without any interaction (even sites with both neighbors frozen in the unexcited positions). So the effective model at first order is that of a free paramagnet. Let us denote the by $\mathcal{K}$ the set of all non-frozen (even) sites, then we can recast the Hamiltonian as
\begin{equation}
	\hat{H}^{(1)}=-\kappa \sum_{j\in \mathcal{K}} \hat{s}^x_{j}.
\end{equation}

The same transformation can be done is the vicinity of the resonance $\chi=-2\mu+\gamma$ as long as $\gamma$ is not much larger than $\kappa$. The expression of $h_0$ becomes $h_0=(2\chi-\gamma)(n_o-\frac{M}{2})$ and the odd sites are still frozen, but now the first order Hamiltonian gains a diagonal contribution as
\begin{equation}
	\label{eq:H_param}\hat{H}^{(1)}=-\kappa \sum_{j\in \mathcal{K}} \hat{s}^x_{j}-\sum_{j=0}^{M}\gamma \hat{s}^z_{2j}=E_0-\sum_{j\in \mathcal{K}} \kappa \hat{s}^x_{j}+\gamma \hat{s}^z_j,
\end{equation}
where $E_0$ is the (constant) contribution of all frozen even sites.
This Hamiltonian is still integrable, but the energy of the eigenstates and the revival period from the initial Fock states is altered.

If $\chi=\; +2\mu$, the same analysis can be done but the even sites are frozen and the odd ones can flip freely if their neighbors are unexcited.

\subsubsection*{\texorpdfstring{$n=1$: $\chi=0$}{n=1: chi=0}}

If $n=1$, we get $\mu=\pm\infty \chi$, meaning that $\mu$ can take any value but that $\chi$ must be 0. That regime is quite special, as it does not have confinement and the only thing slowing down the dynamics is the large mass of the fermionic matter. Nonetheless, we can treat it like the non-resonant case, and we find that $h_0=2\mu(n_o+n_e-M)$. The $\hat{S}^{(1)}$ operator and its commutators with respect to $\hat{V}$ are the same, but the projector $\hat{\mathcal{P}}_0$ is changed to reflect the new $h_0$. It now leaves $\hat{P}_{j-1}\hat{\sigma}^+_{j}\hat{\sigma}^-_{j+1}\hat{P}_{j+2}$ unaffected as even and odd sites are on an equal footing. So moving an excitation between neighbors preserves $h_0$, which only depends on the total number of excitations, and there is an emergent $\mathrm{U}(1)$ symmetry. The effective Hamiltonian at 2nd order is then
\begin{equation}\label{eq:Heff2_res}
	\begin{aligned}
		\hat{H}_\mathrm{eff}^{(2)}&{=} {-
		}\frac{\kappa^2}{8\mu}\sum_{j=1}^{L_m}\Big(2\hat{P}_{j-1}\hat{s}^z_{j}\hat{P}_{j+1}\\
		&{+}\hat{P}_{j-1}\hat{\sigma}^+_{j}\hat{\sigma}^-_{j+1}\hat{P}_{j+2}{+}{\rm h.c.}\Big).
	\end{aligned}
\end{equation}
We note that in this regime there is no Hilbert space fragmentation. Indeed, in each $\mathrm{U}(1)$ sector all states are connected.  

It is interesting to note that the effective Hamiltonian in Eq.~\ref{eq:Heff2_res} resembles the $\mathcal{M}_1$ supersymmetric model of lattice fermions introduced in Ref.~\cite{FendleySUSY}, which is an integrable model that can be mapped to an XXZ-type chain. The two terms that appear in the Hamiltonian are the same, but their prefactors are different. However, the same mapping to the XXZ model can be done, albeit with a different value of $\Delta$. Consequently, in each of the $\mathrm{U}(1)$ sectors the effective Hamiltonian is integrable, as was already derived in Ref.~\cite{FSS}.

We can also study what happens if we add a small $\chi$ such that $\kappa,\chi\ll \mu$. In this case we have
\begin{align}
	\hat{H}_0 &{=-}\sum_{l=1}^{L_m} 2\mu \hat{s}^z_{l}, \\ 
	\hat{V}&{=-}\,\sum_{l=1}^{L_m}\left[ \hat{P}_{l-1}\hat{s}^x_{l}\hat{P}_{l+1}+(-1)^l\frac{\chi}{\kappa} \hat{s}^z_{l} \right].
\end{align}
The PXP term will lead to the same 2nd order term, while the $(-1)^l\hat{s}^z_{l}$ term will add a first order term. We end up with 
\begin{equation}\label{eq:Heff12_res_mu_sm}
	\begin{aligned}
		\hat{H}_\mathrm{eff}^{(1,2)}&{=}{-}\sum_{j=1}^{L_m} (-1)^l\chi \hat{s}^z_{l} {-
		}\frac{\kappa^2}{8\mu}\sum_{j=1}^{L_m}\Big(2\hat{P}_{j-1}\hat{s}^z_{j}\hat{P}_{j+1}\\
		&+\hat{P}_{j-1}\hat{\sigma}^+_{j}\hat{\sigma}^-_{j+1}\hat{P}_{j+2}
		+{\rm h.c.}\Big)
	\end{aligned}
\end{equation}
This Hamiltonian no longer maps to the XXZ model, and when $\chi$ and $\frac{\kappa^2}{8\mu}$ are of comparable strength we find level spacing statistics close to Wigner-Dyson in each $\mathrm{U}(1)$ sector.

\subsubsection*{\texorpdfstring{$n=2$: $\mu=\pm \frac{3}{2}\chi$}{n=2: mu=-3/2 chi}}

For $n=2$, we have $\mu=\pm \frac{3}{2}\chi$.
The sign just changes which sublattice is affected, so we will only focus on the case $\mu=\frac{3}{2}\chi$
In that case,
\begin{equation}\label{eq:d0_res32}
	h_0=(-2\chi)(n_o+2n_e-\frac{3}{2}M).
\end{equation}
We then find that the effective 3rd order Hamiltonian is
\begin{equation}\label{eq:Heff3_res}
	\begin{aligned}
		\hat{H}_\mathrm{eff}^{(3)}&{=}{-
		}\frac{\mu\kappa^3}{3}\sum_{j=1}^{L_m}\frac{\hat{P}_{2j{-}2}\hat{\sigma}^+_{2j{-}1}\hat{\sigma}^-_{2j}\hat{\sigma}^+_{2j{+}1}\hat{P}_{2j{+}2}{+}{\rm h.c}}{(4\mu^2-\chi^2)(2\mu-(-1)^j\chi)} \\
		&=-\frac{\kappa^3}{32 \chi^2}\sum_{j=1}^{M}\hat{P}_{2j{-}2}\hat{\sigma}^+_{2j{-}1}\hat{\sigma}^-_{2j}\hat{\sigma}^+_{2j{+}1}\hat{P}_{2j{+}2}{+}{\rm h.c}.
	\end{aligned}
\end{equation}
This corresponds to projecting a term of the form $\hat{P}_{j-2}\hat{s}^x_{j-1}\hat{s}^x_{j}\hat{s}^x_{j+1}\hat{P}_{j+2}$ into the constrained Hilbert space of the PXP model.
It is also important to keep in mind that even if this is the leading dynamical term, the diagonal Hamiltonian obtained at second order -- see Eq.~\ref{eq:H2} -- is still present here.
The Hilbert space of the third order Hamiltonian is further fragmented. This can be seen by recognizing that the configuration ${\circ}_{2j-4}{\circ}_{2j-3}{\bullet}_{2j-2} {\circ}_{2j-1}{\bullet}_{2j}{\circ}_{2j+1}{\circ}_{2j+2}$ is frozen. Indeed, the only term that can remove an excitation on an even site is dressed with projectors on the two neighboring even sites. This effectively creates an equivalent of the PXP constraint between even sites. Every pair of excitations on consecutive even sites cannot be removed or created by the Hamiltonian in Eq.~(\ref{eq:Heff3_res}).

\begin{figure*}[htb!]
	\centering
	\includegraphics[width=\linewidth]{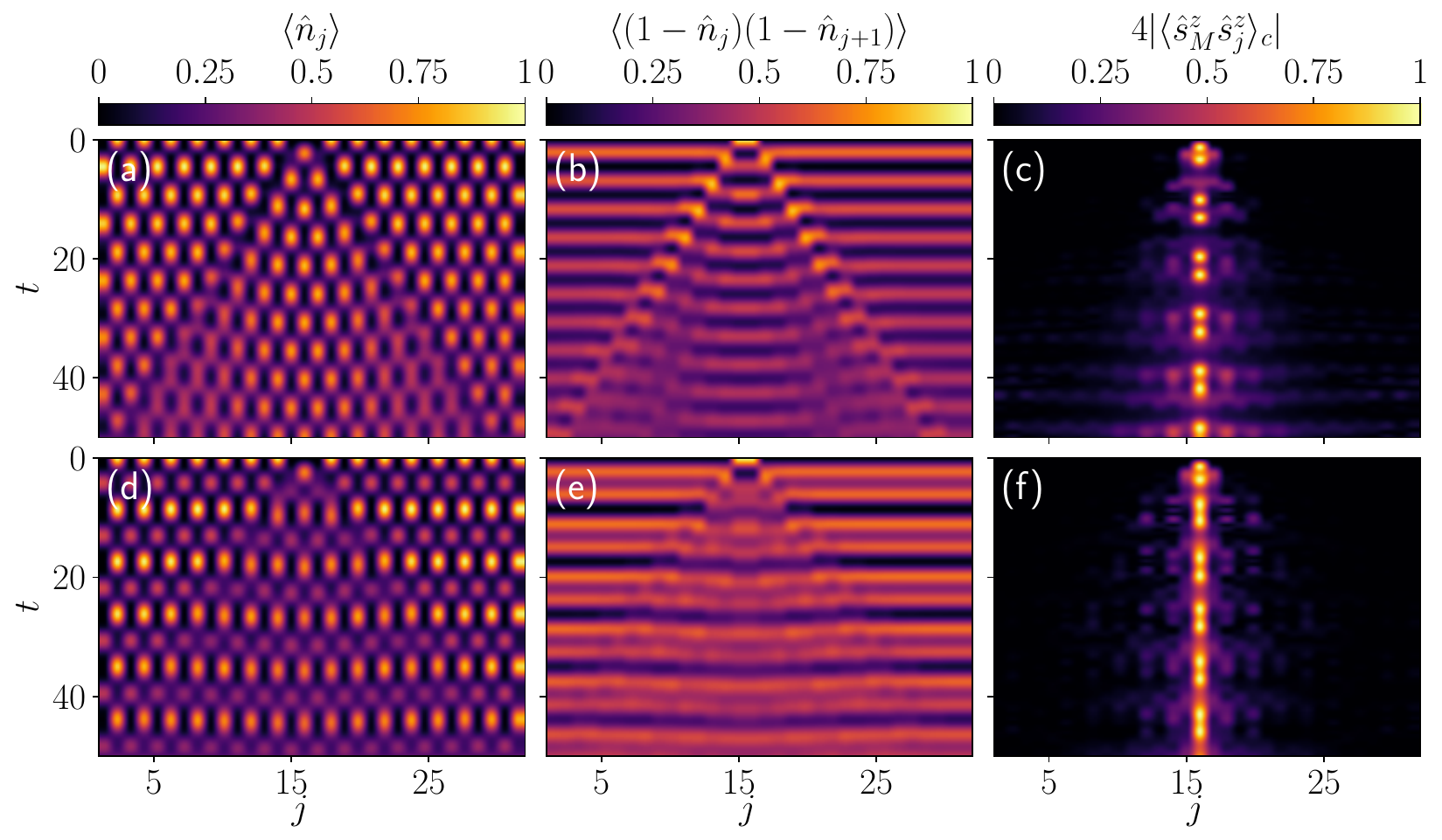}
	\caption{Observables dynamics after a quench from the N\'eel states with a defect with $L_m=32$. (a)-(c) $\chi=0$ (d)-(f) $\chi=0.35$. The effect of confinement is visible for the ZZ correlator in panels (c) and (f). At late times, the N\'eel and anti-N\'eel domains are still clearly distinguishable in panels (a) and (b) but not in panels (c) and (d).}
	\label{fig:defect_chi_ED}
\end{figure*}

The largest fragment is the one containing the state $\ket{\mathbb{Z}_4^\prime}=\ket{{\circ}{\circ}{\circ}{\bullet}{\circ}{\circ}{\circ}{\bullet}{\circ}{\circ}\cdots{\circ}{\bullet}{\circ}}$. This fragment can actually be mapped exactly onto the PXP model with $L_m/2$ sites. The mapping is relatively straightforward, as the PXP state simply corresponds to the configuration of the even sites. We recognize that the $\ket{\mathbb{Z}_4^\prime}$ state is thus the N\'eel state. The odd sites still play a role at enforcing the effective PXP constraint. If there is an excitation on an even site, is is always possible to remove it as  ${\circ}{\bullet}{\circ}\to {\bullet}{\circ}{\bullet}$. However, if there is no excitation on an even site, to create one both neighbors must be excited in order to do the process ${\bullet}{\circ}{\bullet}\to {\circ}{\bullet}{\circ}$. But excitations on the neighboring odd sites are only created when deexciting the neighboring even sites. This means that an even site can only be excited if both of its even neighbors are not, exactly as expected from the PXP constraint. Here we provide a quick example corresponding to going from ${\bullet}{\circ}{\bullet}\to{\circ}{\bullet}{\circ}$ (in PXP terms):
\begin{equation}
	\begin{aligned}
		{\circ}{\bullet}{\circ}{\circ}{\circ}{\bullet}{\circ} \to {\circ}{\bullet}{\circ}{\circ}{\bullet}{\circ}{\bullet} \to {\bullet}{\circ}{\bullet}{\circ}{\bullet}{\circ}{\bullet} \to {\bullet}{\circ}{\circ}{\bullet}{\circ}{\circ}{\bullet}.
	\end{aligned}
\end{equation}

We stress that this effective PXP constraint is not simply a consequence of the emergent conserved quantity. Indeed, as shown in Eq.~(\ref{eq:d0_res32}), this quantity only depends on the number of excitations on even and odd sites, and does not take into account their positions beyond that. So the emergent PXP constraint is a symptom of true Hilbert space fragmentation. It can be associated with a set of $L_m/2$ independent conserved quantities $\Big\{\hat{n}_{2j}\hat{n}_{2j+2} \Big\}_{j=1}^{L_m}$.

We also note here that the leading order Hamiltonian is still the second order one in Eq.~(\ref{eq:Heff2_res}). Furthermore, no process generated by Eq.~(\ref{eq:Heff3_res}) preserves the expectation value of $\hat{H}^{(2)}_\mathrm{eff}$. As such $\hat{H}^{(3)}_\mathrm{eff}$ is further suppressed, and one would need to perform a nested Schrieffer-Wolf transformation with $\hat{H}^{(2)}_\mathrm{eff}$ as $\hat{H}_0$ and $\hat{H}^{(3)}_\mathrm{eff}$ as $\hat{V}$ in order to get the right leading terms and prefactors.

\subsubsection*{\texorpdfstring{$n=-1$: $\mu=0$}{n=-1: mu=0}}
This resonance happens when $\mu=0$, and has been treated extensively in Ref.~\cite{Chen2021}. While it produces an effective Hamiltonian at 2nd order, this Hamiltonian is also purely diagonal. So there is no additional term due to this resonance. As removing (or adding) an excitation on both sublattices does not change $h_0$, we could expect a term of the form $\hat{P}_{j-1}\hat{s}^+_j\hat{s}^+_{j+1}\hat{P}_{j+2}+{\rm h.c.}$. However it is clear that such a term is forbidden by the PXP constraint. Any longer range term  $\hat{P}_{j-1}\hat{s}^+_j\hat{P}_{j+1}\hat{P}_{j+2r}\hat{s}^+_{j+2r+1}\hat{P}_{j+2r+2}+{\rm h.c.}$ with $r$ integer is also identically zero. Indeed, both sites are far enough apart so that they do not interact through the constraint. As a consequence the order in which the excitations are added does not matter. But both possible paths have opposite signs, and so they exactly cancel each other. This is the same reason why there are generically no long-range processes in the effective Hamiltonians obtained for these models.

The same observation can be made for the 4th order Hamiltonian, as it does not contain any additional term when compared to the one in the incommensurate case.

\subsubsection*{\texorpdfstring{$n=-2$, $n=-3$ and $n=3$}{n=-2, n=-3 and n=3}}
As for $n=-1$, $n=-2$, $n=-3$ and $n=3$ do not lead to additional terms in the effective Hamiltonian up to 4th order. This can be seen in the expressions in Eqs.~(\ref{eq:S1c1})-\ref{eq:S1c3}, as none of the terms that get projected out by $\hat{\mathcal{P}}_0$ in the incommensurate case survive here. For $n=-2$, we would need terms with 2 $\hat{s}^+$ on one sublattice and one $\hat{s}^+$ on the other sublattice.  For $n=-3$, we would need terms with 3 $\hat{s}^+$ on one sublattice and one $\hat{s}^+$ on the other sublattice. For $n=3$, we would need terms with 3 $\hat{s}^+$ on one sublattice and one $\hat{s}^-$ on the other sublattice. It is straightforward to see so none of these terms show up, and as such there is no meaningful difference with the incommensurate case up to 4th order.

\subsection*{Resonances at intermediate values of \texorpdfstring{$\mu$ and $\chi$}{mu and chi}}
For values of $\chi$ and $\mu$ which are equal to only a few time $\kappa$, the system already starts to form fragments with a definite value of $h_0$. However, they are still weakly connected and multiple of them contribute to the dynamics.  At the resonances discussed, the sectors have an energy spacing that is always commensurate and so this can lead to revivals. This occurs for example for the N\'eel state around $\chi=0.6$ and $\mu=-0.9$.

\section*{S5 \quad MPS methods}

In order to reach large systems to assess confinement in the PXP model and Bose-Hubbard model, we use matrix product state (MPS) techniques as implemented in the TeNPy package \cite{tenpy}. 
\subsection*{PXP model}
\begin{figure}[htb!]
	\centering
	\includegraphics[width=\linewidth]{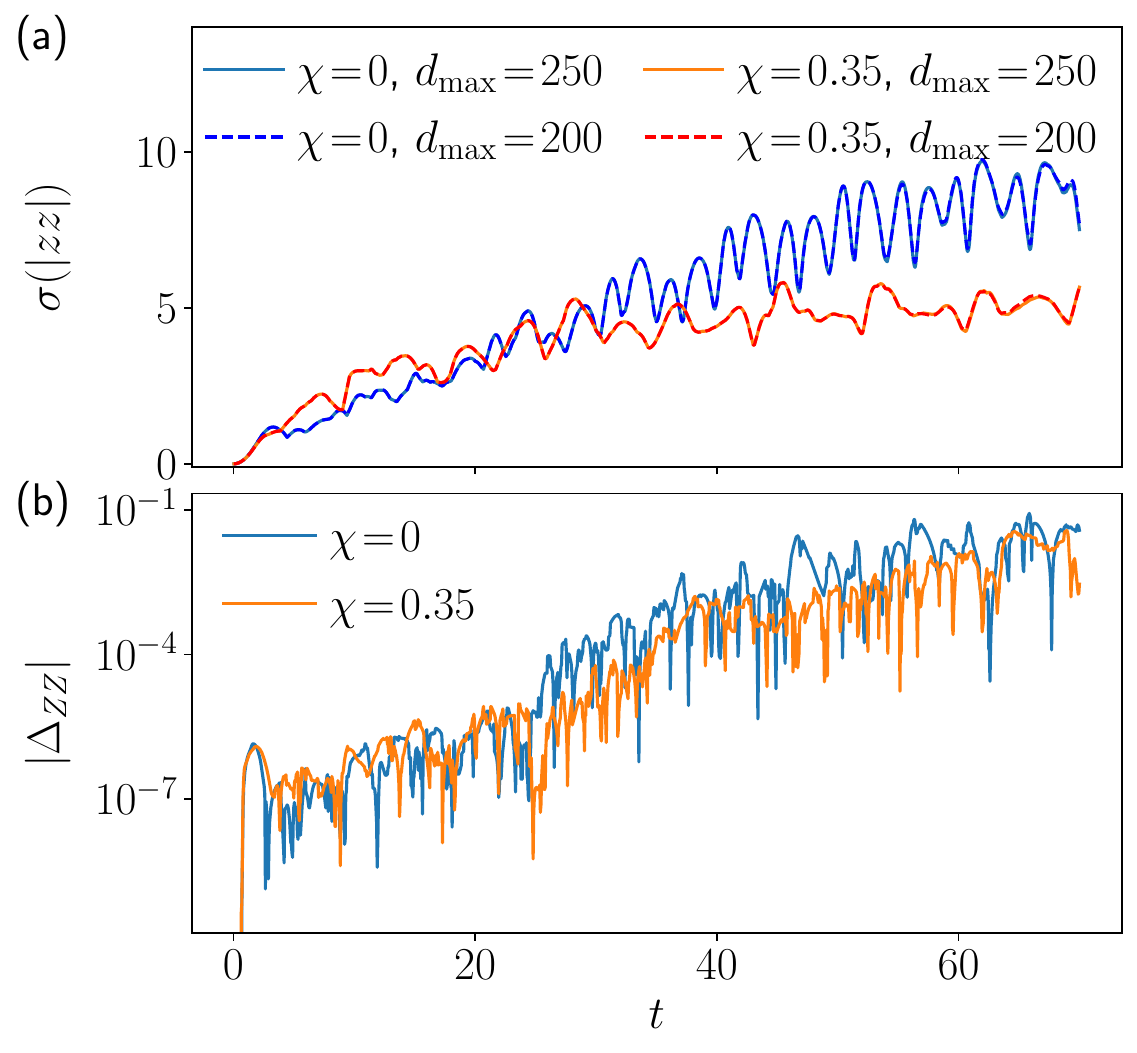}
	\caption{(a) Root mean square of the spread of the connected correlator between Pauli $Z$ matrices at the middle site and other sites in the chains for various values of $\chi$ for two different bond dimensions. (b) Absolute value of the difference of the $ZZ$ correlator between $d_\mathrm{max}=250$ and $d_\mathrm{max}=200$.}
	\label{fig:defect_ZZ_RMS_dmax}
\end{figure}

For PXP, we use the time-dependent variational principle (TDVP). We use two-site TDVP at first to grow the bond-dimension until it saturates to a desired value. At that point, we switch to one-site TDVP as it is faster and exactly conserves the energy. We also use OBC instead of PBC to avoid correlations spanning the whole length of the chain. The initial state chosen is then ${\bullet}{\circ}{\bullet}{\circ}\cdots {\circ}{\bullet}$ as it leads to suppressed boundary effects.  The results were compared to exact ones for 29 sites, and different bond-dimensions were used to assess convergence. Fig.~\ref{fig:defect_ZZ_RMS_dmax} shows the data for the RMS ZZ correlator, introduced in the main text, for two different bond dimensions $d_\mathrm{max}$, as well as their difference. The results show that convergence has been achieved for the larger $d_\mathrm{max}$ used.

The results in Fig.~5 in the main text can also be compared with their equivalent using exact time-evolution for $L_m=32$ and PBC in Fig.~\ref{fig:defect_chi_ED}.

Times as long as 70 units cannot be reached as the particle and antiparticle have scattered with each other already at time $\approx 50$. Nonetheless, we qualitatively see the same as in the larger system with OBC. Not only do we see confinement, we also see the same synchronization of oscillations in the case with $\chi=0.35$.

\subsection*{Bose-Hubbard model}
\begin{figure}[htb!]
	\centering
	\includegraphics[width=\linewidth]{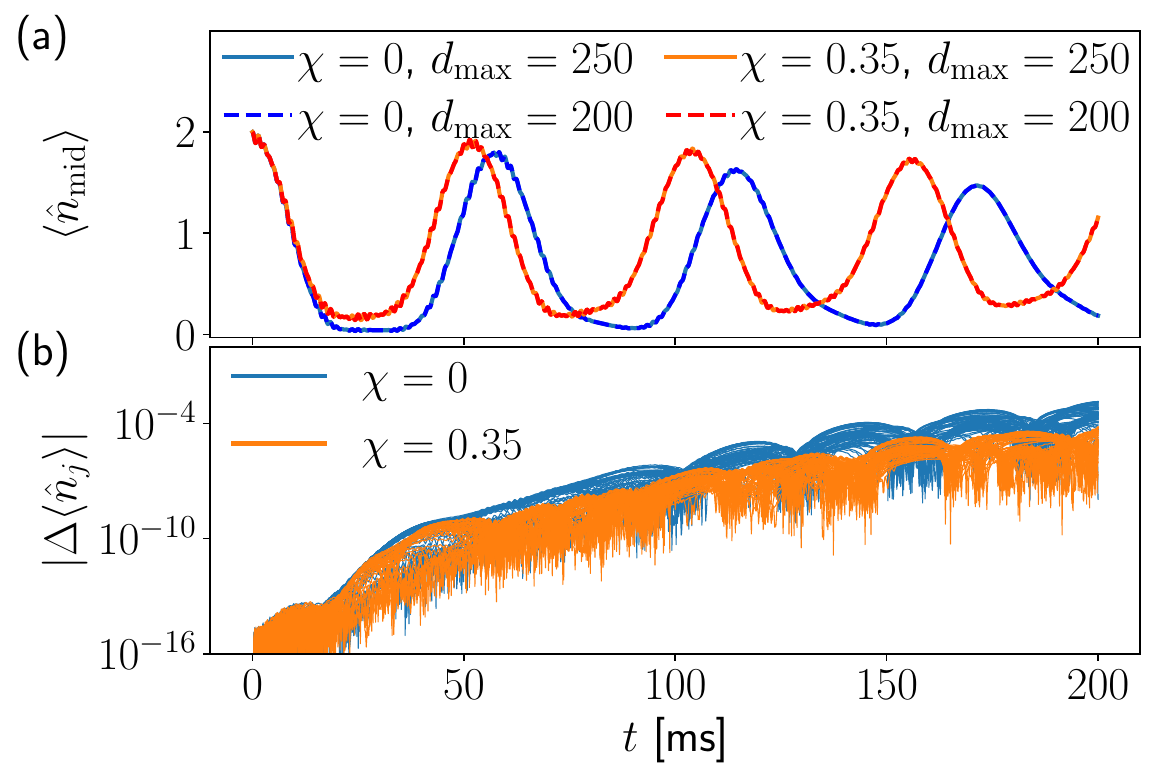}
	\caption{TEBD simulation of quenches from the N\'eel state in the Bose-Hubbard model with $L=87$. (a) Occupation of the middle (gauge) site for various values of $\chi$ for two different bond dimensions. No difference is visible upon visual inspection, showing the convergence with maximum bond dimension. (b) Absolute value of the difference in the occupation of each site between $d_\mathrm{max}=250$ and $d_\mathrm{max}=200$.}
	\label{fig:BHM_dmax}
\end{figure}

The simulations for the Bose-Hubbard model were done using time-evolving block decimation (TEBD) with two-site unit cells. Fig.~\ref{fig:BHM_dmax} shows a comparison of the occupation of the sites between two different maximum bond dimensions $d_\mathrm{max}$. Even at late times, the difference are small compared to the occupation values.

\section*{S6 \quad Symmetric subspace and TDVP semi-classical limit}

In order to study the PXP model for large systems, on can use the symmetric subspace approximation~\cite{Turner21}. The `symmetric subspace' is defined as a set of states that are fully symmetric under the exchange of atoms on the even (or odd) sublattice, but not necessarily symmetric under the permutations that exchange excitations between the two sublattices. This is a form of ``mean-field" theory for the dynamics of the N\'eel state in the PXP model, as it retains only the information about the total number of excitations on each sublattice (as opposed to the detailed positions of the excitations). 
The basis states defining the symmetric subspace approximation are, by construction, independent of $\chi$ and $\mu$. 

\begin{figure}[tb]
	\centering
	\includegraphics[width=\linewidth]{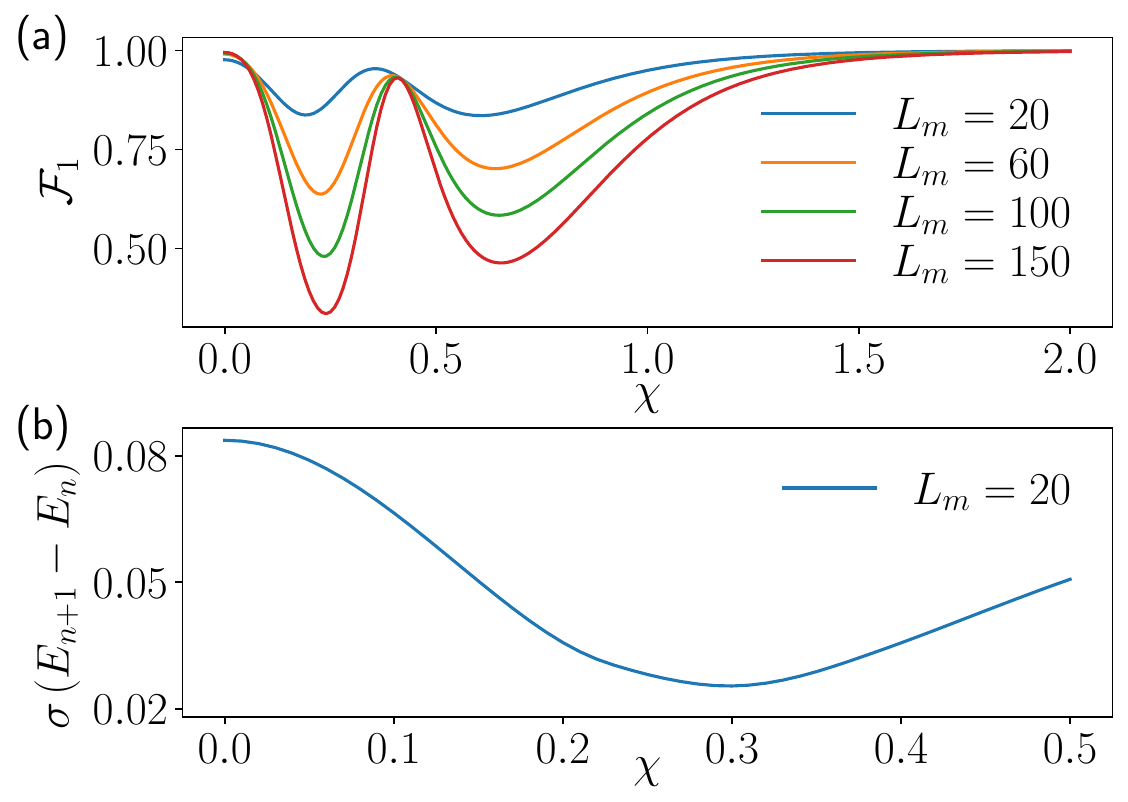}
	\caption{Effect of the confining potential in the symmetric subspace approximation. (a) Revival fidelity from the N\'eel state for various system sizes. (b) Standard deviation of the energy spacings of the scarred eigenstates.
		Both the best revival fidelity and the most equal energy spacing show optimal points between 0.3 and 0.5. The only values of $\chi$ for which the fidelity does not decay with system size are $\chi=0$, $\chi=\infty$ and $\chi=0.41$.}
	\label{fig:chi_sym}
\end{figure}

The fidelity after the quench from the N\'eel state within the symmetric subspace is computed in Fig.~\ref{fig:chi_sym}. The nonmonotonic behavior is well captured within this approximation. For generic values of $\chi$, the fidelity decays as $L_m$ increases. The only values of $\chi$ for which this does not happen are $\chi=0$, $\chi=\infty$ and $\chi=0.41$, hinting that these are all special points. This is expected for $\chi=0$ and $\chi=\infty$, as in these two cases one of the two terms in the Hamiltonian is completely removed. However it is quite a strong result for $\chi=0.41$ as it implies that the fidelity improvement is not a mere finite-size effect but a more fundamental change.

\begin{figure}[H]
	\centering
	\includegraphics[width=\linewidth]{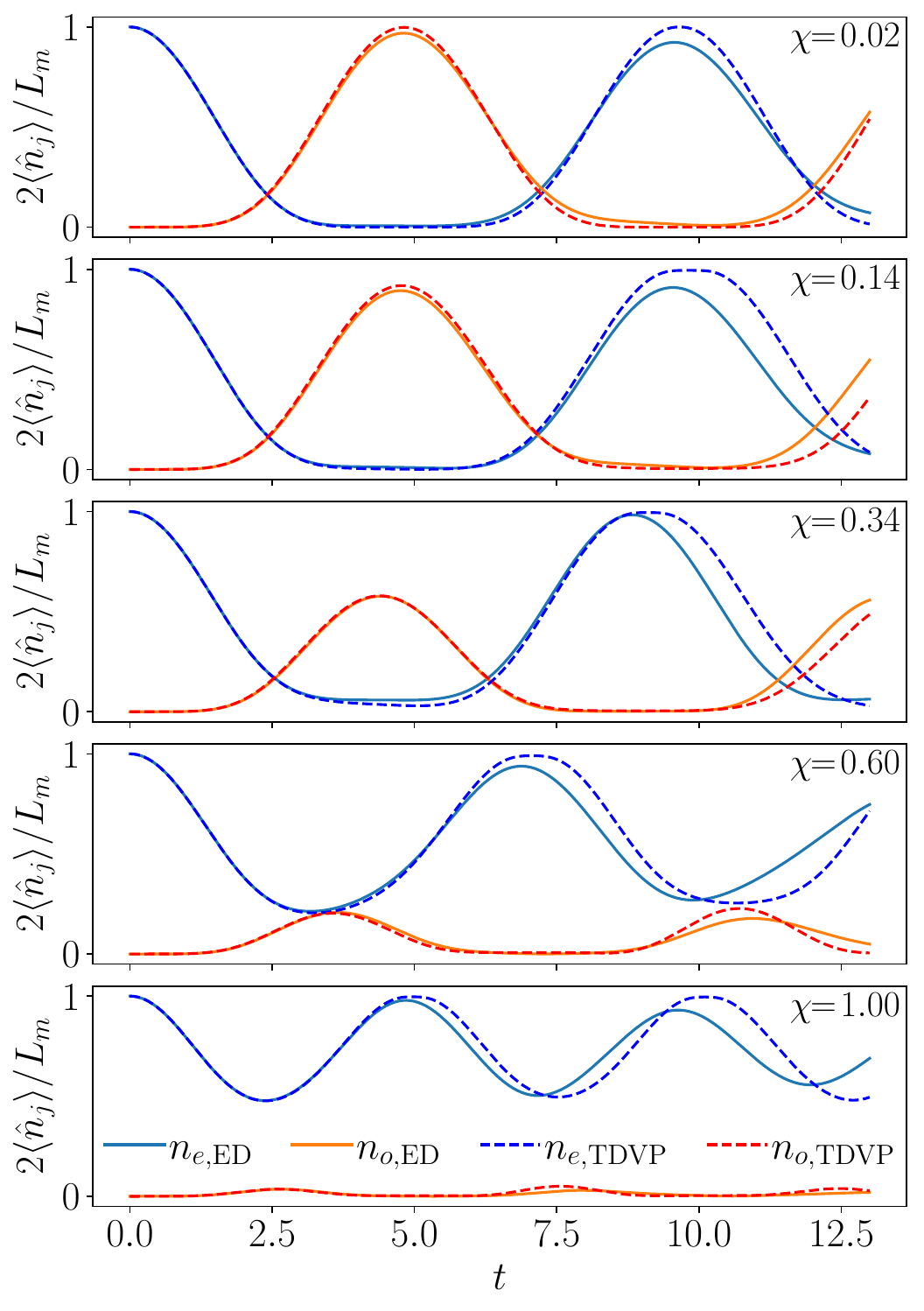}
	\caption{Sublattice occupation after a quench from the N\'eel state as obtained with ED and $L_m=20$ and with TDVP. The data agrees well up to the first revival for all values of $\chi$.}
	\label{fig:TDVP_comp}
\end{figure}

For an infinite-system, the dynamics in the symmetric subspace admits a classical description using the time-dependent variational principle (TDVP) Ansatz \cite{Ho2019,Michailidis2020,Turner21}. The quantum state is defined using four classical variational parameters: two $\theta$ angles that relates to the excitation density on each sublattice ($\theta_o$ and $\theta_e$) and two $\phi$ angles that describe phases ($\phi_o$ and $\phi_e$). This corresponds to two spin-coherent states (one on each sublattice), on which a projector was applied to eliminate all states that violate the PXP constraint. This Ansatz was developed to give a semi-classical limit for the N\'eel state revivals at $\chi=\mu=0$, but it also holds for $\chi\neq 0$. The equations of motion are the same as in Eqs.~(9a) and (9b) of Ref.~\cite{Michailidis2020} for $C=2$, except that now for $\mu_z=+\chi$ for $\dot{\phi_1}$ and $\mu_z=-\chi$ for $\dot{\phi_2}$. 
We integrate numerically the equations of motion starting from the N\'eel state, and compute the excitation density on the even ($\langle \hat{n}_e \rangle$) and odd ($\langle \hat{n}_o \rangle$) sublattices from them. Fig.~\ref{fig:TDVP_comp} compares the TDVP and exact diagonalization (ED) results in a system with $L_m=20$. The TDVP Ansatz assumes a system of infinite size, so we do not expect it to match ED result at longer times where finite-size effects kick in.  Nonetheless, we see a good agreement at shorter times, for all values of $\chi$. 

\begin{figure}[tb]
	\centering
	\includegraphics[width=\linewidth]{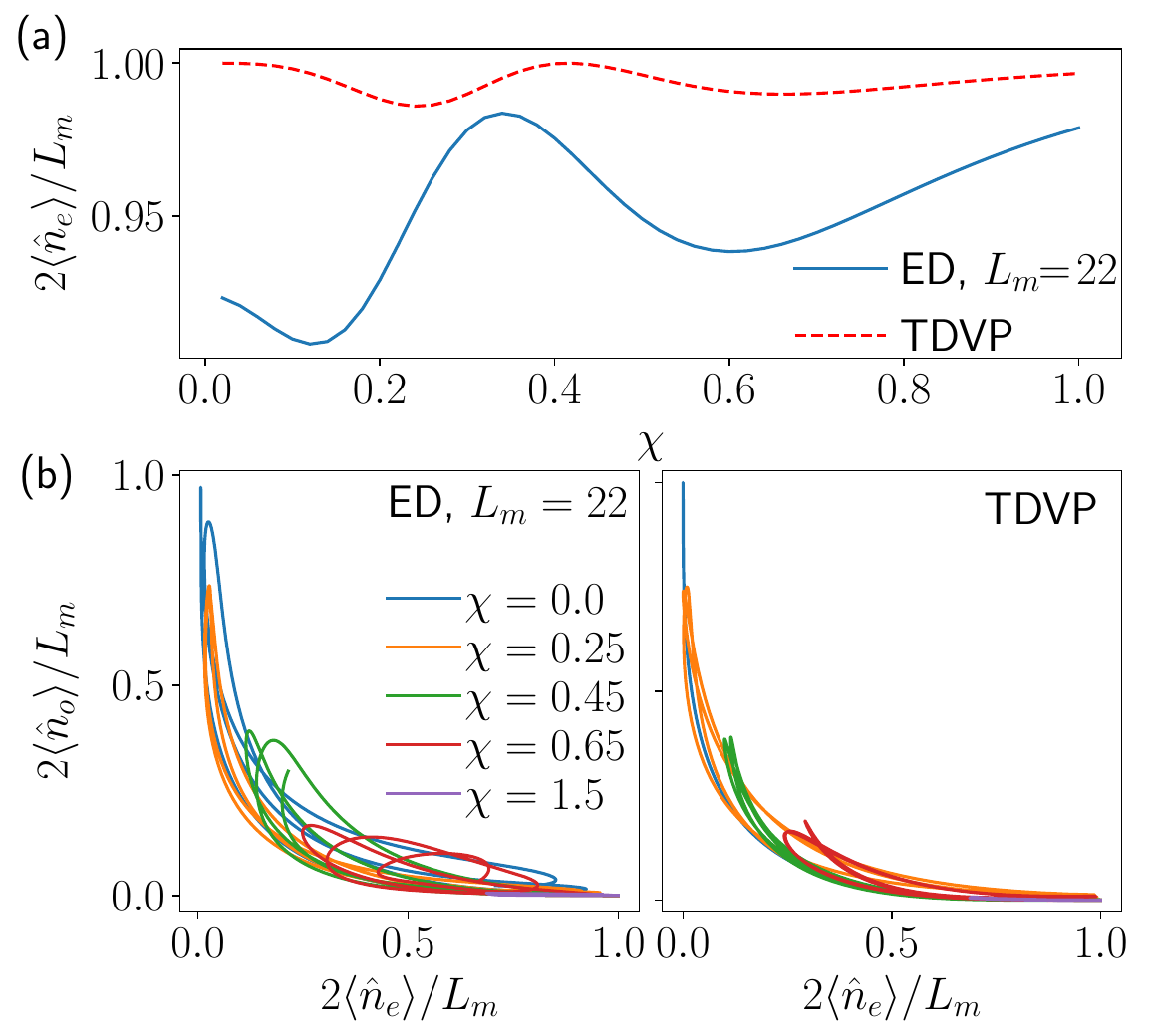}
	\caption{Comparison between TDVP and ED data for sublattice excitation density. Top: value of the even sublattice excitation density after one period. The same qualitative behavior can be seen in ED and TDVP data. Bottom: trajectory of the two sublattice excitation densities. The TDVP data shows trajectories that are closer to periodic.}
	\label{fig:TDVP_all}
\end{figure}

We note that there is a major difference between the cases $\chi=0$ and $\chi\neq 0$. For the former, the two $\phi$ angles are dynamically frozen when starting from the N\'eel state and there are only two independent degrees of freedom. As a consequence, chaos cannot develop and every trajectory is either periodic (like the N\'eel and anti-N\'eel one) or it goes asymptotically towards a fixed point. For $\chi\neq 0$, the $\phi$ angles are no longer frozen and the system has 3 independent degrees of freedom (taking into account energy conservation) and thus chaos can develop. We find that for $\chi\neq 0$ the trajectory is no longer exactly periodic, highlighting the instability of the N\'eel periodic orbit at $\chi=0$.
Despite this, we still find that the excitation density shows persistent oscillations. The expectation of $\langle \hat{n}_e \rangle$ after one period (meaning at its first local maximum) shows the same nonmonotonic behavior as in the quantum system, at least qualitatively, as shown in Fig.~\ref{fig:TDVP_all} (a).

In order to have a better understanding of the dynamics between the revivals, we can plot the expectation values of $\langle \hat{n}_o \rangle$ and $\langle \hat{n}_e \rangle$ against each other over time. This is shown on Fig.~\ref{fig:TDVP_all} (b) using data from both ED and TDVP. As expected, both sets of results give similar results. The TDVP data shows better revivals in the sense that successive ``loops'' with the same $\chi$ are almost on top of each other, so the orbit is close to periodic. 

These results provide an interesting example of quantum-many-body scars having a semi-classical with an orbit that is not periodic. This is unlike what has been seen for the N\'eel state with $\chi=0$ and for the polarized state with $\mu\approx \pm 0.4$. In both cases, the TDVP phase space only has two independent variables, and the classical orbits are exactly periodic. For single-particle scarring in quantum billiard, the classical orbits are also periodic, even though the phase space has enough degrees of freedom for chaos to develop. So the $\chi\neq 0$ case offers a different paradigm, and it would be interesting to explore this in more in more details.

\end{document}